\documentclass[onecolumn,preprintnumbers,amssymb,amsmath,superscriptaddress,letterpaper]{revtex4}[12pt]

\usepackage{graphicx}
\usepackage{dcolumn}
\usepackage{bm}
\usepackage{natbib}
\usepackage{pstricks}
\usepackage{epsfig}
\usepackage{rotating}
\usepackage{multirow}

\newcommand{\be}{\begin{equation}}
\newcommand{\ee}{\end{equation}}
\newcommand{\bea}{\begin{eqnarray}}
\newcommand{\eea}{\end{eqnarray}}

\newcommand{\gsim}{ \mathop{}_{\textstyle \sim}^{\textstyle >} }
\newcommand{\lsim}{ \mathop{}_{\textstyle \sim}^{\textstyle <}}

\begin{document}

\title{Extracting limits on dark matter annihilation from gamma-ray observations towards dwarf spheroidal galaxies}

\author{Ilias Cholis}
\email{ilias.cholis@sissa.it}
\affiliation{SISSA, Via Bonomea, 265, 34136 Trieste, Italy}
\affiliation{INFN, Sezione di Trieste, Via Bonomea 265, 34136 Trieste, Italy}
\author{Paolo Salucci}
\email{salucci@sissa.it}
\affiliation{SISSA, Via Bonomea, 265, 34136 Trieste, Italy}
\affiliation{INFN, Sezione di Trieste, Via Bonomea 265, 34136 Trieste, Italy}

\date{\today}

\begin{abstract}
Dwarf spheroidal galaxies compose one of the most dark matter 
dominated classes of objects, making them a set of targets
to search for signals of dark matter annihilation. Recent developments 
in $\gamma$-ray astronomy, most importantly the launch of the 
\textit{Fermi}-LAT instrument, have brought those targets into attention.
Yet, no clear excess of $\gamma$-rays has been confirmed from these targets,
resulting in some of the tightest limits on dark matter annihilation from 
indirect searches. In extracting limits from dwarf spheroidal galaxies, it is 
of great importance to properly take into account all relevant uncertainties.
Those include the dark matter distribution properties of the dwarf 
spheroidals and the uncertainties on the underlying background.
We revisit the limits on dark matter annihilation, from $\gamma$-rays studying
a set of close-by dwarf spheroidal galaxies for which we have good understanding 
of the uncertainties in the dark matter distribution.
For those targets, we perform and compare results for alternative methods in
extracting the background $\gamma$-ray flux. This provides a method to 
discriminate among the dark matter annihilation targets, those that can 
give robust constraints. We finally present our tightest limits on dark 
matter annihilation that come only from the targets that ensure accurate 
understanding of both the $\gamma$-ray background and the dark matter 
distribution uncertainties.
\end{abstract}
\maketitle

\section{Introduction}
\label{sec:intro}

Dark matter (DM) composes approximately $85 \%$ of the matter density 
in the universe, 
yet its particle physics properties in the Weakly Interacting Massive 
Particle (WIMP) case still remain unknown. Measurements of cosmic rays (CR) 
\cite{Boezio:2008mp, Adriani:2008zr, Adriani:2008zq,Collaboration:2008aaa, Aharonian:2009ah, Abdo:2009zk, ATIClatest} 
have generated new model building \cite{ArkaniHamed:2008qn, Cholis:2008qq, Harnik:2008uu, Bai:2008jt, Fox:2008kb, Grajek:2008pg, Pospelov:2008jd, MarchRussell:2008tu, Cirelli:2009uv, Shepherd:2009sa, Phalen:2009xw, Hooper:2009fj, Goh:2009wg, Cholis:2009va, Hooper:2009gm} 
and new constraints on dark matter properties\cite{Cirelli:2008pk, Donato:2008jk, Cholis:2010xb, Evoli:2011id, Bergstrom:2009fa}. 
Also, direct detection experiments 
\cite{Baudis:2007ew, Barbeau:2007qi, Aprile:2009yh, Ahmed:2009zw, Bernabei:2008yh, Angloher:2008jj} 
have provided their own set of constraints (or preferred regions) on the mass and interaction properties 
of DM particles with nucleons 
\cite{Angle:2008we, Kopp:2009qt, Bernabei:2010mq, Aprile:2010um, Aalseth:2010vx,Aalseth:2011wp, Angloher:2011uu} 
(see also \cite{Collar:2010gg, Fox:2011px, Hooper:2011hd}), 
providing possible insights on DM properties 
\cite{Fitzpatrick:2010em, Chang:2010yk, Chang:2010en, Buckley:2010ve, Belikov:2010yi, DelNobile:2011je}).

Recently strong constraints on DM annihilation have been put from observations at 
$\gamma$-rays towards dwarf spheroidal (dSph) galaxies \cite{Abdo:2010ex, Ackermann:2011wa}. 
DSph  galaxies are DM dominated objects, where the production of $\gamma$-rays 
from point sources and from interactions between CRs and the local medium is expected 
to be suppressed. This suppresion are due to the fact that baryonic gas 
densities are very low, star formation responsible for the
producion of CRs is suppressed and also because CR escape timescales from those 
galaxies are expected to be much smaller compared to galaxies of sizes 
similar to that of the Milky Way. 
Thus, dSphs provide some of the best targets to look for signals from DM annihilation
 \cite{2004PhRvD..69l3501E, Colafrancesco:2006he, Strigari:2006rd, Bovy:2009zs, Scott:2009jn, Perelstein:2010at}, 
that could possibly be probed with the current $\gamma$-ray telescopes.

Let us  consider a galaxy in which DM annihilates producing $\gamma$-rays.
In that case the relative flux is given by\footnote{for a Majorana DM particle}
\begin{equation}
\label{eq:GammaFlux}
\frac{d\Phi_{\gamma}}{dE} = \int \int \frac{\langle \sigma v \rangle}{4 \pi}
\frac{dN_{\gamma}}{dE}_{DM} \frac{\rho_{DM}^{2}(l, \Omega)}{2 \, m_{\chi}^2} 
dl d \Omega ,
\end{equation}
where $d \Omega$ is the solid angle within which the measurement is done, 
and $l$ the length along the line of sight to the object.  
For a homogeneous annihilation cross-section eq.~\ref{eq:GammaFlux} can 
be simplified to $d\Phi_{\gamma}/dE = \Phi^{PP} J $ where 
\begin{equation}
\label{eq:PhiPP}
\Phi^{PP} = \frac{\langle \sigma v \rangle}{4 \pi}
\frac{dN_{\gamma}}{dE}_{DM} \frac{1}{2 \, m_{\chi}^2}
\end{equation}
is a factor depending only 
on the elementary particle physics  underlying the annihilation of the DM particles 
with mass $m_{\chi}$, annihilating with a cross-section $\langle \sigma v \rangle$ 
and producing a spectrum of $\gamma$-rays per annihilation 
given by $\frac{dN_{\gamma}}{dE}_{DM}$. 
The $J$-factor: 
\begin{equation}
J= \int \int \rho_{DM}^{2}(l, \Omega) dl d \Omega,
\end{equation}
gives the  line of sight integral of the DM density-squared over the 
solid angle $d \Omega$  under which the measurement is done. 

Since limits from dSphs are on the $d\Phi_{\gamma}/dE$ from each dSph galaxy, 
the limits on DM annihilation cross-section depend strongly on the DM profile
assumptions for those dSphs, i.e the $J$-factors as we describe in 
section~\ref{sec:J-factors}. 
The $\rho_{DM}$ profile inserted for the calculation of the $J$-factors
 must be that  obtained from observations from those dSphs.
Over the past decade, observations have 
provided detailed information about the distribution of DM  within the  
regions of spiral galaxies where the baryons reside 
(\cite{PSS96, Salucci07}), suggesting a preference towards cored profiles
for the dwarf spheroidal galaxies, rather than NFW profiles. Since the 
$J$-factors depend on the $\rho^{2}_{DM}$, the choice on an NFW or in general
a cuspy profile (rather than a cored profile), may have an impact on 
suggesting greater values on the J-factors. 

Also, since no clear excess towards dSphs has been
observed, it is equally important to understand the $\gamma$-ray background
contribution. The $\gamma$-ray background spectrum is composed of the 
isotropic component -which is the sum of the extragalactic $\gamma$-ray 
background and misidentified CRs-, the diffuse $\gamma$-rays produced 
in our Galaxy from CR interaction with the interstellar medium (ISM), and 
also of point sources laying inside the observation window/Region of Interest 
(ROI).
In section~\ref{sec:FermiGamma} we discuss various methods of calculating 
residual $\gamma$-ray spectra at the location of the dSphs 
(that come from subtracting the modeled $\gamma$-ray
backgrounds from the total fluxes), and suggest 
alternative methods. In section~\ref{sec:methods} we further study individually 
eight dSph galaxies and the quality/robustness of the limits on DM annihilation 
based on the alternative methods of extracting residual spectra. 
Finally, in section~\ref{sec:constraints} we present the strongest and most robust 
conservative limits on DM annihilation from dSphs and give our conclusions 
in section~\ref{sec:conclusions}.  

\section{J-factors}
\label{sec:J-factors}

In disk systems  the ordered rotational motions and  the known geometry of 
the tracers of gravitational field  facilitate  the mass modeling. Moreover,
in dwarf galaxies the DM emerges clearly.   
It has been shown  that the fraction  DM/Luminous matter increases with 
radius and at a fixed radius it increases with decreasing luminosity 
\cite{deblok08}.
\cite{PSS96} it has suggested that the stellar disks are ``maximal'' 
and embedded in  
dark matter halos with a {\it cored} density distribution 
 ~\cite{PSS96}(see also~\cite{chemin11,gentile04,g05,g07}. 
Furthermore,  there appear scaling relations between  the structural 
properties of  the luminous  and of the dark mass components,
~\cite{Kormendy85,  Burkert95,KF04}.
In fact, results from the universal Rotaion Curve and individual 
Rotation Curves (RCs) of galaxies,  
\cite{Salucci:2000ps, Salucci07} 
imply that in galaxies the DM halo has  a  Burkert  density profile:
\begin{equation}
   \rho (r)={\rho_0\, r_0^3 \over (r+r_0)\,(r^2+r_0^2)},
\end{equation}
in which the  two free parameters, the core radius $r_0$ and the central halo
density $\rho_0$ are related.  In other words  RCs of spirals  and of Low 
Surface Brightness (LSB) galaxies  yield  a mass towards distribution that can be uniquely modeled by means  of a  luminous and a cored dark matter component.  \cite{Salucci07}.

The  knowledge of the mass distribution in
pressure-supported systems like ellipticals and   dwarf spheroidals  
 is much more uncertain.  The Local Group dwarf 
spheroidals galaxies (dSphs) occupy  the faint end of the luminosity function 
of pressure-supported systems, \cite{pena08, walker09a}.   
Their vicinity to us  makes them the best candidates for the detection of  
possible $\gamma$-ray emission arising from annihilating dark matter, in thier 
densest parts. However, deriving the  dSph  mass model is not an 
easy task  both observationally, in terms of measuring in each dSph a 
sufficiently large number of meaningful relative velocities of stars, and 
also  from a 
 dynamical modeling point of view, due the lack of precise information on the 
dynamical state of the latter.  
On the other hand,  dSphs are simple systems and are DM dominated at all radii 
that are at least two 
orders of magnitudes less luminous 
than the faintest spirals~\cite{kleyna02}. The DM halo typically
outweighs the baryonic matter by a large factor (from a few tens, up
to several hundred).    
No dark-visible (potentially uncertain)  mass decomposition is then  needed to derive the DM properties.   
Also, there is
evidence of universality in the DM halo structural properties of dSph galaxies 
\cite{Mateo98}  \cite{Gilmore07,Koch07a,strigari08,walker09b},   
in tension with the collision free particles  of the naive $\Lambda$CDM  scenario. 

The study of the  kinematics of the Milky Way dSphs has been
revolutionized by  multi-object spectrographs on 4m
and 8m-class telescopes.  Large data sets comprising between several
hundred and several thousand individual stellar velocities per galaxy,
have now been acquired for all the most  luminous dSphs surrounding the
Milky Way \cite{Munoz06,walker09a,Koch07a,Koch07b,Battaglia08}.
For these objects  reliable measurements of the (projected) dispersion 
velocity profile are  now available, so that  a mass model can be attempted 
\cite{Walker07,Koch07a,Koch07b,Battaglia08}. It is well known that  the   dSph  kinematics  can be made  
compatible  with the  gravitational potential of the cuspy  DM halos out of   cosmological
N-body simulations in the $\Lambda$CDM scenario (e.g., \cite{navarro97}). 
On the other hand, the same  kinematics are also compatible with shallower DM 
profiles \cite{walker09b}. The point being that in dSphs the kinematics 
alone (e.g. the velocity dispersions)  cannot discriminate among the various 
DM density profiles. 
The Jeans equation that relates the density and the velocity dispersion of the
stellar component to the mass profile of the dark matter halo
has a well-known degeneracy between the mass profile  and the  velocity 
anisotropy profile (see for instance\cite{Koch07a, Battaglia08,Evans09}).
In spite of these difficulties,  cored DM  profiles  seem to be  favored  in  very  recent  studies, 
\cite{Gilmore07,kleyna02,Goerdt06,Battaglia08,Amorisco11,Walker11}. 
However, also in these cases,  cuspier NFW profiles cannot be ruled out.

The situation for dwarf spirals is very different. These objects, DM 
dominated down to their inner regions,  have very reliable  
kinematics from their stellar and gaseous disks that, unambiguously point 
towards centrally flat density  dark matter  halos. The kinematics commonly  
suggest a cored  profile as is a Burkert profile or a similarly shallow one. 
As a matter of fact, for dwarfs spirals   we can  state that the NFW  halo velocity profile under no circumstances  can reproduce their rotation curve \cite{Gentile:2005de}.
 
Of equal importance is the fact that in the evolution of DM density in dSphs 
strong outflow events occur 
(due to some strong early supernovae explosions), that suppress the
baryonic infall at later stages.
This mass loss and input in the dSph galaxies is claimed to have largely 
modified the cosmological NFW  DM halo profile that has emerged from N-Body 
DM-only simulations in the  $\Lambda$CDM scenario  
 \cite{Maccio':2011eh, RagoneFigueroa:2012vc}, to shallower profiles.

For these reasons, in the estimate of  the $J$ factor in  nearby  dSphs, 
which is the first step  of constraining the mass and cross section of 
annihilating dark matter, it is  unjustified to assume that  the dSphs have 
dark matter halos  following  the  NFW profile. 
On this line \cite{2012MNRAS.tmp.2161S} have investigated the internal 
kinematics of the  Milky Way dSphs, by first taking the usual assumption that
the luminous component consists of a 
single pressure-supported stellar population in dynamical
equilibrium; therefore tracing the underlying gravitational
potential.  \cite{2012MNRAS.tmp.2161S} also assumed that the latter 
is dominated by the dark matter halo, leading to uts density profile 
to be easily  traced by the  Jeans equation, given by:
 
\begin{equation}
  \nu(r)v^2_{r}=Gr^{-2\beta}\displaystyle\int_r^{\infty}s^{2\beta-2}\nu(s)M(s)ds.
  \label{eq:jeanssolution}
\end{equation}
In  order to use  observables, the stellar component must be projected  along the 
line of sight \citep{bt08}:
\begin{equation}
  \sigma_p^2(R)=\frac{2}{I(R)}\displaystyle \int_{R}^{\infty}\biggl (1-\beta(r)\frac{R^2}{r^2}\biggr ) \frac{\nu(r) v_r^2(r)r}{\sqrt{r^2-R^2}}dr,
  \label{eq:jeansproject}
\end{equation}
where $I(R)$ is the projected stellar density profile and
$\sigma_p(R)$ is the projected velocity dispersion profile which are both known. 
$M(r)$ is the mass profile of DM, $\nu(r)$ describes the 3-dim stellar distribution
and $v_{r}(r)$ is the radial velocity of the stars.  
The orbital anisotropy $\beta(r)$ is
not constrained, as all information about the velocity distribution is
restricted to the component along the line of sight. 
\cite{2012MNRAS.tmp.2161S} did  the 
simplifying assumption that $\beta=\mathrm{constant}$, which  provides
the following solution to eq.~\ref{eq:jeansproject} \cite{mamon05},
and then use in eq.~\ref{eq:jeansproject}
rather than  the NFW mass/density profile,  that 
derived for a Burkert halo \cite{Salucci07}.
\begin{eqnarray}
  M(r)&=&4\pi\displaystyle\int_{0}^{r}s^2\rho(s)ds\nonumber\\
 & =&2\pi\rho_{0}r_{0}^{3}\left(\ln[(1+r/r_0)]+0.5\ln[(1+r^2/r_0^2)]-\tan^{-1}[r/r_0]\right).
  \label{eq:jeansmass}
\end{eqnarray}
Here the two DM structural parameters are its  central density and  its core
radius. In short \cite{2012MNRAS.tmp.2161S} {\it assumed} a cored  profile 
to represent the DM halo,   allowed a  radially constant velocity anisotropy 
different for every object and aimed to reproduce the observed dispersion profiles as closely as
possible. In detail \cite{2012MNRAS.tmp.2161S} obtain (marginalized)
1-D posterior probability distribution functions for each of the two DM profile free
parameters using a Markov-Chain Monte Carlo (MCMC) algorithm leading to  ``best fits'' 
values for these parameters. These results demonstrate that Burkert profiles can provide an 
excellent description of dSph velocity dispersion profiles and that the available data 
quite well 
constrain  the structural parameters. In Table 2 of \cite{2012MNRAS.tmp.2161S} 
the authors provide  the  individual values  of $r_{0}$ , $\rho_{0}$ and their best 
fit uncertainties for eight galaxies, which we study in the following.

The  reliability of the derived mass  model is  also  given by the fact that  
although the DM densities in dSphs are about two orders of magnitude higher 
than those found in (larger) disc systems (by comparison to the stellar ones), 
their  DM  halos best fit values for the structural parameters lay on the extrapolation 
(at lower masses), of the same $r_0$ and $\rho_0$  relationship 
found  to hold for  spirals and ellipticals \cite{Donato:2009ab}: 
\begin{equation}
   r_{0} = 8.6  (\rho_0/(10^{23} g cm^{-3})^{-1} kpc
\label{eq:r0Overrho0}
\end{equation} 

We note that based on eq.~\ref{eq:r0Overrho0} the error propagation of the two 
quantities are not statistically independent. 
The product of $\rho_{0} r_{0}$ varies at most by $50\%$ in galaxies 
\cite{Donato:2009ab}.
This uncertainty is the major source of uncertainty in the evaluation of the 
$J$-factors for the targets calculated at $\alpha_{c}$ 
(given in table~\ref{tab:J-factors}) and defined as:
\begin{equation}
\alpha_{c} = \frac{2 \overline{r_{half}}}{D},
\end{equation}
with $\overline{r_{half}}$  being the half light radius of the 
stellar population in the dwarf and $D$
it's distance to us. The induced 
uncertainty in the $J$-factor from uncertainty of the distance $D$ from us 
is subdominant by comparison.

\begin{table}[t]
\begin{tabular}{|c|c|c|c|c|c|c|c|c|}
\hline
dSph& $D$ (kpc)& $\delta D$ (kpc)& $l$ & $b$ & $\overline{J} \times 10^{17}$ ($GeV^{2} cm^{-5}$)& $\delta J_{high} \times 10^{17}$ ($GeV^{2} cm^{-5}$) & $\delta J_{low} \times 10^{17}$ ($GeV^{2} cm^{-5}$) & $\alpha_{c}$ \\
\hline \hline
Carina & 103 & 4 & 260.1 & -22.2 & 2.69 & 0.47 & 0.54 & $0.27^{\circ}$ \\
\hline
Draco & 84 & 8 & 86.4 & 34.7 & 29.2 & 7.52 & 5.84 & $0.27^{\circ}$ \\
\hline
Fornax & 138 & 9 & 237.1 & -65.7 & 5.66 & 1.38 & 1.51 & $0.56^{\circ}$ \\
\hline
LeoI & 247 & 19 & 226.0 & 49.1 & 3.07 & 1.00 & 1.51 & $0.11^{\circ}$ \\
\hline
LeoII & 216 & 9 & 220.2 & 67.2 & 3.98 & 8.11 & 3.52 & $0.08^{\circ}$ \\
\hline
Sculptor & 87 & 5 & 287.5 & -83.2 & 18.3 & 4.08 & 3.85 & $0.42^{\circ}$ \\
\hline
Sextans & 88 & 4 & 243.5 & 42.3 & 23.6 & 58.3 & 17.2 & $0.89^{\circ}$ \\
\hline
Ursa Minor & 74 & 12 & 105.0 & 44.8 & 25.0 & 18.9 & 17.7 & $0.49^{\circ}$ \\
\hline \hline
\end{tabular}
\caption{Dwarf Spheroidal galaxies used in this analysis. For the $J$-factors
we give the mean value from the fit of \cite{2012MNRAS.tmp.2161S} $\overline{J}$
and the upper and lower 1 $\sigma$ uncertainties $\delta J_{high}$, $\delta J_{low}$,
all evaluated within an angle of $\alpha_{c}$.}
\label{tab:J-factors}
\end{table}

We refer to the original paper of \cite{2012MNRAS.tmp.2161S} for details but it 
is worth to discuss explicitly the role played by velocity
anisotropy in the derived modeling. Under the assumption of
Burkert haloes for dSphs the inclusion of velocity anisotropy as a
free parameter improves the quality of the dispersion profile fits
relative to those obtained for isotropic models. However, \cite{2012MNRAS.tmp.2161S} 
we also find that the scatter in the $\rho_0$-$r_0$ relation is smaller for
anisotropic models, thus the better we reproduce the observed
dispersion profiles using Burkert haloes, the tighter is the
correlation between the halo parameters.  

\begin{figure*}[t]
\centering\leavevmode
\includegraphics[width=4.00in,angle=0]{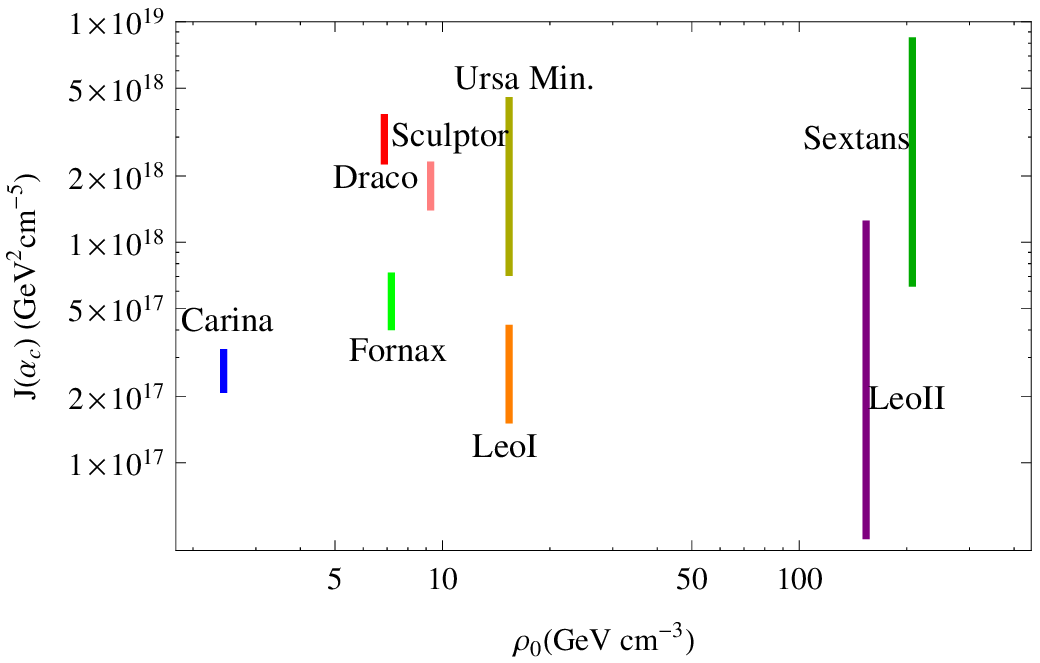}
\includegraphics[width=4.00in,angle=0]{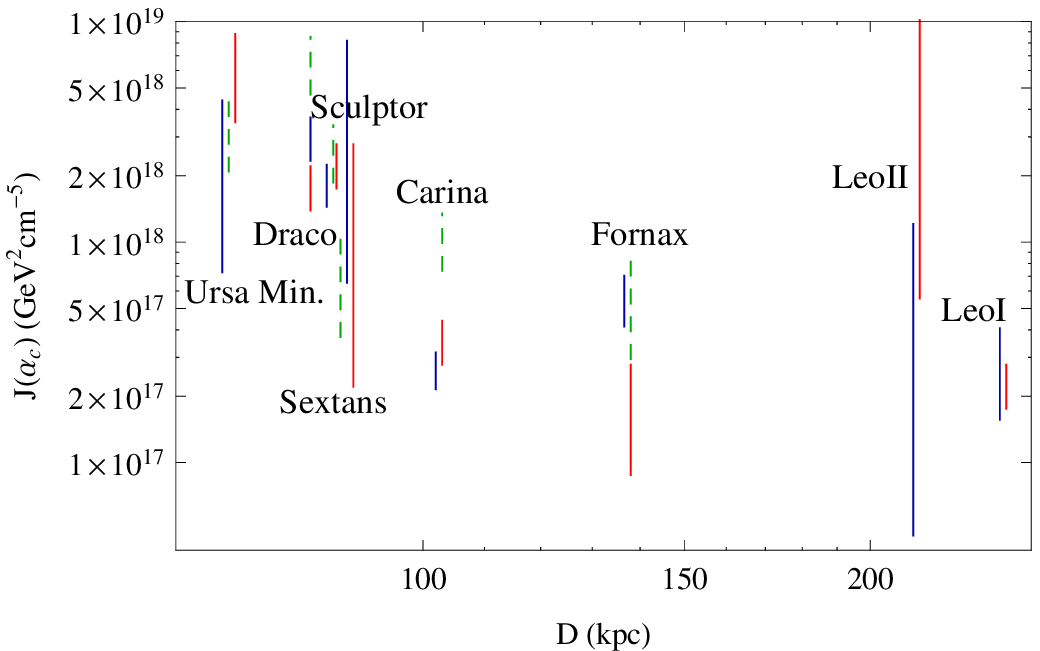}
\caption{\textit{Top}: The $J$-factor at $\alpha_{c}$ vs the best fit value for $\rho_{0}$ 
from \cite{2012MNRAS.tmp.2161S}. \textit{Bottom}: Comparison between J-factors in that work: 
\textit{dark blue} 
(from \cite{2012MNRAS.tmp.2161S}),in the work of \cite{Charbonnier:2011ft}:
\textit{red} and the assumptions of \cite{Ackermann:2011wa} (recalculated within $\alpha_{c}$): 
\textit{dashed green}.}
\label{fig:JfactorsComparison}
\end{figure*}

In Fig.~\ref{fig:JfactorsComparison} (upper plot) we give the results on the 
$J$-factors from \cite{2012MNRAS.tmp.2161S} versus the best fit value for 
the central density. There is a clear correlation between the uncertainty on 
the value of the $J$-factor and the best fit value for $\rho_{0}$, where larger values 
for $\rho_{0}$, are related to larger uncertainty on
the evaluated $J(\alpha_{c})$. We note that both the uncertainties and 
the $\rho_{0}$ value, come from the MCMC 
fit to the data, i.e.  the velocity dispersion profiles for the eight dwarf 
spheroidal galaxies. Yet, such a correlation is not a complete coincidence.
Given that all dSphs lay at $\rho_{0} r_{0} \approx \, const.$ (our 
eq.~\ref{eq:r0Overrho0}), larger $\rho_{0}$ result in lower values of
$r_{0}$ and $\overline{r_{half}}$. Thus smaller in size DM halos. For the 
larger DM halos the stars distributed in the inner part and generally within
 the inner kpc (see \cite{2012MNRAS.tmp.2161S})
\footnote{with the exception of Fornax and Draco where velocity dispersion data
extend out to $\approx$1.5 kpc see Fig. 1 of \cite{2012MNRAS.tmp.2161S}}, 
probe better the inner part of the actual DM halo profile, which results in 
smaller uncertainties for the $J$-factors.

In Fig.~\ref{fig:JfactorsComparison} (bottom) we also compare the results of
\cite{2012MNRAS.tmp.2161S} which are our reference results for the $J$-factors,
 to those of \cite{Ackermann:2011wa} and \cite{Charbonnier:2011ft}.
As is clear, \cite{Ackermann:2011wa} has a tendency in assuming smaller 
uncertainties in the $J$-factors than \cite{2012MNRAS.tmp.2161S} and
 \cite{Charbonnier:2011ft} do. Such an assumption can influence the strength of 
the limits on DM annihilation rates, on top of the fact that in \cite{Ackermann:2011wa} 
a cuspy NFW profile has been used.

Finally, we want to emphasize that the results of \cite{2012MNRAS.tmp.2161S}, 
do not {\it require} the presence of cored haloes in dSphs, nor do they constrain the
density and scale lengths of their haloes in a model-independent way.
On the other hand, the fact that the dSph kinematics can be reproduced
using Burkert DM halo profiles whose structural parameters lie on the
same scaling relations as those of spirals,  provides  support for the 
  assumption   that the mass distributions in dSph galaxies has the  same 
framework of  those of spirals. This can replace the idea that the dSph follow 
the profiles arising from N-Body simulations in LCDM scenario.
\footnote{It is usual to stress that while many cored models are as really 
indistinguishable when compared to the observed  kinematics in galaxies, 
for instance the Plummer profile, or some cases the pseudo-isothermal profile;
 the Zhao model sometimes used in modeling dSph is not the correct 
extrapolation of a NFW profile that gets a inner flat core. In fact, 1) 
3 DM free parameters are to many in view of  the qualitatively limited  
amount of information data available,  the best fit solution is prone to bad 
degeneracies of the parameter values 2) in cored models (e.g. in spiral) 
there no support to the fact  that the large scale and central  DM  density 
profiles are somewhat related, an unavoidable consequence of the Zhao model.} 

Let us stress that in previous works, faint objects like Segue I have been 
used to constrain WIMP masses and cross sections. However, for these objects, 
presently, we do not have a dispersion velocity profile of their stellar 
component, but only a measure of an ``average'' dispersion, that, in addition 
cannot even be attributed to a particular radius. Therefore, for 
them, an analysis of investigating different mass profiles has no 
meaning. A single measurement at an  unspecified radius does not allow us to 
build a reliable mass model, even by taking a number of assumptions.  
Since we have no way of estimating the total dark mass, let alone the  
DM distributions, of Segue I, Ursa Major II and Coma Berenices, we leave them 
for future work.

\section{Fermi gamma-ray data}
\label{sec:FermiGamma}

In \cite{Abdo:2010ex, Charbonnier:2011ft, :2010zzt, Ackermann:2011wa},  
it has been suggested that no clear excess of $\gamma$-rays  between 
200 MeV and 100 GeV (above the expected background) has been measured 
towards known dwarf spheroidal galaxies. Thus strong constraints on the 
DM annihilation can be placed. Yet, since the expected DM signal is smaller 
than the modeled background, the exact assumptions made to calculate the 
background $\gamma$-rays can be crucial in finding or hiding a DM signal. 

Since we want to set conservative limits on DM annihilation, we will ignore
a possible contribution to the $\gamma$-ray flux from each dSph from other
sources of $\gamma$-rays in the dSph, considering that the entire
$\gamma$-ray flux from each dSph comes from annihilating DM. Yet in the
same angular window there is background  flux from unrated sources unrated 
to the dSphs.
This $\gamma$-ray background originates from the galactic 
diffuse $\gamma$-rays, the isotropic $\gamma$-ray background, the contribution 
of nearby point sources and finally from CRs misidentified for $\gamma$-rays. 
At lower energies different sources can overlap with the possible 
contribution from the dSph due to the poorer angular resolution of the 
$\textit{Fermi}$ LAT instrument at low energies. The CRs contribution 
can vary based on different selection criteria of $\gamma$-ray events (event class). 
Using tighter criteria on the selection of $\gamma$-ray events, a 
cleaner sample can be achieved at the expense of lower statistics 
\footnote{See: http://www.slac.stanford.edu/exp/glast/groups/canda/lat\_Performance.htm}.
As mentioned in the introduction, we study the effects on the induced
limits on the DM annihilation, from different assumptions around in modeling 
the $\gamma$-ray background around each target, and in choosing different 
classes of $\gamma$-ray events.

According to the \textit{Fermi} LAT site
\footnote{http://fermi.gsfc.nasa.gov/ssc/data/analysis/documentation/Pass7\_usage.html}, 
there are three classes of events that can be used for $\gamma$-ray analysis, the 
"SOURCE", the "CLEAN" and the "ULTRACLEAN" classes of events, with the SOURCE class 
having the highest contamination of CR events (among those three classes) and the 
ULTRACLEAN the lowest. We will use all three classes to derive excess $\gamma$-ray 
fluxes from the direction of the targets. We will also study the effects in choosing 
different combinations of Regions Of Interest (ROI), for the $\gamma$-ray flux
of each dSph target and its relevant $\gamma$-ray background.
In our analysis we used 3 years of data taken between Augist 2008 and August 2011
with the PASS7 classes criteria.

We will first use the SOURCE class of events as was done by the \textit{Fermi} Collaboration 
in \cite{Abdo:2010ex, Ackermann:2011wa}, with the analysis provided in the Fermi 
ScienceTools\footnote{http://fermi.gsfc.nasa.gov/ssc/data/analysis/scitools/}.
In \cite{Ackermann:2011wa} for every dSph of interest an initial ROI of $10^{\circ}$ 
radius centered at each target was selected. Then a binned Poisson likelihood 
fit was performed "to both spatial and spectral information in the data"
\cite{Ackermann:2011wa}, using $10^{\circ}$ square spatial maps at energies 
between 200 MeV and 100 GeV. In those fits, the normalizations of the isotropic 
diffuse and the galactic diffuse $\gamma$-ray components 
\textit{were left free in all ROIs}, and so where the normalizations of the point sources 
within $5^{\circ}$ from the dSph.

We point out that in a region of $10^{\circ} \times 10^{\circ}$ where as we show in 
Table~\ref{tab:J-factors} (see $\alpha_{c}$), all dSph are practically
point sources\footnote{For energies 100 MeV $<E_{\gamma}<$ 10 GeV the containment angle
at normal incidence for SOURCE and CLEAN class is between $0.2^{\circ} - 6^{\circ}$ at $68\%$ containment and 
$0.7^{\circ} - 20^{\circ}$ at $95\%$ containment; being smaller at the higher energies.
Between 10 and 100 GeV, the $68\%$ ($95\%$) containment angle varies much less, between 
 $0.2^{\circ} - 0.3^{\circ}$ ($0.7^{\circ} - 0.8^{\circ}$), being again smaller at higher 
energies. See:  http://www.slac.stanford.edu/exp/glast/groups/canda/lat\_Performance.html for a plot and http://fermi.gsfc.nasa.gov/ssc/data/analysis/documentation/Cicerone/Cicerone\_LAT\_IRFs/IRF\_PSF.html  for parametric description for all classes and for front/backcoverted events. } 
and thus their size is defined by the PSF at each relevant energy, 
choosing to vary the normalizations of the two diffuse components and the close-by 
point sources, can lead in hiding any DM signal and thus lead to strong constraints on DM annihilation.
In fact, the \textit{Fermi} Collaboration has released the Isotropic Diffuse Gamma-Ray 
spectrum \cite{Abdo:2010nz}, with an analysis on the expected contamination from CRs 
for different cuts on events. Thus instead of allowing the isotropic diffuse component 
to vary, one should use a \textit{fixed} spectrum that would be the sum of the Extra 
Galactic Background $\gamma$-ray spectrum as has been also given in \cite{Abdo:2010nz}, 
and the CR contamination spectrum relevant for the SOURCE class of events.
Regarding the galactic diffuse component, we know it is the sum of $\gamma$-rays 
produced from $\pi^{0}$ decays and to a smaller extend decays of other mesons produced 
at $p-p$, $p-He$, $He-p$, $He-He$ collisions between CRs and the ISM targets, bremsstrahlung 
off CR $e^{\pm}$ interacting with the ISM gas and up-scattering of CMB and the interstellar 
radiation field photons from CR $e^{\pm}$. Its contribution within each ROI should also 
come from a fixed physical model such as those done in \cite{Strong:2004de, Cholis:2011un}
where also agreement with CR measurements has been confirmed and not with the galactic 
diffuse model of "gal\_2yearp7v6\_v0" that is let also to have a free normalization in the minimization 
fit. The same arguments apply also to the point sources normalizations and pawer-laws 
(which are also let free).

Moreover, these targets of annihilating DM are expected to be strongly 
subdominant components of $\gamma$-rays below $E_{\gamma} \sim$ 
GeV for the case of WIMP masses $m_{\chi}\approx 10$ GeV. 
As a result including in the analysis data bellow 1 GeV that have by far more statistics than above 1 GeV
will let the fit to be dominated by the low energy data. Just the $1 \sigma$ errors on the flux at low energies
whether statistical fluctuations or of systematic origin, will be significant in suggesting the 
presence or absence of residuals at high energies where the DM signal may lie. 
An example of a possible systematic error at low energies would be, a dim point source with a hard 
spectrum and a cut-off at $E_{\gamma} \sim 1$ GeV, as are millisecond pulsars (MSPs) 
\cite{2009Sci...325..848A} that has not been included (being not a known source), 
such a source could cause 
a fluctuation to the low energy data. MSPs has been observed to have a spectrum given by:
\be
\frac{dN_{\gamma}}{dE} \sim E^{-\Gamma} e^{-E/E_{c}},
\ee
with $\Gamma = 1.5 \pm 0.4$ and $E_{c} = 2.8 \pm 1.9$ GeV and luminosity in $\gamma$-rays of 
$L = 10^{33.9 \pm 0.6}$~erg/s \cite{2009Sci...325..848A}. 
In the Galaxy their contribution to the EGBR is expected to be $O(0.1)$ \cite{SiegalGaskins:2010mp, 
Calore:2011bt} between 0.5 and few GeV giving a flux of:
\be
E^{2} \frac{dN_{\gamma}}{dE} \approx 5\times 10^{-4} \, MeV \, cm^{-2} s^{-1} sr^{-1},\; \textrm{between 0.1 and 10 GeV.}
\ee
Following the assumptions of \cite{Malyshev:2010xc} that would lead to a population of $\sim 10^4$ MSPs 
in the entire sky. That makes the probability of one unidentified MSP within a region of 
$10^{\circ} \times 10^{\circ}$ to be $O(1)$ even if their distribution in the sky
is not truly isotropic since they are galactic sources.
We clarify also that the large scale distribution properties and the impact 
of the averaged spectral properties of the MSP has been included in the combination 
of the Isotropic and galactic diffuse template components that are fitted to 
the actual $\gamma$-ray data. Yet the combination of these components can not 
account for all posible structures in the $\gamma$-ray events, which is the 
reason why, after all they have been used in the searches for unknown point
sources \cite{Collaboration:2010ru, Collaboration:2011bm}.

The case for detecting a DM signal may be even more difficult for light DM 
cases where both the DM and the backgrounds have their main contribution at low energies, with the
backgrounds though, being dominant. We note also that below 1 GeV the containment angle of the 
\textit{Fermi} LAT instrument is $> 1^{\circ}$ ($> 3^{\circ}$) at 68$\%$ (95$\%$) at normal 
incidence for PASS7 SOURCE and CLEAN  class data (P7SOURCE\_V6, P7CLEAN\_V6), thus the dSphs are dimmed even further at
low energies by having their luminosity below 1 GeV being spread at a wider angle. 
 
For all these reasons we are concerned that the contribution of dim DM point-like sources can be
hidden by a background with many available degrees of freedom in the minimization procedure, and 
thus lead to very tight constraints on annihilating DM from each dSph. 
For a significantly brighter source as those detected by the \textit{Fermi} LAT, 
\cite{Abdo:2009mg, Collaboration:2010ru, Collaboration:2011bm} that 
may not be as much of an issue, since also for their joint optimization analysis 
there are many more ROIs (there are 1873 point sources in the 2 yr catalogue 
\cite{Collaboration:2011bm}) sharing data and the information of already known 
point sources \cite{Collaboration:2011bm}.
  
\begin{figure*}[t]
\centering\leavevmode
\includegraphics[width=3.20in,angle=0]{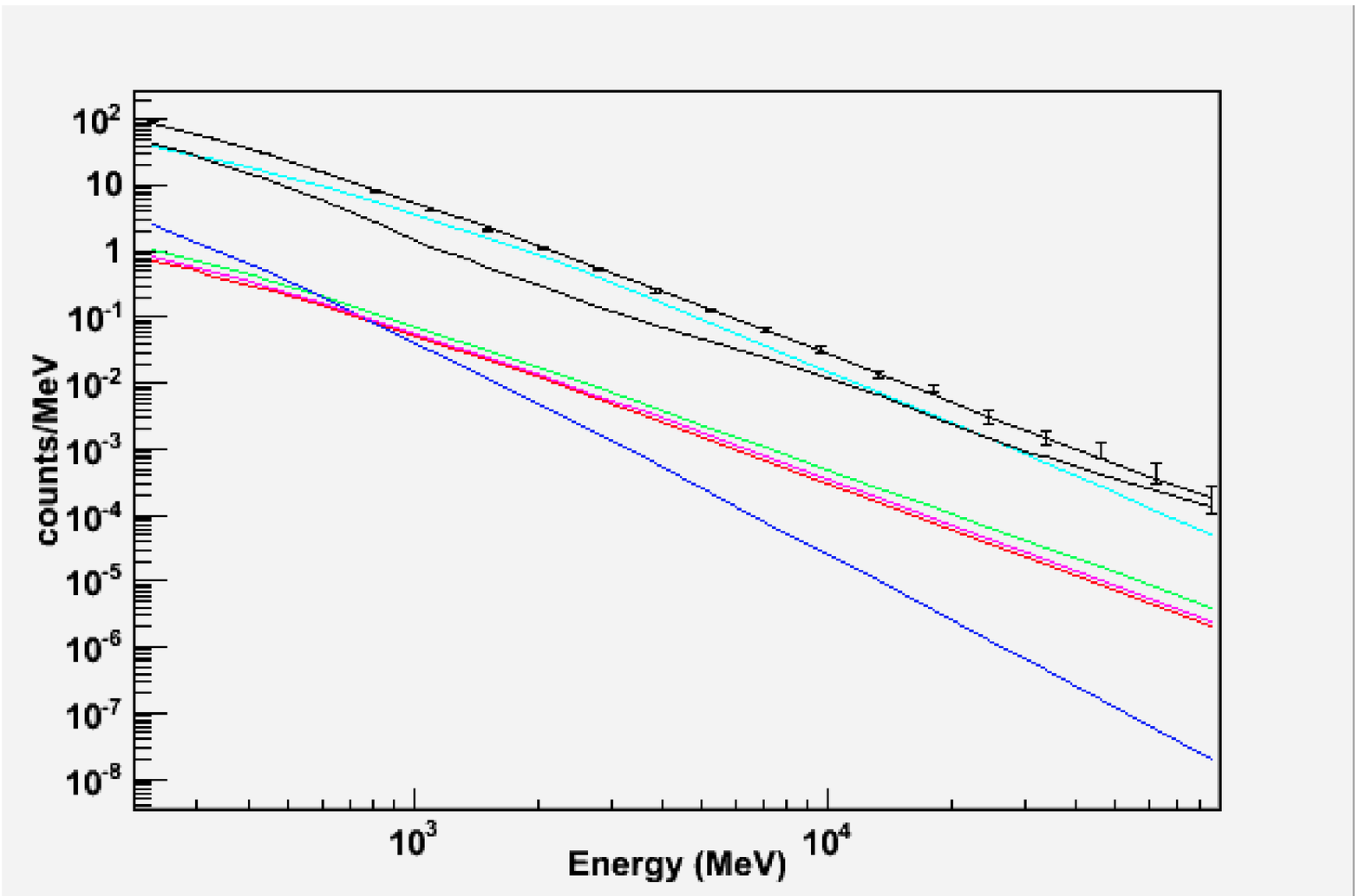}
\hspace{-0.1cm}
\includegraphics[width=3.20in,angle=0]{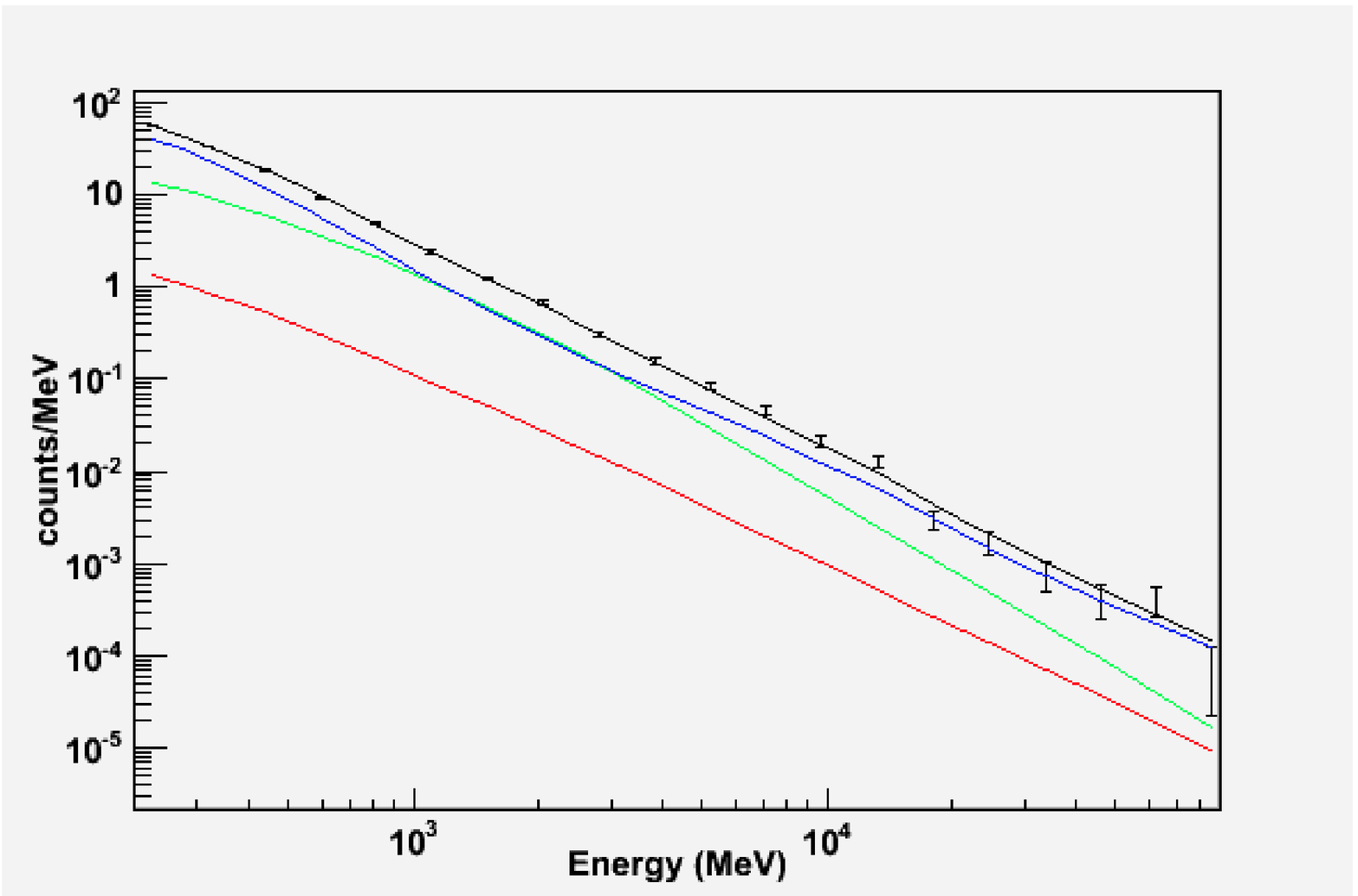}\\
\includegraphics[width=3.20in,angle=0]{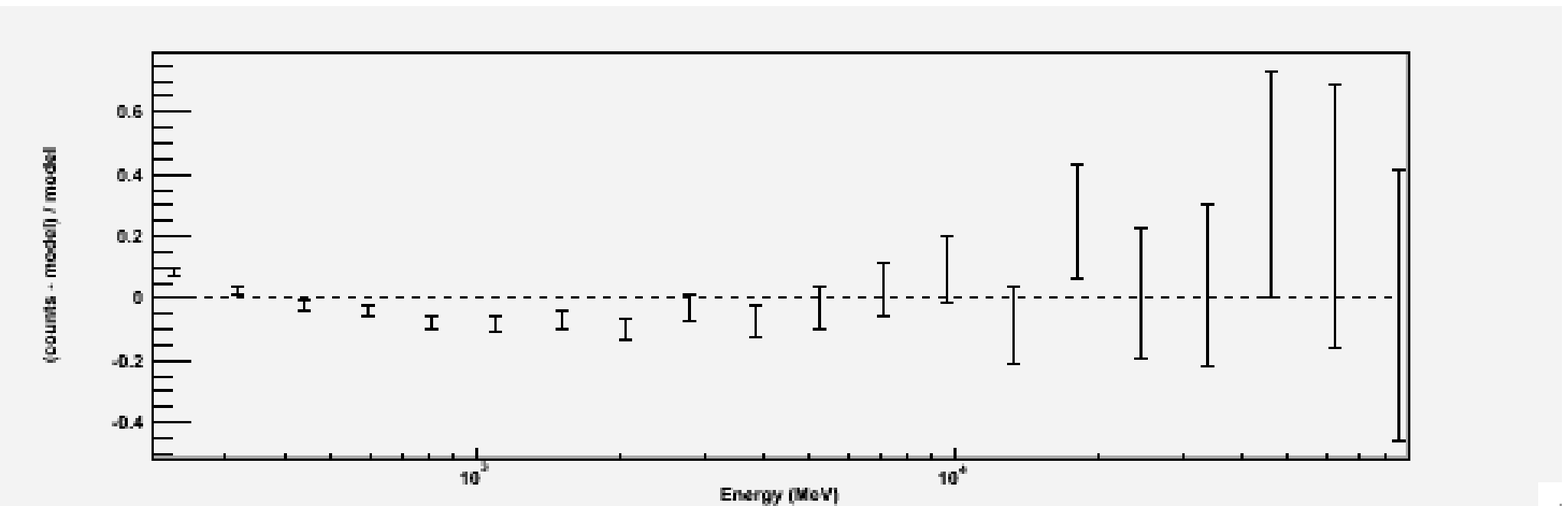}
\hspace{-0.1cm}
\includegraphics[width=3.20in,angle=0]{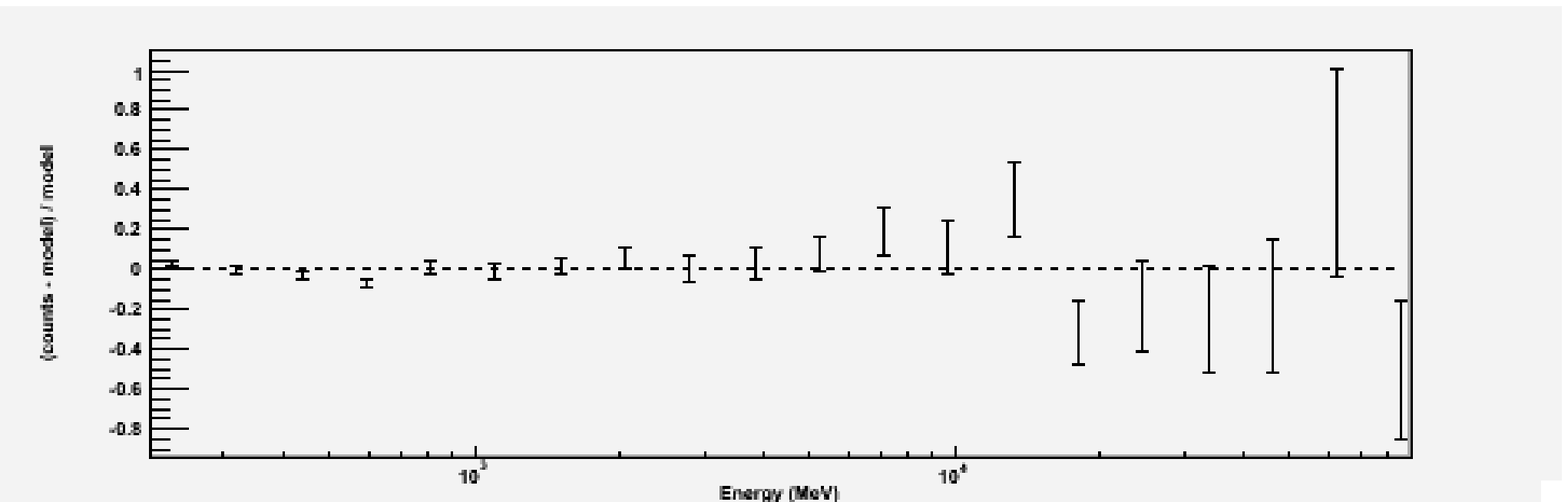}
\caption{Total (counts/MeV) and residual (counts-model)/model spectra from the 
minimization procedure of Fermi tools P7. We used SOURCE class data within $5^{\circ}$ 
from the relevant dSph position, and within energies of 200 MeV and 100 GeV (separated 
in 20 energy bins). \textit{Left}: Draco, where apart from the isotropic and the galactic
diffuse components, four known point sources contribute within $5^{\circ}$. \textit{Right}: 
Sculptor, where one point source has been detected.}
\label{fig:Fermi_dwSp_Spectra}
\end{figure*}

We want to also note that in \cite{Ackermann:2011wa} a joint likelihood analysis was done, 
where 10 ROIs where combined. As is written in the "Data Analysis" section of \cite{Ackermann:2011wa},
the normalizations of the nearby point sources and of the diffuse gamma-ray sources as well as 
the $J$-factors of the dSphs are among the "ROI-dependent model parameters", i.e. 
\cite{Ackermann:2011wa} still allow for freedom in the normalizations of the diffuse components,
something that is certainly wrong for the isotropic one which is also the most important at high 
latitudes where many of these targets lay (see Fig.~\ref{fig:Fermi_dwSp_Spectra}).    

Finally we point out that the joint likelihood analysis is made to account for uncertainties in the $J$-factors of 
the dSphs, which in fact can still not correct for a possible systematic bias towards higher 
values of $J$-factors coming from assuming an NFW DM profile.

\begin{figure*}[t]
\centering\leavevmode
\includegraphics[width=3.40in,angle=0]{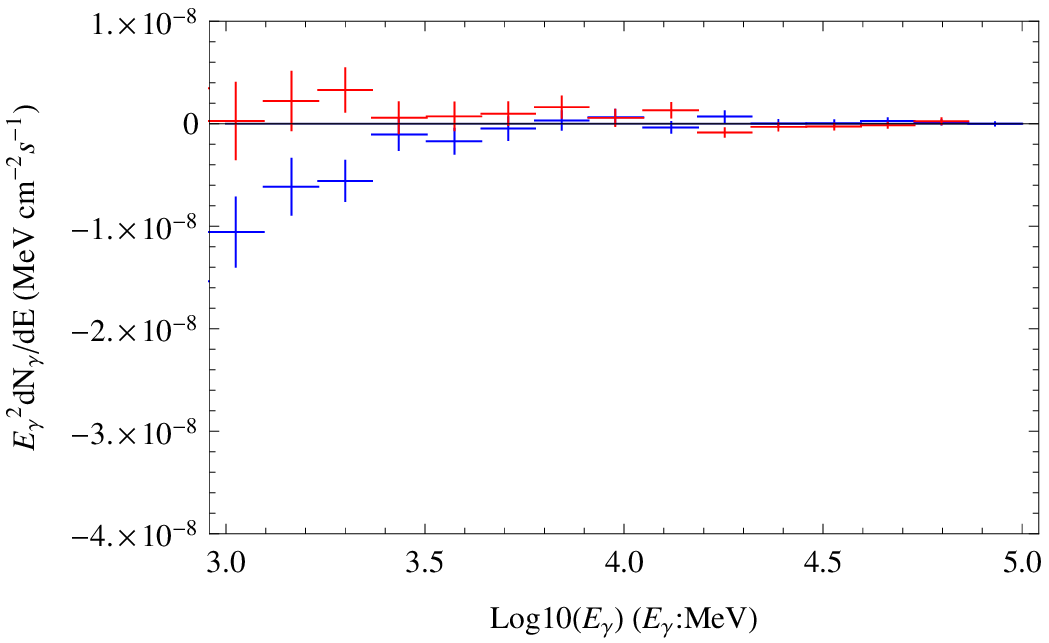}
\hspace{-0.1cm}
\includegraphics[width=3.40in,angle=0]{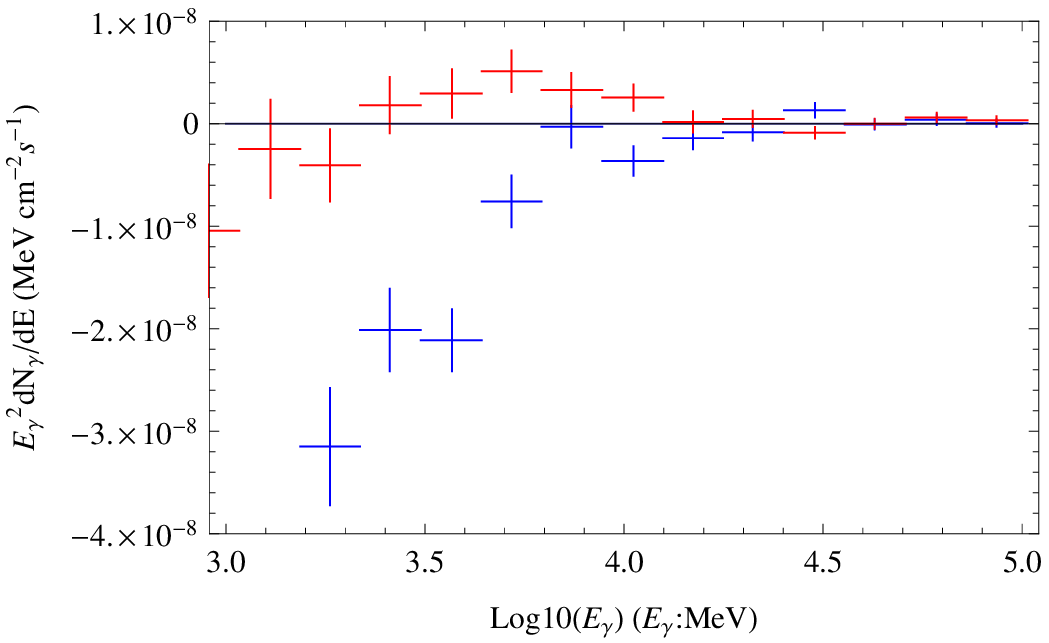}
\caption{Excess gamma-ray differential flux for Draco (dark blue) and Sculptor (red) dwarf spheroidal galaxies, using
\textit{Fermi} data ``SOURCE'' class. We present the spectra between 1 GeV and 100 GeV.
\textit{Left:} Using in the minimization procedure all data between 200 MeV and 100 GeV, within a radius of $5^{\circ}$, as presented in Fig.~\ref{fig:Fermi_dwSp_Spectra}.
\textit{Right:} Using all data between 100 MeV and 100 GeV, within a radius of $10^{\circ}$}
\label{fig:ExcessGammasFermiColl}
\end{figure*}

As an example using the publicly available \textit{Fermi} ScienceTools we calculated 
the residual spectra shown in Fig.~\ref{fig:Fermi_dwSp_Spectra} and~\ref{fig:ExcessGammasFermiColl}
for Draco and Sculptor with all 
the relevant point sources within $5^{\circ}$ from each dSph galaxy. For Draco which is one of 
the targets that provides the strongest limits on DM annihilation cross-sections, there are 4 
relevant point sources (see top left of Fig.~\ref{fig:Fermi_dwSp_Spectra}). 

As a test to the sensitivity of the minimization procedure that gives 
residual spectra we used the region of the sky centered at Draco
(not including the DM contribution from Draco) and decided to add 
one by one the known point sources from brightest to dimmest (defined by the fit when all
known sources are included). Such a choice of a test can be motivated by the possibility 
of some yet undetected point source contributing to the total $\gamma$-ray flux within the ROI.
We also decided to change the angle where the minimization is carried from $5^{\circ}$ to $10^{\circ}$,
where 15 point sources have been discovered. We also increased minimally 
the energy range for the 
minimization, from 200-10$^{5}$ MeV to 100-10$^{5}$ MeV within the $10^{\circ}$ radius case; 
in order to check the sensitivity of the results on varying the relevant assumptions. 
Finally we have fixed the normalization of the isotropic component to 1 and for different 
choices of ROI and energy range redone the fit.   
For every case we give in Table~\ref{tab:Draco_FermiColl} the normalizations of the isotropic 
diffuse, the galactic diffuse and the normalization and index for the closest to 
Draco detected point source J1725.2$+$5853 (``J1725'')\footnote{J1725 is not the brightest 
point source defined from the minimization procedure, as the one with the highest integrated flux 
for each case of assumptions on ROI and energy range.} that come from the fit,
using the ScienceTools. As can be seen the normalization of the galactic diffuse and the isotropic 
diffuse can have a change between fits of $O(0.1)$, which is far more than the residual signal.
\begin{table}[t]
\begin{tabular}{|c|c|c|c|c|c|c|c|c|}
\hline
ROI & $E_{min}$ (MeV) & \# of P.S. & $N_{GDM}$ & $F_{GDM}$ ($\times 10^{-4}$) & $N_{iso}$ & $F_{iso}$ ($\times 10^{-4}$) & $N_{J1725}$ & $\alpha_{J1725}$ \\
\hline \hline
$5^{\circ}$ & 200 & 1 p.s. & 1.458 $\pm$ 0.041 & 4.160 $\pm$ 0.118 & 0.830 $\pm$ 0.032 & 0.697 $\pm$ 0.028 & - & - \\
\hline
$5^{\circ}$ & 200 & 4 p.s. & 1.403 $\pm$ 0.042 & 4.001 $\pm$ 0.121 & 0.803 $\pm$ 0.036 & 0.674 $\pm$ 0.030 & 4.00 $\pm$ 0.58 & -2.526 $\pm$ 0.14 \\
\hline
$10^{\circ}$ & 200 & 1 p.s. & 1.574 $\pm$ 0.021 & 4.489 $\pm$ 0.060 & 0.795 $\pm$ 0.016 & 0.669 $\pm$ 0.014 & - & - \\
\hline
$10^{\circ}$ & 200 & 10 p.s. & 1.426 $\pm$ 0.021 & 4.068 $\pm$ 0.060 & 0.781 $\pm$ 0.017 & 0.655 $\pm$ 0.014 & 4.12 $\pm$ 0.59 & -2.31 $\pm$ 0.14 \\
\hline
$10^{\circ}$ & 200 & 15 p.s. & 1.406 $\pm$ 0.007 & 4.012 $\pm$ 0.020 & 0.775 $\pm$ 0.005 & 0.650 $\pm$ 0.005 & 4.00 $\pm$ 0.43 & -2.25 $\pm$ 0.10 \\
\hline
$10^{\circ}$ & 100 & 1 p.s. & 1.582 $\pm$ 0.017 & 7.617 $\pm$ 0.084 & 0.903 $\pm$ 0.011 & 1.871 $\pm$ 0.022 & - & - \\
\hline
$10^{\circ}$ & 100 & 10 p.s. & 1.462 $\pm$ 0.018 & 7.058 $\pm$ 0.086 & 0.871 $\pm$ 0.012 & 1.804 $\pm$ 0.024 & 3.49 $\pm$ 0.53 & -2.40 $\pm$ 0.16 \\
\hline
$10^{\circ}$ & 100 & 15 p.s. & 1.474 $\pm$ 0.018 & 7.120 $\pm$ 0.087 & 0.842 $\pm$ 0.014 & 1.744 $\pm$ 0.029 & 3.29 $\pm$ 0.58 & -2.19 $\pm$ 0.17 \\
\hline
$5^{\circ}$ & 200 & 4 p.s. & 1.192 $\pm$ 0.018 & 3.401 $\pm$ 0.052 & 1.000 $\pm$ 0.000 & 0.839 $\pm$ 0.000 & 3.69 $\pm$ 0.58 & -2.12 $\pm$ 0.13 \\
\hline
$10^{\circ}$ & 200 & 15 p.s. & 1.157 $\pm$ 0.003 & 3.301 $\pm$ 0.008 & 1.000 $\pm$ 0.000 & 0.839 $\pm$ 0.000 & 3.70 $\pm$ 0.21 & -2.10 $\pm$ 0.04 \\
\hline
$10^{\circ}$ & 100 & 15 p.s. & 1.274 $\pm$ 0.009 & 6.150 $\pm$ 0.043 & 1.000 $\pm$ 0.000 & 2.071 $\pm$ 0.000 & 3.03 $\pm$ 0.57 & -2.02 $\pm$ 0.14 \\
\hline \hline
\end{tabular}
\caption{Galactic Diffuse Model (GDM):"gal\_2yearp7v6\_v0", Isotropic Model (iso):"iso\_p7v6source", J1725 refers to the specific point source (see text). $F_{GDM}$ and $F_{iso}$ are in units of ph $\textrm{cm}^{-2}$$\textrm{s}^{-1}$, from the specific region of interest. $N_{GDM}$, $N_{iso}$ and $N_{J1725}$ are normalizations for the equivalent spectral components.}
\label{tab:Draco_FermiColl}
\end{table}
Thus we conclude that the method for analyzing the residual spectra from dSphs as is done by the 
\textit{Fermi} Collaboration makes assumptions that are not generically valid, and there is a need
for alternative methods.

The target of our analysis is to use an alternative method to study the possible DM signal 
from dSph galaxies addressing the following issues:
\begin{itemize}
\item{Model the background for each dSph in a method that does not allow for many degrees 
of freedom that could result as stated above in hiding any small excess.}
\item{Minimize the dependence on the possibility of other undetected point sources existing in the ROI 
(unless they happen to overlap with the dSph location).}
\item{Avoid having low energy $\gamma$-rays dominate our results.}
\item{Include the fact that the \textit{Fermi} LAT instrument has a PSF that depends on 
energy.}
\item{Minimize the significance of CR contamination, which at high latitudes and energies 
can be important.}
\item{Avoid having too low statistics.}
\end{itemize}

\begin{figure*}[t]
\centering\leavevmode
\includegraphics[width=3.20in,angle=0]{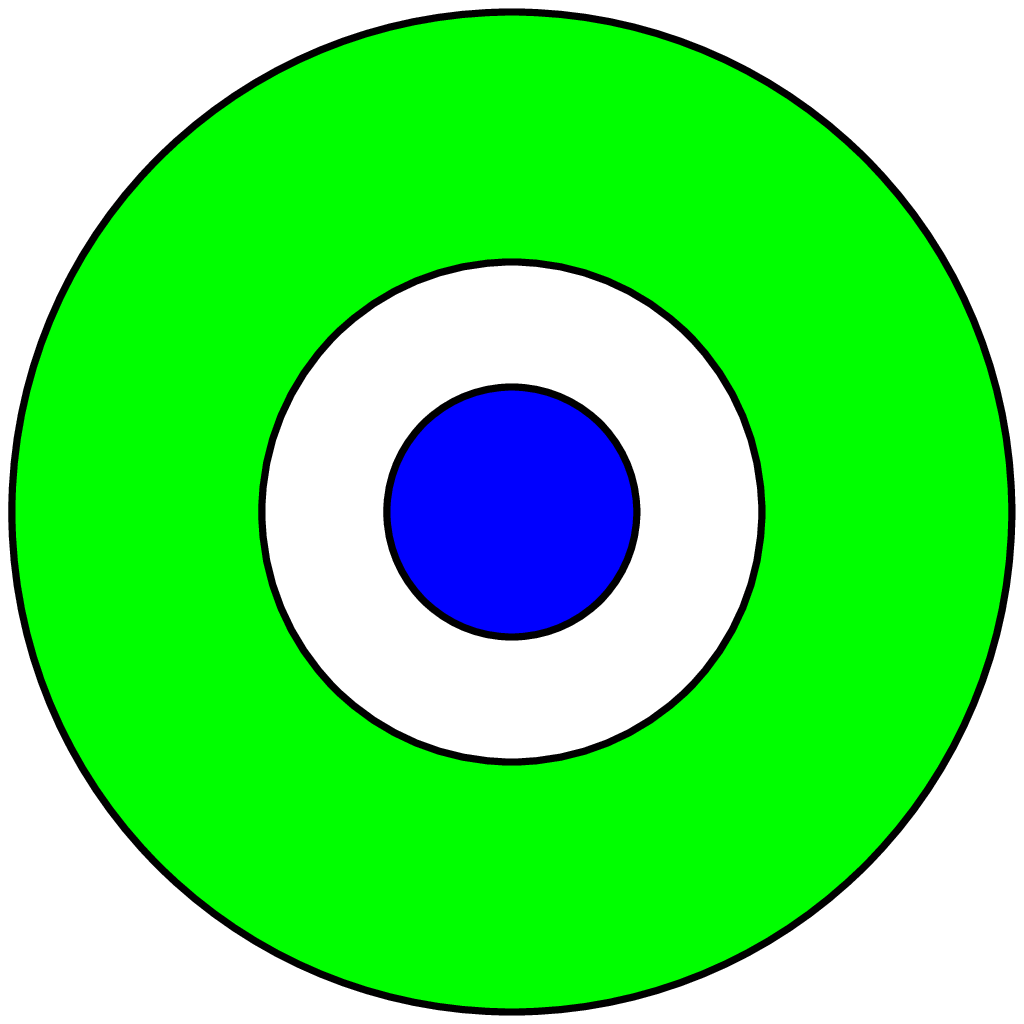}
\hspace{-0.1cm}
\includegraphics[width=3.20in,angle=0]{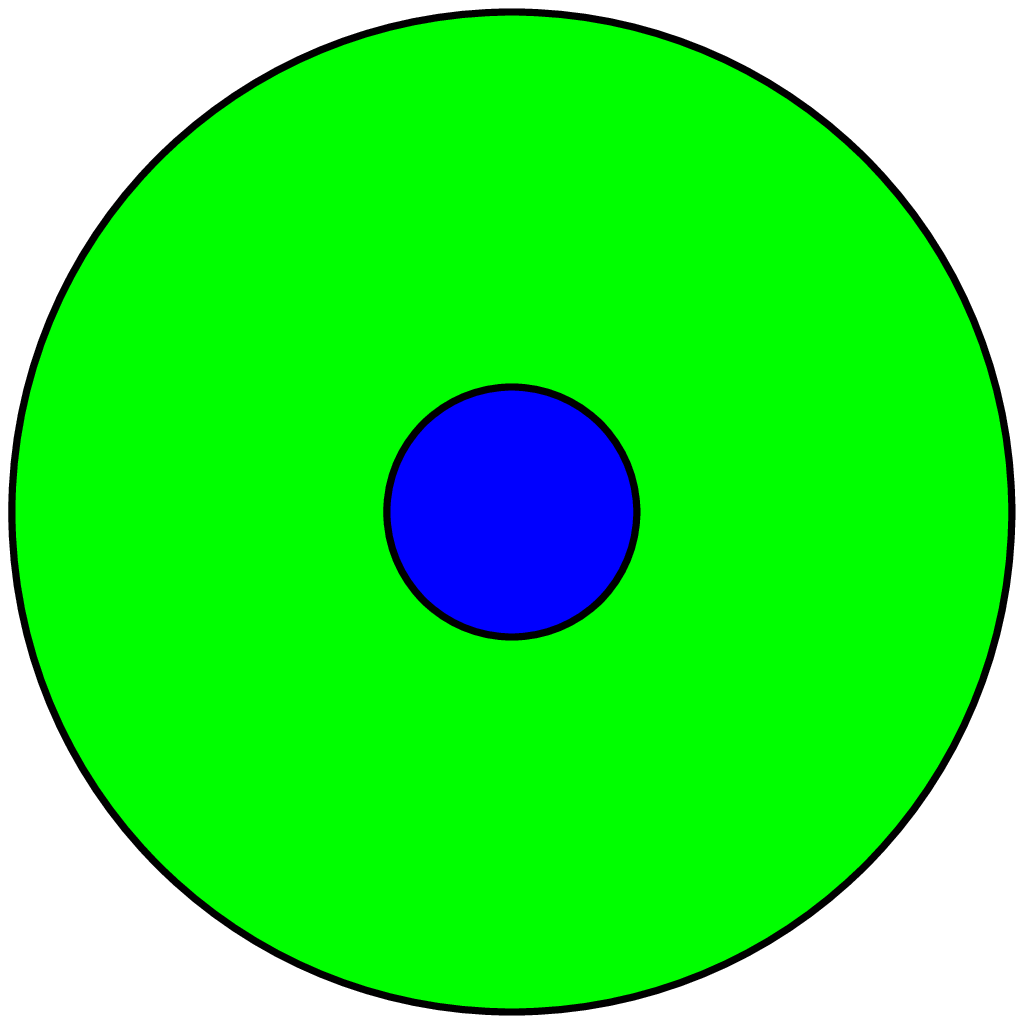}
\caption{Signal \textit{blue}(central disk) and Background 
\textit{light green}(outer ring) Regions Of Interest for versions 1 to 4
\textit{left}, and versions 5 to 6 \textit{right}.}
\label{fig:ROIs}
\end{figure*}

In our method for every dSph galaxy and energy bin we choose 2 regions of interest.
One contains the dSph centered at it's location which for simplicity we will refer to 
as "signal ROI(s)". The other in that energy bin does not contain the dSph but is 
used to measure the 
$\gamma$-ray background of the dSph referred to as "background ROI(s)". The signal ROIs 
include the region of the sky defined by a radius of angle $\alpha_{1}$. The background ROIs 
are defined in the most general case by two radii, $\alpha_{2}$ and $\alpha_{3}$ and include
the region of the sky defined as $\alpha_{2}< \alpha <\alpha_{3}$ from the center of each dSph.
As we will describe later, we will test various combinations of  $\alpha_{1}$, $\alpha_{2}$, 
$\alpha_{3}$ in order to ensure that our criteria set above are met. 
We have tested cases where $\alpha_{1} \leq \alpha_{2}$ 
as shown in Fig.\ref{fig:ROIs}. For all dSphs and at each energy bin we subtract 
from the averaged $\gamma$-ray flux of the signal ROI the  averaged $\gamma$-ray flux of the 
equivalent background ROI.

We stress that the  background $\gamma$-ray flux to each dSph at a given energy bin is the 
$\gamma$-ray flux at that energy bin between $\alpha_{2}< \alpha <\alpha_{3}$, and is kept 
fixed thus no additional d.o.f. for the background model are allowed. Similar methods are 
common in gamma-ray astronomy \cite{Abramowski:2011hc, Aharonian:2006au, Aharonian:2006wh} and in general\cite{Planck, 2011A&A...536A...7P, Haakonsen:2009mg, Shirahata:2009ui}.
Furthermore the isotropic flux contribution which is the dominant component in most cases, 
and includes the EGBR and the CR contamination is in our method subtracted\footnote{apart 
from its Poisson noise} (without adding a freedom in it's normalization).
Also, since our choices for the sizes of the ROIs are similar to that of \cite{Abdo:2010ex, 
Ackermann:2011wa} any contribution from undetected point sources is taken into account. 
An other characteristic of our analysis is that regarding claims of positive (excess) or 
negative residual $\gamma$-ray fluxes, each energy bin is independent from the others. This 
does not let the lower energy bins with the higher statistics influence the search for a 
possible residual $\gamma$-ray flux at higher energies.     

We clarify that for background ROIs extending too far away from the dSphs one includes 
into the background flux the contribution of sources (e.g. point sources
or emission from ISM gas) that are irrelevant to the actual background 
$\gamma$-ray flux at the location of the dSphs. Thus while we show also 
some results from more extended background ROIs we will focus on these choices
where proximity to the target is ensured. 

Since we want to take into account the PFS vs $E_{\gamma}$ information, while at the same 
time avoid having our analysis on the DM limits be dominated by low energy $\gamma$-rays, 
we choose to use $E_{\gamma} \geq 1$ GeV and up to 100 GeV. At $E>100$ GeV the low statistics 
and the noise of the contamination of CRs becomes a concern. To minimize contamination from 
CRs, we
use for the rest of our analysis either the CLEAN or the ULTRACLEAN class of events.
At $E<1$ GeV the PSF becomes too large, which in our method has the disadvantage 
of having the region of the sky where the dSph does not contribute being too far from the 
center of the dSph. That results in the signal ROI being very large, and thus the background 
ROI is too far from the center of the dSph, to provide a proper model for the background 
$\gamma$-ray spectrum. We break the 1-100 GeV range in 6 logarithmically equally 
spaced energy bins, giving in theory a total of 12 ROIs for each target and for the entire 
energy range. As we will explain later though, the ROIs for the last three energy bins 
are always the same, setting the actual number of ROIs per target to be 8. 

We clarify that such an analysis is not optimized for low mass DM. As we explained earlier 
in the text, in our opinion the low mass 
DM cases can not be probed in dSphs with an instrument whose PSF containment angle is significantly 
larger than the target sizes at the relevant energies where the DM annihilation signal lies. 
 
For each energy bin we will take as the relevant angular size of the object in the sky $\alpha_{dSph}(E)$, either $\alpha_{c}$, or the $95\%$ containment angle at normal incidence at approximately the center of 
the energy bin, depending on which of the two is the largest. Thus we treat the smallest in size 
(on the sky) dSphs as practically point sources and the largest ones as extended sources.
For our selection of dSphs apart from Sextans at energies above 10 GeV  $\alpha_{dSph}$ is defined 
by the containment angle.  

\begin{figure*}[t]
\centering\leavevmode
\includegraphics[width=3.20in,angle=0]{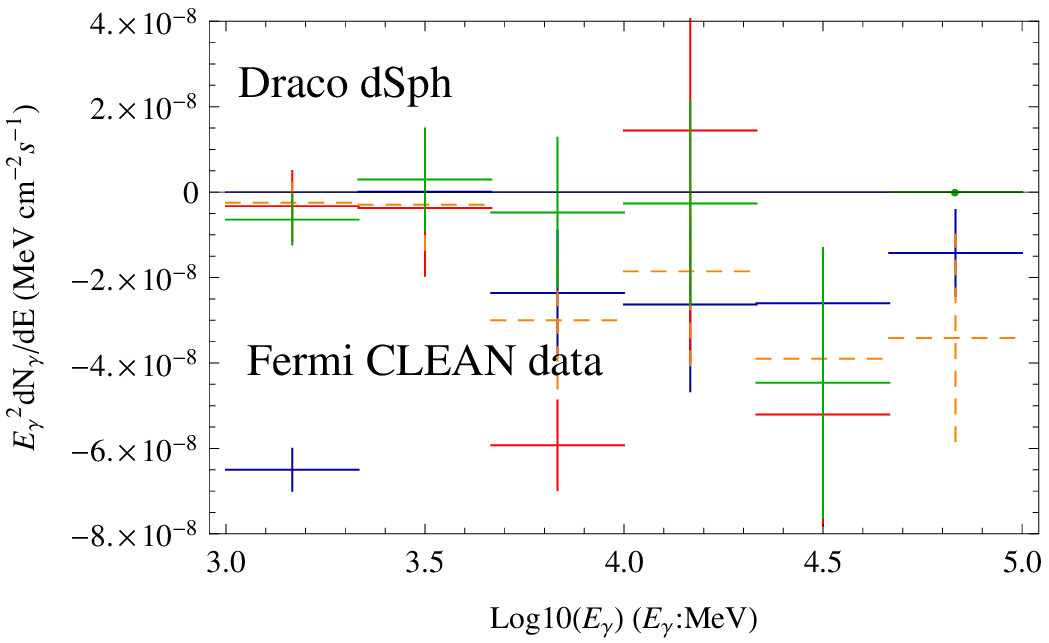}
\hspace{-0.1cm}
\includegraphics[width=3.20in,angle=0]{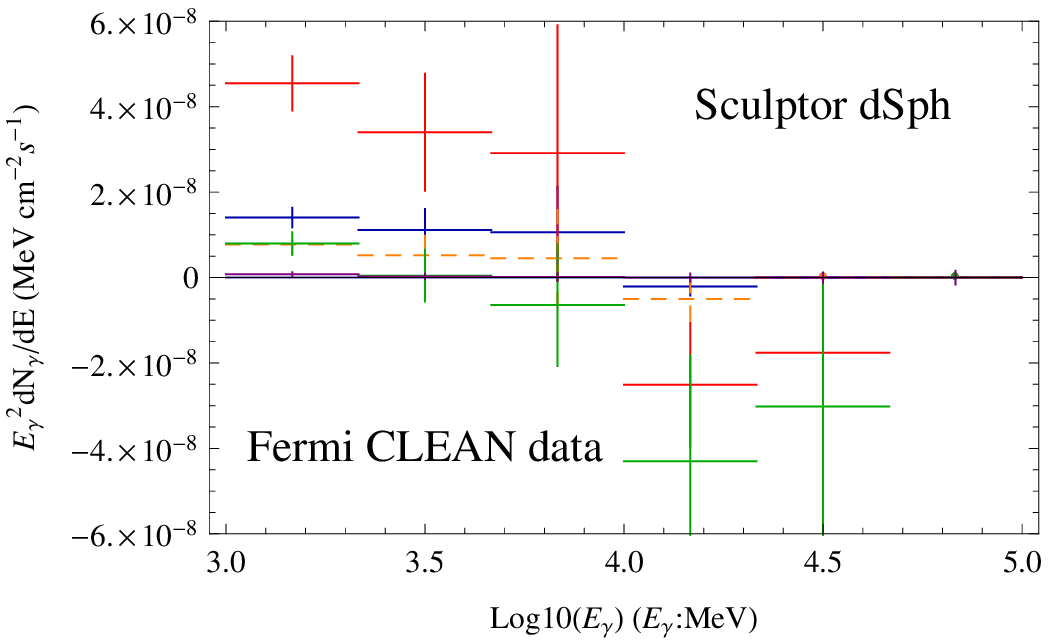}\\
\includegraphics[width=3.20in,angle=0]{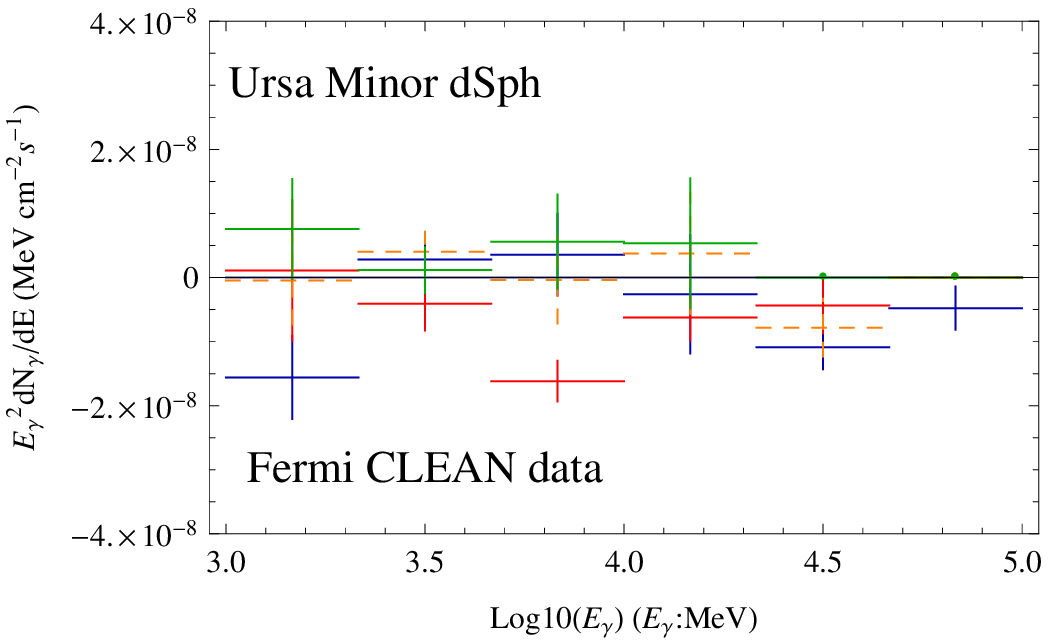}
\hspace{-0.1cm}
\includegraphics[width=3.20in,angle=0]{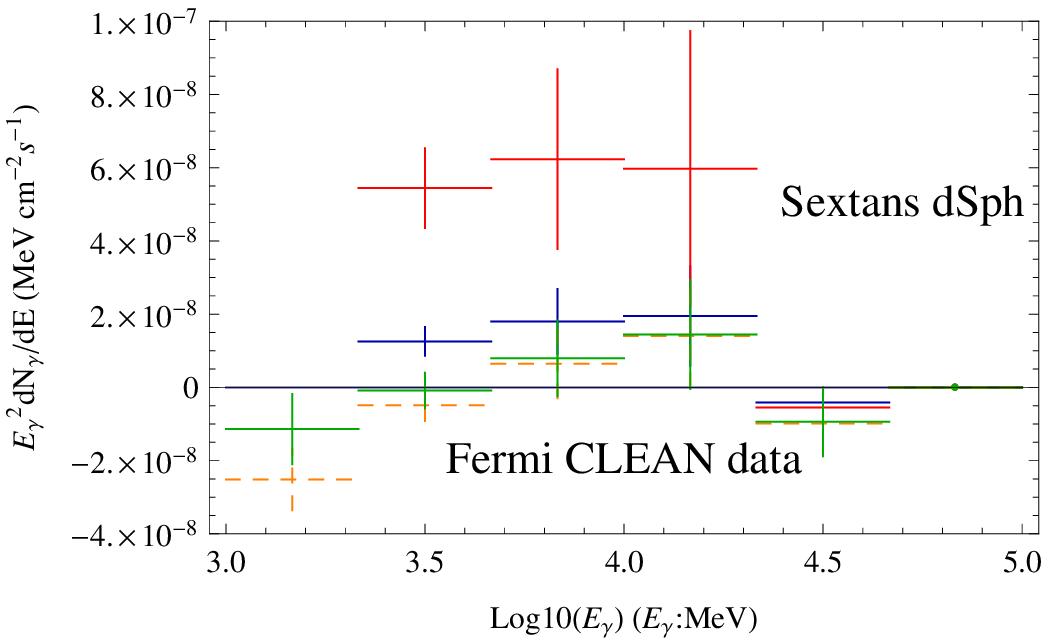}
\caption{Excess gamma-ray differential flux for dwarf spheroidal galaxies, using \textit{Fermi} data ``Clean'' class. \textit{Upper left:} Draco, \textit{upper right:} Sculptor, \textit{lower left:} Ursa Minor, \textit{lower right:} Sextans. Dark blue: version 1, red: version 2, dashed orange: version 3, green: version 4.}
\label{fig:ExcessGammasVersions}
\end{figure*}

For our analysis $\alpha_{dSph}$ will set either $\alpha_{1}$ or $\alpha_{2}$ at each bin, given in 
Table~\ref{tab:versionsGammas}.
The most important remaining issue is that of low statistics at higher energy bins.
In Table~\ref{tab:versionsGammas} we present our assumptions for $\alpha_{1}$, $\alpha_{2}$ and $\alpha_{3}$.
We show in Fig.~\ref{fig:ExcessGammasVersions} the residual spectra for the four targets that give the 
tightest limits on DM annihilation rates, for the case where $\alpha_{1} < \alpha_{2}$ (versions 1 to 4), 
using the CLEAN data set.

\begin{figure*}[t]
\centering\leavevmode
\includegraphics[width=3.20in,angle=0]{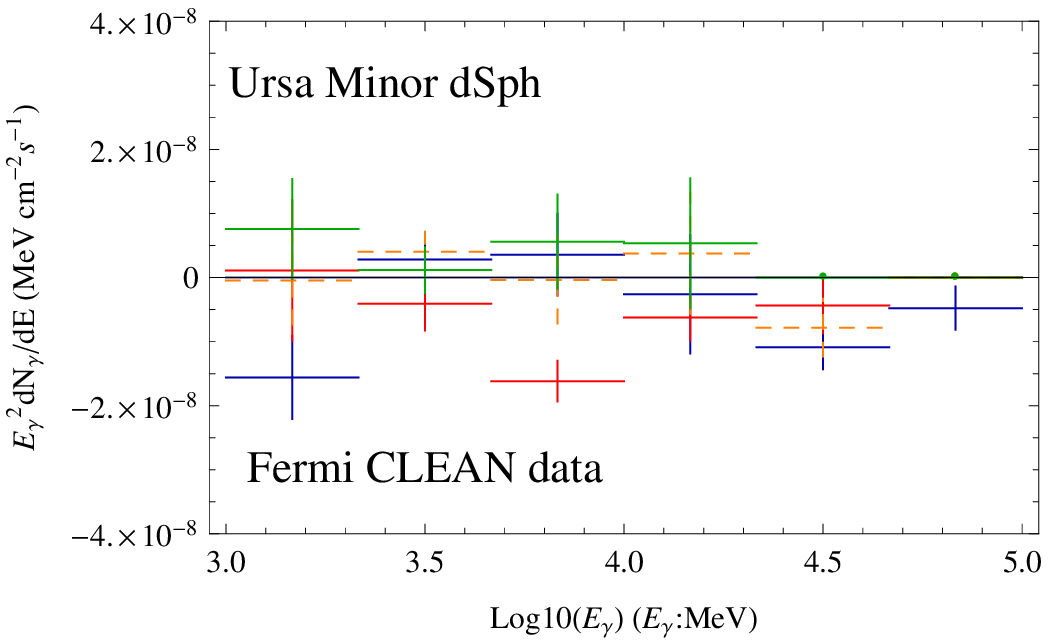}
\hspace{-0.1cm}
\includegraphics[width=3.20in,angle=0]{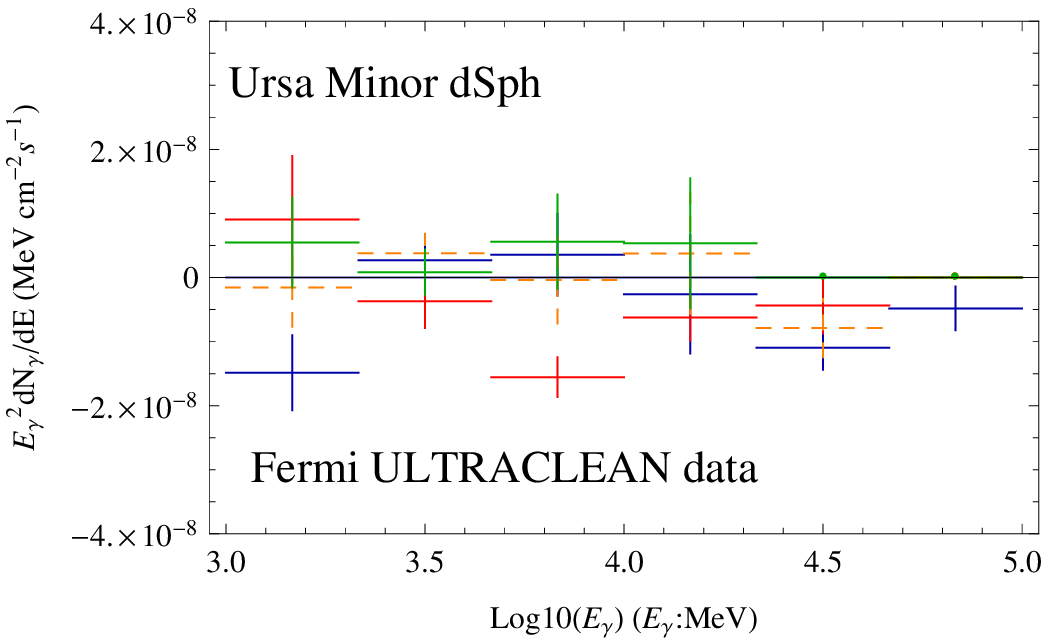}
\caption{Residual spectra from ROIs versions 1 to 4  for Ursa Minor using 
CLEAN \textit{left} and ULTRACLEAN \textit{right} event selection. Colors as in 
Fig.~\ref{fig:ExcessGammasVersions}.}
\label{fig:UltraClean_VS_Clean}
\end{figure*}

In Fig.~\ref{fig:UltraClean_VS_Clean} we present for Ursa Minor the residual spectra using,
equivalently CLEAN and ULTRACLEAN data. 
There is not much gain in using the cleanest of the two samples and thus to avoid having too 
low statistics we use for the remaining work the CLEAN selection class.

\begin{table}[t]
\begin{tabular}{|l|c|c|c|c|c|c|}
\hline
case & $E_{\gamma}^{mean}$ (GeV)&$E_{\gamma}^{min}$ (GeV)&$E_{\gamma}^{max}$ (GeV)&$\alpha_{1}$&$\alpha_{2}$&$\alpha_{3}$ \\
\hline \hline
\multirow{6}[4]*{\begin{sideways}Version 1 \end{sideways}} & 1.47 & 1.00 & 2.15 & $2.5^{\circ}$ & $5.0^{\circ}$ & $10.0^{\circ}$ \\
\cline{2-7}
 & 3.16 & 2.15 & 4.64 & $1.5^{\circ}$ & $3.0^{\circ}$ & $6.0^{\circ}$ \\
\cline{2-7}
 & 6.81 & 4.64 & 10.0 & $1.0^{\circ}$ & $2.0^{\circ}$ & $4.0^{\circ}$ \\
\cline{2-7}
 & 14.7 & 10.0 & 21.5 & $0.8^{\circ}$ & $1.6^{\circ}$ & $3.2^{\circ}$ \\
\cline{2-7}
 & 31.6 & 21.5 & 46.4 & $0.8^{\circ}$ & $1.6^{\circ}$ & $3.2^{\circ}$ \\
\cline{2-7}
 & 68.1 & 46.4 & 100  & $0.8^{\circ}$ & $1.6^{\circ}$ & $3.2^{\circ}$ \\
\hline \hline
\multirow{6}[4]*{\begin{sideways}Version 2 \end{sideways}} & 1.47 & 1.00 & 2.15 & $1.5^{\circ}$ & $2.5^{\circ}$ & $5.0^{\circ}$ \\
\cline{2-7}
 & 3.16 & 2.15 & 4.64 & $0.9^{\circ}$ & $1.5^{\circ}$ & $3.0^{\circ}$ \\
\cline{2-7}
 & 6.81 & 4.64 & 10.0 & $0.6^{\circ}$ & $1.0^{\circ}$ & $2.0^{\circ}$ \\
\cline{2-7}
 & 14.7 & 10.0 & 21.5 & $0.48^{\circ}$ & $0.8^{\circ}$ & $1.6^{\circ}$ \\
\cline{2-7}
 & 31.6 & 21.5 & 46.4 & $0.48^{\circ}$ & $0.8^{\circ}$ & $1.6^{\circ}$ \\
\cline{2-7}
 & 68.1 & 46.4 & 100  & $0.48^{\circ}$ & $0.8^{\circ}$ & $1.6^{\circ}$ \\
\hline \hline
\multirow{6}[4]*{\begin{sideways}Version 3 \end{sideways}} & 1.47 & 1.00 & 2.15 & $2.5^{\circ}$ & $5.0^{\circ}$ & $7.5^{\circ}$ \\
\cline{2-7}
 & 3.16 & 2.15 & 4.64 & $1.5^{\circ}$ & $3.0^{\circ}$ & $4.5^{\circ}$ \\
\cline{2-7}
 & 6.81 & 4.64 & 10.0 & $1.0^{\circ}$ & $2.0^{\circ}$ & $3.0^{\circ}$ \\
\cline{2-7}
 & 14.7 & 10.0 & 21.5 & $0.8^{\circ}$ & $1.6^{\circ}$ & $2.4^{\circ}$ \\
\cline{2-7}
 & 31.6 & 21.5 & 46.4 & $0.8^{\circ}$ & $1.6^{\circ}$ & $2.4^{\circ}$ \\
\cline{2-7}
 & 68.1 & 46.4 & 100  & $0.8^{\circ}$ & $1.6^{\circ}$ & $2.4^{\circ}$ \\
\hline \hline
\multirow{6}[4]*{\begin{sideways}Version 4 \end{sideways}} & 1.47 & 1.00 & 2.15 & $2.5^{\circ}$ & $3.75^{\circ}$ & $5.0^{\circ}$ \\
\cline{2-7}
 & 3.16 & 2.15 & 4.64 & $1.5^{\circ}$ & $2.25^{\circ}$ & $3.0^{\circ}$ \\
\cline{2-7}
 & 6.81 & 4.64 & 10.0 & $1.0^{\circ}$ & $1.5^{\circ}$ & $2.0^{\circ}$ \\
\cline{2-7}
 & 14.7 & 10.0 & 21.5 & $0.8^{\circ}$ & $1.2^{\circ}$ & $1.6^{\circ}$ \\
\cline{2-7}
 & 31.6 & 21.5 & 46.4 & $0.8^{\circ}$ & $1.2^{\circ}$ & $1.6^{\circ}$ \\
\cline{2-7}
 & 68.1 & 46.4 & 100  & $0.8^{\circ}$ & $1.2^{\circ}$ & $1.6^{\circ}$ \\
\hline \hline
\multirow{6}[4]*{\begin{sideways}Version 5 \end{sideways}} & 1.47 & 1.00 & 2.15 & $2.5^{\circ}$ & $2.5^{\circ}$ & $5.0^{\circ}$ \\
\cline{2-7}
 & 3.16 & 2.15 & 4.64 & $1.5^{\circ}$ & $1.5^{\circ}$ & $5.0^{\circ}$ \\
\cline{2-7}
 & 6.81 & 4.64 & 10.0 & $1.0^{\circ}$ & $1.0^{\circ}$ & $5.0^{\circ}$ \\
\cline{2-7}
 & 14.7 & 10.0 & 21.5 & $0.8^{\circ}$ & $0.8^{\circ}$ & $5.0^{\circ}$ \\
\cline{2-7}
 & 31.6 & 21.5 & 46.4 & $0.8^{\circ}$ & $0.8^{\circ}$ & $5.0^{\circ}$ \\
\cline{2-7}
 & 68.1 & 46.4 & 100  & $0.8^{\circ}$ & $0.8^{\circ}$ & $5.0^{\circ}$ \\
\hline \hline
\multirow{6}[4]*{\begin{sideways}Version 6 \end{sideways}} & 1.47 & 1.00 & 2.15 & $2.5^{\circ}$ & $2.5^{\circ}$ & $10.0^{\circ}$ \\
\cline{2-7}
 & 3.16 & 2.15 & 4.64 & $1.5^{\circ}$ & $1.5^{\circ}$ & $10.0^{\circ}$ \\
\cline{2-7}
 & 6.81 & 4.64 & 10.0 & $1.0^{\circ}$ & $1.0^{\circ}$ & $10.0^{\circ}$ \\
\cline{2-7}
 & 14.7 & 10.0 & 21.5 & $0.8^{\circ}$ & $0.8^{\circ}$ & $10.0^{\circ}$ \\
\cline{2-7}
 & 31.6 & 21.5 & 46.4 & $0.8^{\circ}$ & $0.8^{\circ}$ & $10.0^{\circ}$ \\
\cline{2-7}
 & 68.1 & 46.4 & 100  & $0.8^{\circ}$ & $0.8^{\circ}$ & $10.0^{\circ}$ \\
\hline \hline
\end{tabular}
\caption{Combinations of $\alpha_{1}$, $\alpha_{2}$, $\alpha_{3}$ for versions 1 to 4 
($\alpha_{1} < \alpha_{2}$) of
evaluating residual spectra, and versions 5 and 6 ($\alpha_{1} = \alpha_{2}$).}
\label{tab:versionsGammas}
\end{table}

\begin{figure*}[t]
\centering\leavevmode
\includegraphics[width=3.20in,angle=0]{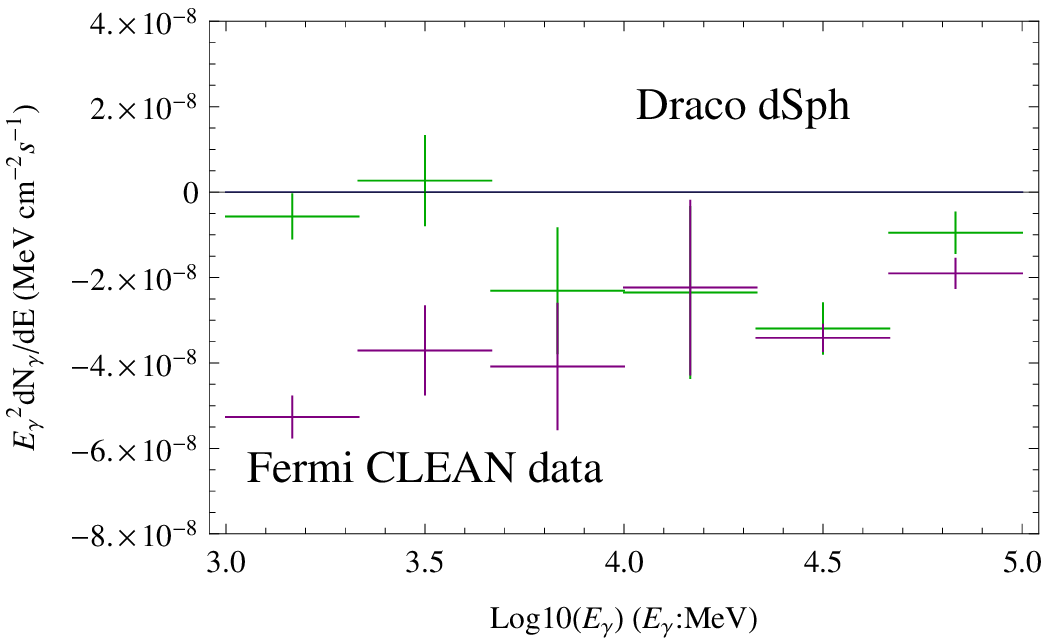}
\hspace{-0.1cm}
\includegraphics[width=3.20in,angle=0]{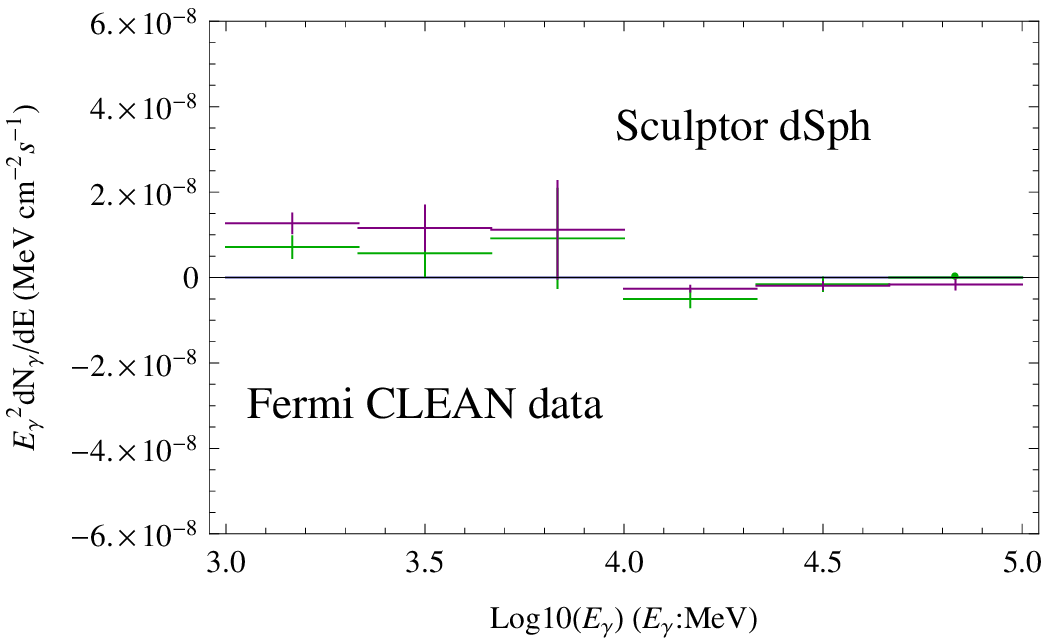}\\
\includegraphics[width=3.20in,angle=0]{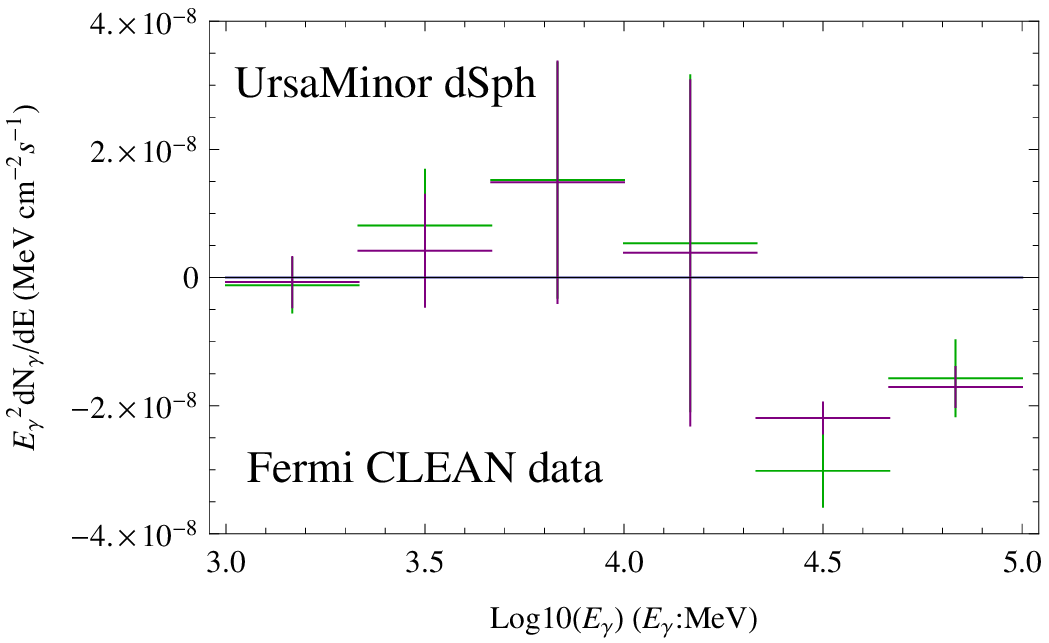}
\hspace{-0.1cm}
\includegraphics[width=3.20in,angle=0]{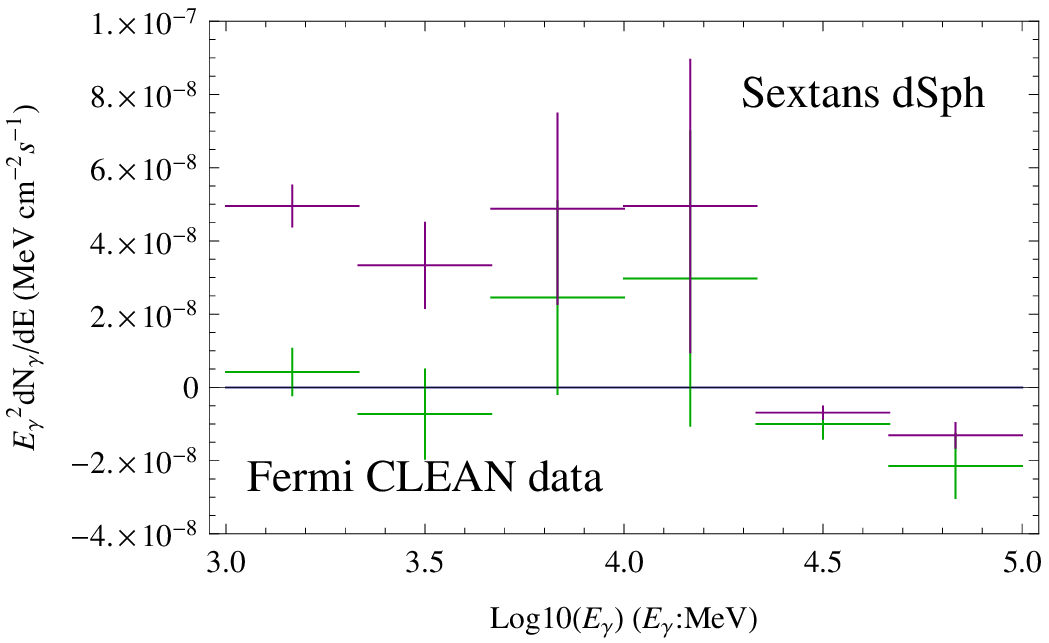}
\caption{Excess gamma-ray differential flux for dwarf spheroidal galaxies, using
\textit{Fermi} data ``Clean'' class.
 \textit{Upper left:} Draco, \textit{upper right:} Sculptor, \textit{lower left:} Ursa Minor, \textit{l\
ower right:} Sextans. Green: version 5, dark purple: version 6.}
\label{fig:ExcessGammas_v5v6}
\end{figure*}

In Fig.~\ref{fig:ExcessGammas_v5v6} we show the residual spectra for 
the case of $\alpha_{1}(E) = \alpha_{2}(E) = \alpha_{dSph}(E)$, 
with $\alpha_{3}$ being $5^{\circ}$ (version 5) or $10^{\circ}$(version 6). 
While with version 6 we extend the background ROI out to $10^{\circ}$, 
ensuring smaller statistical errors for the background, the additional 
angular area ($5^{\circ}$-$10^{\circ}$ from the dSph center) used in 
calculating the background results in decreasing the relevance of the 
calculated background to the actual background flux at the location of the
dSph. That can result in significant changes of the residual spectra at the
lower energy bins, as is most clear for the cases of Draco and Sextans dSphs.
Since the signal ROIs for both versions 5 and 6 are identical, the origin is 
strictly from the background calculation. That difference is mainly due to 
the fact that for these targets the relevant flux from the point sources 
centered within $5^{\circ}$ from the center of
the dSph at study to the flux from the point sources centered within $10^{\circ}$ 
 changes significantly\footnote{Draco from 4 p.s. to 15 p.s., 
Sextans from 4 p.s. to 13 p.s., Sculptor 1 p.s. for both cases, and Ursa Minor 
from 1 p.s. to 3 p.s.; with both cases though the total p.s. flux being a 
subdominant part of the background flux in the energy range of study.}.
Point sources are expected to be more important at lower energies due to their 
softer spectrum (on average) compared to the isotropic and the galactic diffuse 
spectral components of the $\gamma$-ray background. Also at high energies where the PSF is bellow 
$1^{\circ}$ extending the background region out to $10^{\circ}$  can only be 
correct if the background at these energies is almost isotropic.
 
For these reasons we consider version 5 preferable to version 6, and thus 
while we show the residual spectra from version 6 in Fig.~\ref{fig:ExcessGammas_v5v6}, 
we avoid including them further into our analysis. 
For the same arguments
version 4 is preferable to versions 1 and 3. Version 2 assumes very small
signal ROIs resulting in too low statistics, which in return gives systematically
weaker constraints on annihilation rates. We will include versions 1-3 in some
of our results, since at least at high energies they all use for the background
regions that are not further than $5^{\circ}$ away from the center of the dSphs.

\section{Getting different limits}
\label{sec:methods}

Each of the methods described in section~\ref{sec:FermiGamma}, results 
in different calculated residual spectra, and thus in 
different limits on DM annihilation rates. 
The flux from DM annihilation is calculated as
\footnote{For the Sommerfeld enhancement case \cite{ArkaniHamed:2008qn, Hisano:2006nn,Lattanzi:2008qa, SommerfeldRef},
where the cross-section depends on the relative velocity between the 
DM particles, decoupling the annihilation cross-section from the DM profile, 
is not obvious. Yet since dSphs have low velocity dispersions, it may be the case 
that the cross-section has already reached saturation.}:
\begin{equation}
\label{eq:DMannih_Signal}
\frac{d\Phi_{\gamma}}{dE}^{i}_{dSph_{DM}} = \frac{\langle \sigma v \rangle}{4 \pi}
\frac{dN_{\gamma}}{dE}_{ _{DM}} \frac{J^{i}}{2 \, m_{\chi}^2},
\end{equation}
where the index $i$, refers to the energy range/bin at interest since the $J$-factors 
differ between bins:
\begin{equation}
\label{eq:Jfactor}
J^{i} =  \int \int dl d\Omega^{i} \rho^{2}_{DM}(l,\Omega).
\end{equation}
$d\Omega^{i}$ is different for energy bins with different signal ROIs, in our limits 
from versions 1 to 5 (see discussion in sections~\ref{sec:J-factors} and
~\ref{sec:FermiGamma}).  

We clarify that the condition of Boost Factor, BF = 1 is for a thermally averaged
annihilation cross-section of $\langle \sigma v\rangle = 3 \times 10^{-26} cm^{3} s^{-1}$
in eq.~\ref{eq:DMannih_Signal} with the $J$-factors being \textit{calculated for every energy bin}
based on the $\alpha_{dwarf} = \alpha_{1}$. We give in Table~\ref{tab:J-factorsDraco} the 
$J$-factors of Draco dSph within different containment angles $\alpha$.

\begin{table}[t]
\begin{tabular}{|c|c|c|c|}
\hline
$\overline{J} \times 10^{17}$ ($GeV^{2} cm^{-5}$)& $\delta J_{high} \times 10^{17}$ ($GeV^{2} cm^{-5}$) & $\delta J_{low} \times 10^{17}$ ($GeV^{2} cm^{-5}$) & $\alpha$ \\
\hline \hline
74.4 & 45.8 & 33.8 & $2.50^{\circ}$ \\
\hline
72.7 & 39.6 & 32.3 & $1.50^{\circ}$ \\
\hline
69.0 & 29.5 & 29.3 & $1.00^{\circ}$ \\
\hline
65.3 & 22.3 & 26.3 & $0.80^{\circ}$ \\
\hline
51.7 & 9.70 & 16.4 & $0.48^{\circ}$ \\
\hline
29.2 & 7.52 & 5.84 & $0.27^{\circ}$ \\
\hline \hline
\end{tabular}
\caption{The $J$-factor for Draco dSph with its upper and lower 
uncertainties for different containment angles $\alpha$. The first 
four rows are relevant for versions 1, 3, 4 and 5 (see 
Table~\ref{tab:versionsGammas}). The fifth row is relevant for 
version 2, and the last row refers to the $J$-factor calculated at 
$\alpha_{c}$ (see Table~\ref{tab:J-factors}).}
\label{tab:J-factorsDraco}
\end{table}

For a reference model of DM annihilation channel relevant for $\frac{dN_{\gamma}}{dE}_{_{DM}}$ 
in eq.~\ref{eq:DMannih_Signal} ($\frac{dN_{\gamma}}{dE}_{_{DM}}$ is the differential
$\gamma$-ray spectrum per annihilation event) we consider the case of $\chi \chi \longrightarrow W^{+}W^{-}$, 
We take into account only the "prompt $\gamma$-rays" i.e. those that come from final 
state radiation and the hadronization processes after the decay of Ws. We used PYTHIA 6.4
\cite{Sjostrand:2006za} event generator to derive those $\gamma$-ray spectra. Ws can also decay into 
leptons by Branching Ratios of $0.1075 \pm 0.0013$ to $e^{+}e^{-}$, $0.1057 \pm 0.0015$ 
to $\mu^{+}\mu^{-}$, $0.1025 \pm 0.0020$ to $\tau^{+}\tau^{-}$ \cite{PDG2010}. Some $\gamma$-rays will 
also be produced from inverse Compton scattering (ICS) by and bremsstrahlung
radiation off the highly 
energetic $e^{\pm}$, that are among the stable final products. Yet these $e^{\pm}$ propagate 
from their original production point. Thus the relevant $\gamma$-ray components are more 
diffused, not directly related to the $J$-factors and less straightforward to calculate 
since one needs to add the information on the radiation field and the baryonic mass 
distribution in the form of gas and dust at the actual location of the dSphs.
Since we know that dSphs have in mass a suppressed baryonic component, we expect the 
gas and dust number densities and the energy density of the radiation field at infrared and 
optical wavelengths to be suppressed compared to that in our Galaxy. Yet unless one carries a 
detailed analysis, for each dwarf spheroidal, one can not be confident that the hadronic channel
giving prompt $\gamma$-rays is always the dominant part; especially for masses significantly 
larger ($\gsim 20$ times) than the energy range of the $\gamma$-ray data used
\footnote{It has been shown in calculations of DM annihilating in our Galaxy for cases as
$\chi \chi \longrightarrow W^{+}W^{-}$, the $\gamma$-ray spectra of DM origin at energies 
close to the mass of the DM particle ($E_{\gamma} \sim m_{\chi}/3$) are dominated by  
the prompt $\gamma$-rays while for energies $E_{\gamma} \lsim m_{\chi}/30$ the ICS and 
bremsstrahlung contribution become dominant (see for instance Fig. 1 of \cite{Malyshev:2010xc}).}.
Thus the reader should consider those limits as conservative ones. We also clarify that in 
our PYTHIA  simulation we did not consider the decays of $\tau$s into $\pi^{0}$ and $K^{0}$s,
which happens about $\approx 50 \%$ of the time \cite{PDG2010}. That last component is also 
a source of prompt $\gamma$-rays, which can though not change our results by
 more than $\approx 10\%$. For the case of the $\chi \chi \longrightarrow W^{+}W^{-}$ 
channel that we use here as reference, EW corrections \cite{Ciafaloni:2010ti} do 
not change our results (see also discussion in section~\ref{sec:constraints}).
Yet for more model dependent cases
\cite{Hryczuk:2011vi, Ciafaloni:2012gs}, additional corrections are necessary. 

\begin{figure*}[t]
\centering\leavevmode
\includegraphics[width=3.40in,angle=0]{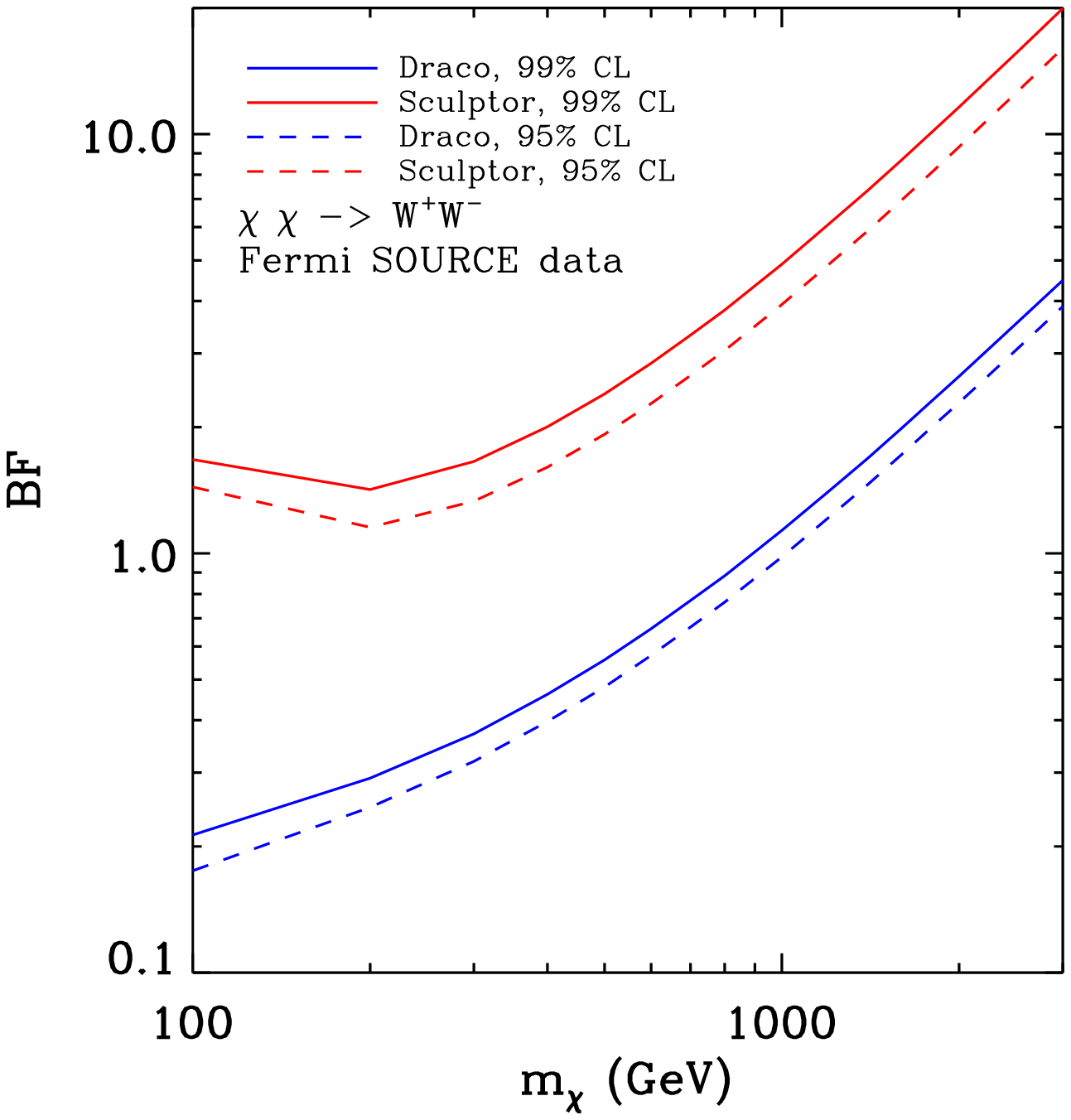}
\hspace{-0.1cm}
\includegraphics[width=3.40in,angle=0]{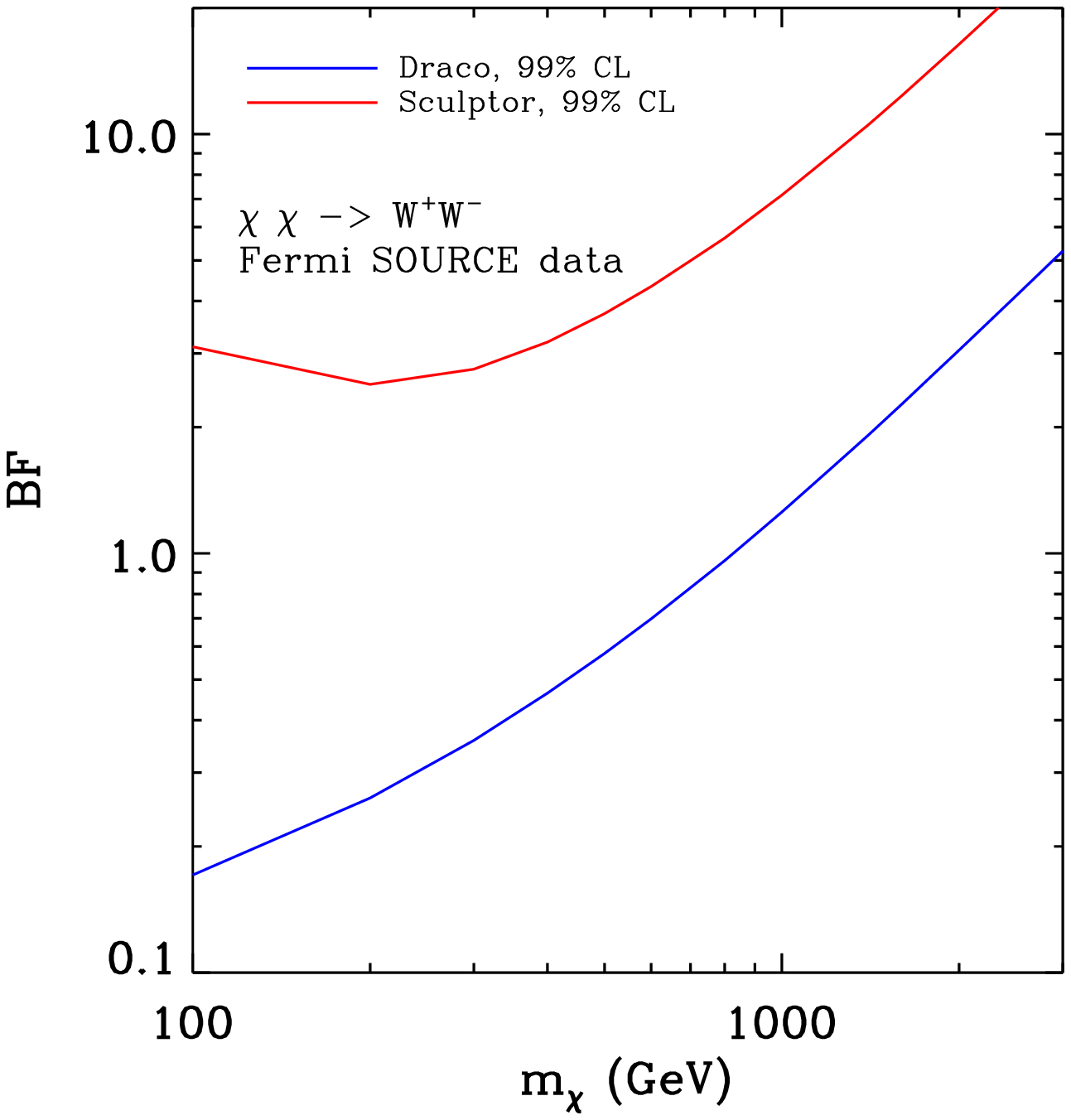}
\caption{99$\%$ and 95$\%$ CL limits on annihilation rate for Draco (blue) and Sculptor 
(red) dwarf spheroidal galaxies. \textit{Left}: using residual spectra of 
Fig.~\ref{fig:ExcessGammasFermiColl}(left), with energy between 1 GeV and 100 GeV within 
$5^{\circ}$. \textit{Right}: using residual spectra of Fig.~\ref{fig:ExcessGammasFermiColl}
(right), within $10^{\circ}$.}
\label{fig:DMLimits_FermiColl}
\end{figure*}

In Fig.~\ref{fig:DMLimits_FermiColl}, we show the 95$\%$ and 99$\%$ CL limits on 
annihilation rate BF from Draco and Sculptor residual spectra (using the Fermi 
Tools only for the residual spectra) as shown in Fig.~\ref{fig:Fermi_dwSp_Spectra} 
and~\ref{fig:ExcessGammasFermiColl}. To have a more direct comparison of the impact
that our methods for the calculation of the residual spectra have on the DM BF limits, 
we used only the residual spectra between 1 and 100 GeV in 15 energy bins given 
also in Fig.~\ref{fig:ExcessGammasFermiColl}. We calculate limits from both the 
cases where the data lay within $5^{\circ}$ (Fig.~\ref{fig:ExcessGammasFermiColl} 
and~\ref{fig:DMLimits_FermiColl} (left)) and within $10^{\circ}$ 
(Fig.~\ref{fig:ExcessGammasFermiColl} and~\ref{fig:DMLimits_FermiColl} (right)) from 
each dSph. For the limits of Fig.~\ref{fig:DMLimits_FermiColl} since the ROI 
did not vary with $E_{\gamma}$ we used the mean values and uncertainties from the 
case where $\alpha = 2.5^{\circ}$.

\begin{figure*}[t]
\centering\leavevmode
\includegraphics[width=2.35in,angle=0]{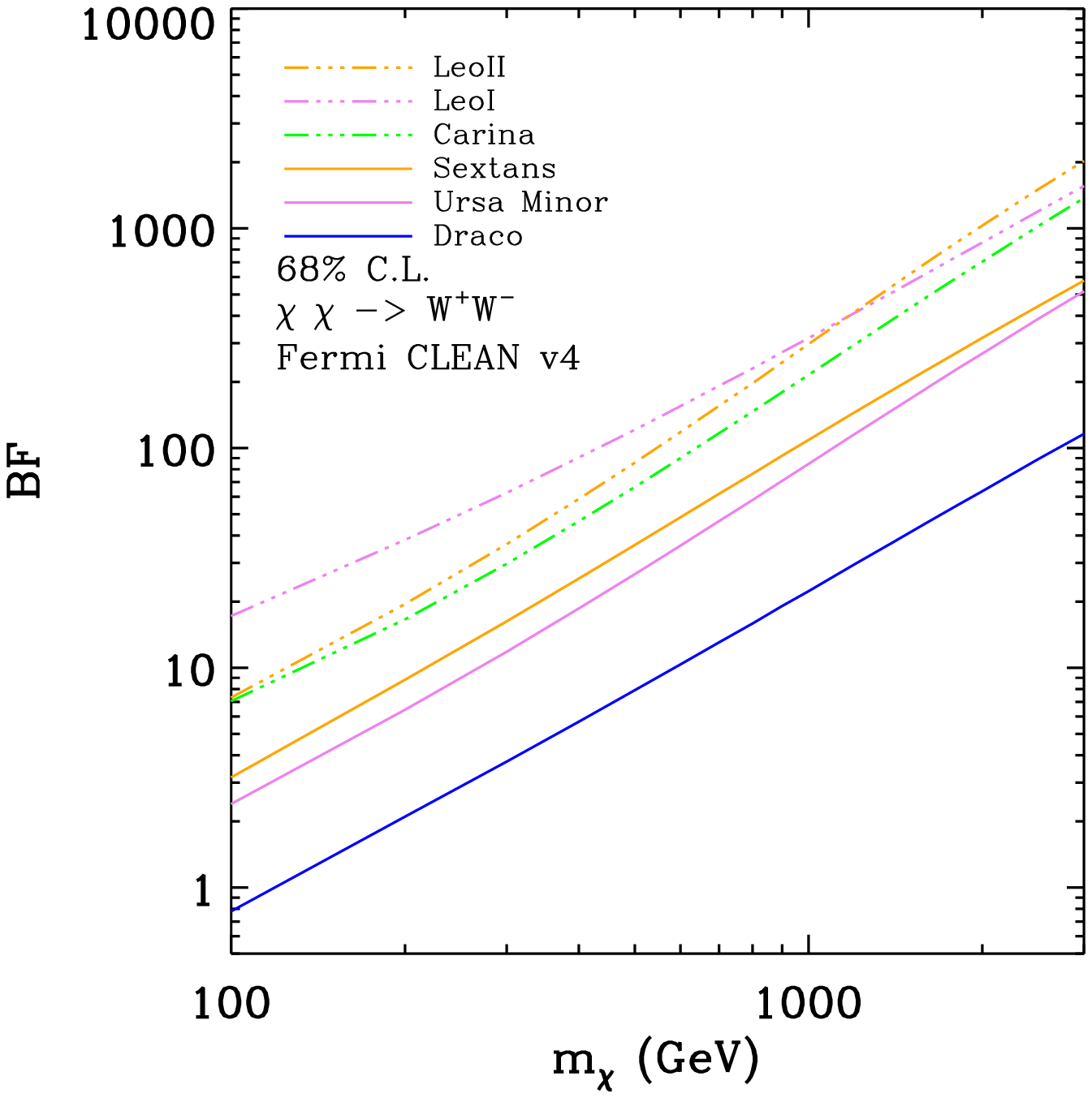}
\hspace{-0.35cm}
\includegraphics[width=2.35in,angle=0]{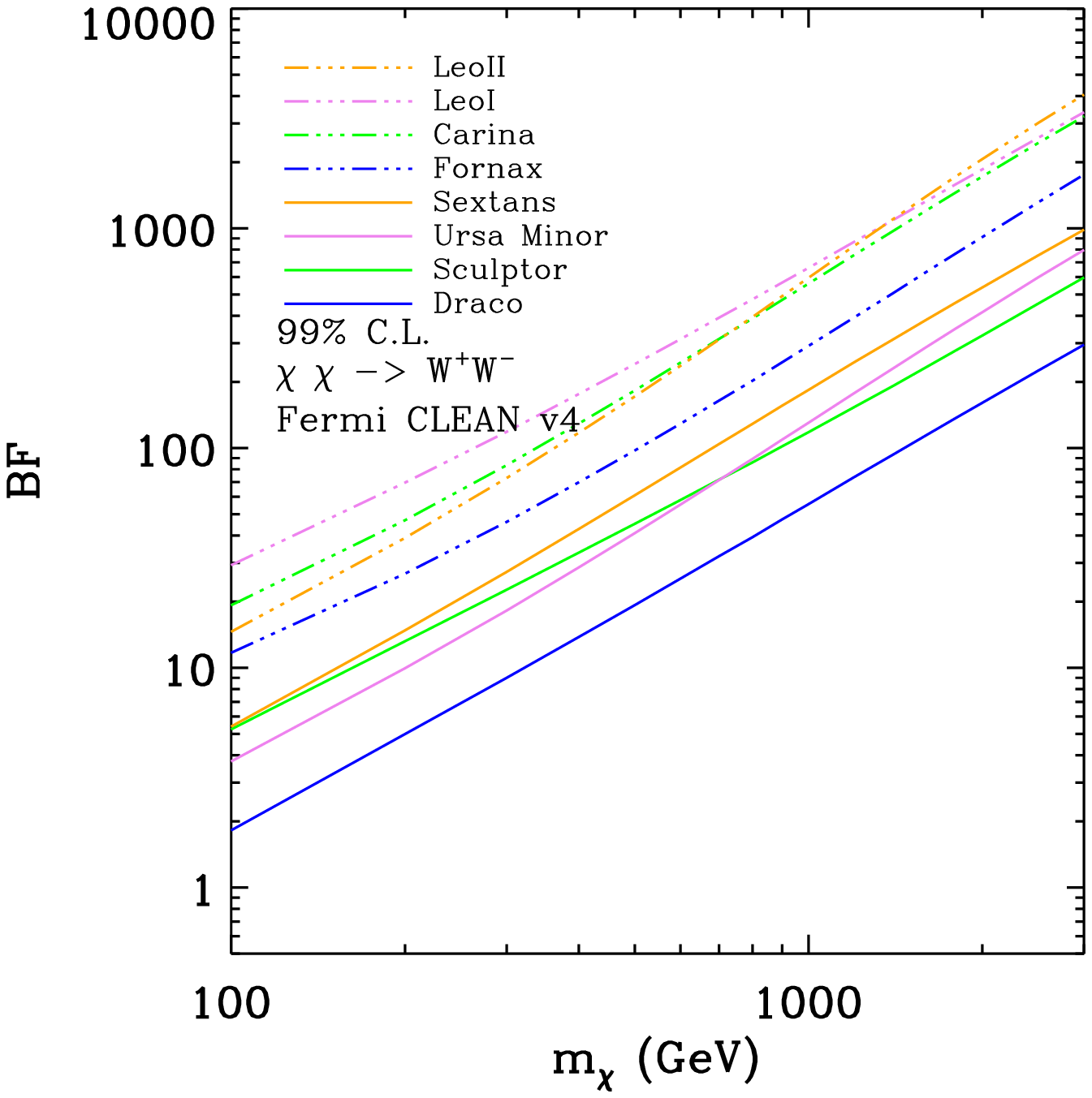}
\hspace{-0.35cm}
\includegraphics[width=2.35in,angle=0]{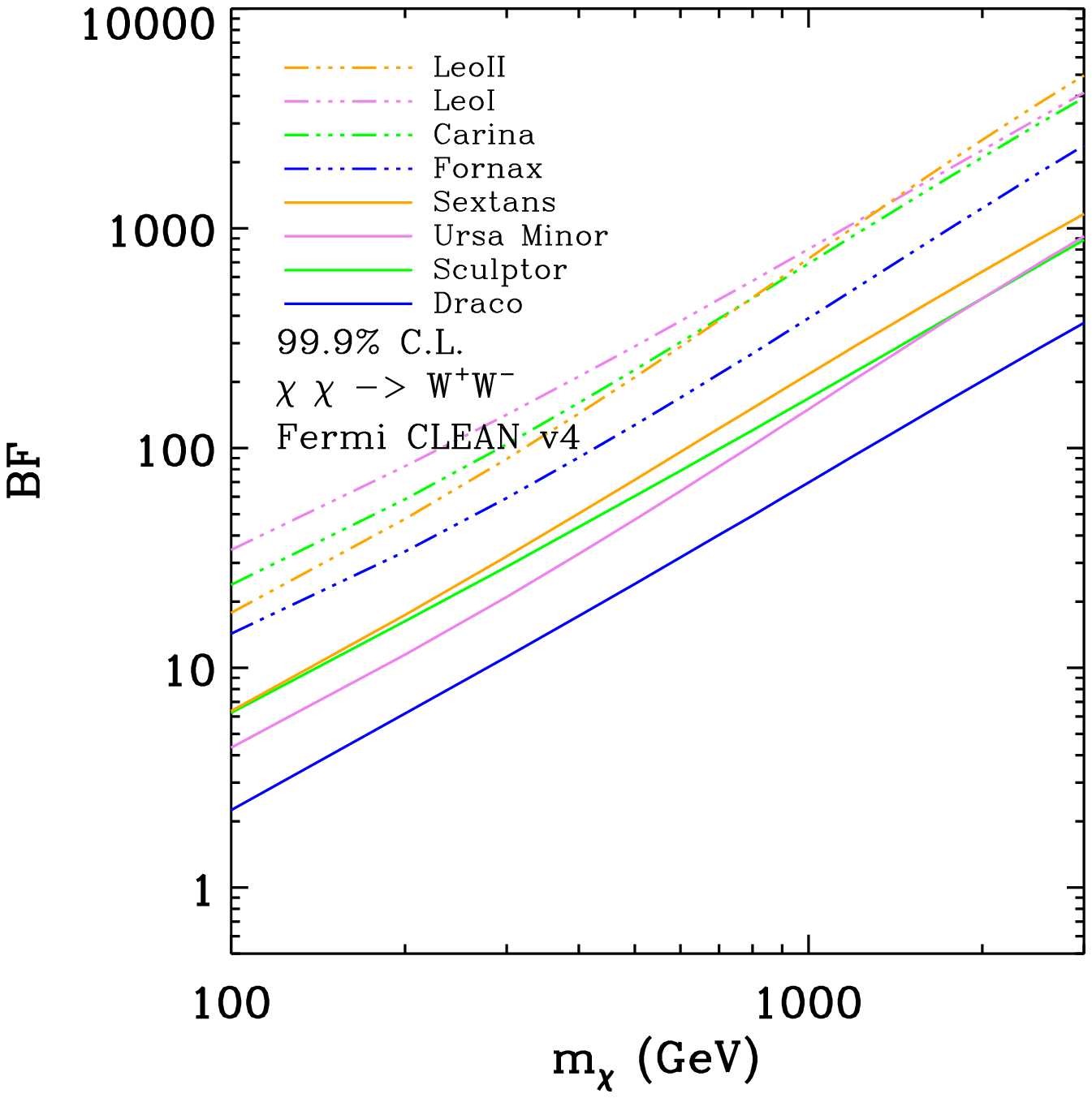}\\
\includegraphics[width=2.35in,angle=0]{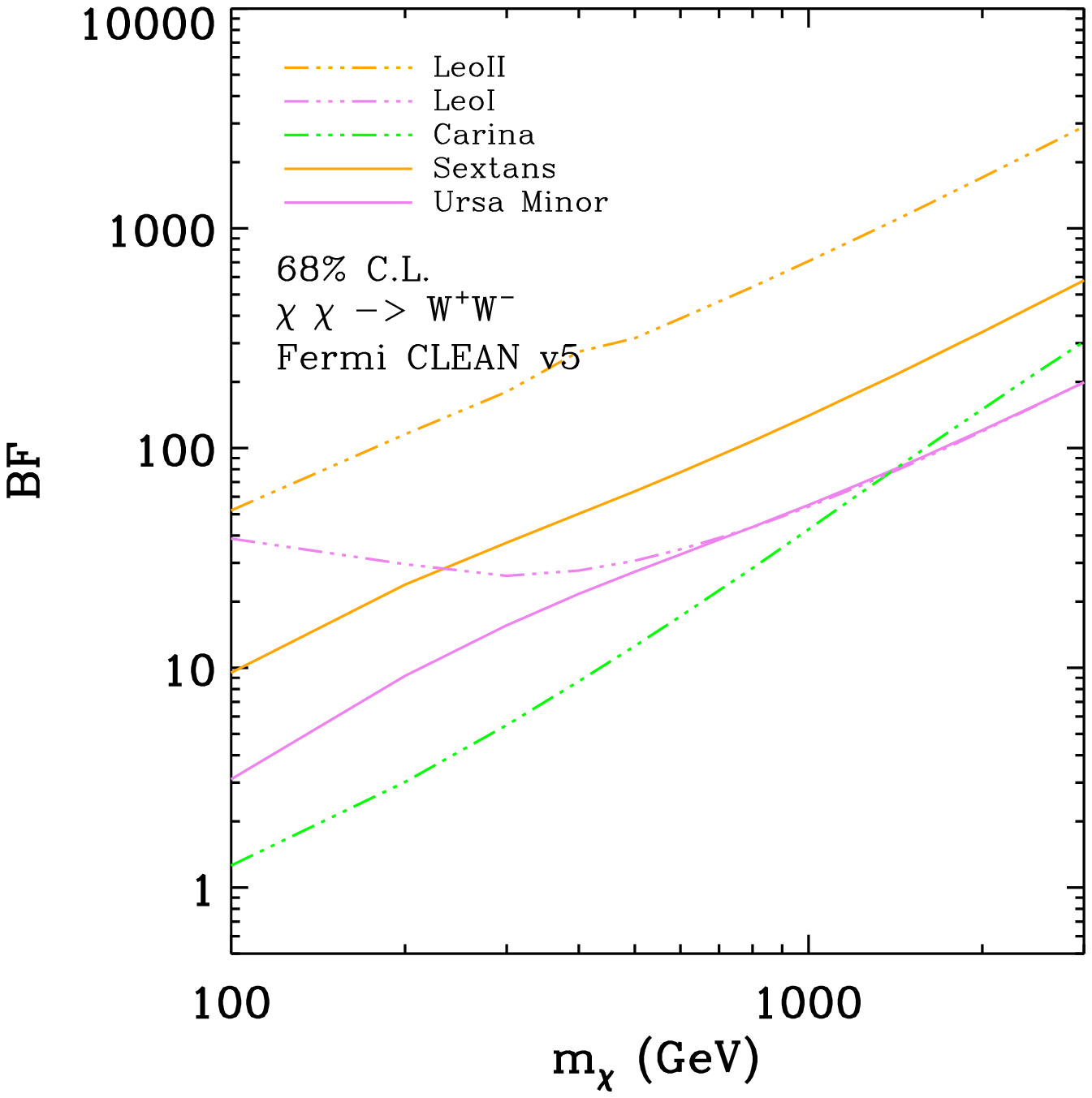}
\hspace{-0.35cm}
\includegraphics[width=2.35in,angle=0]{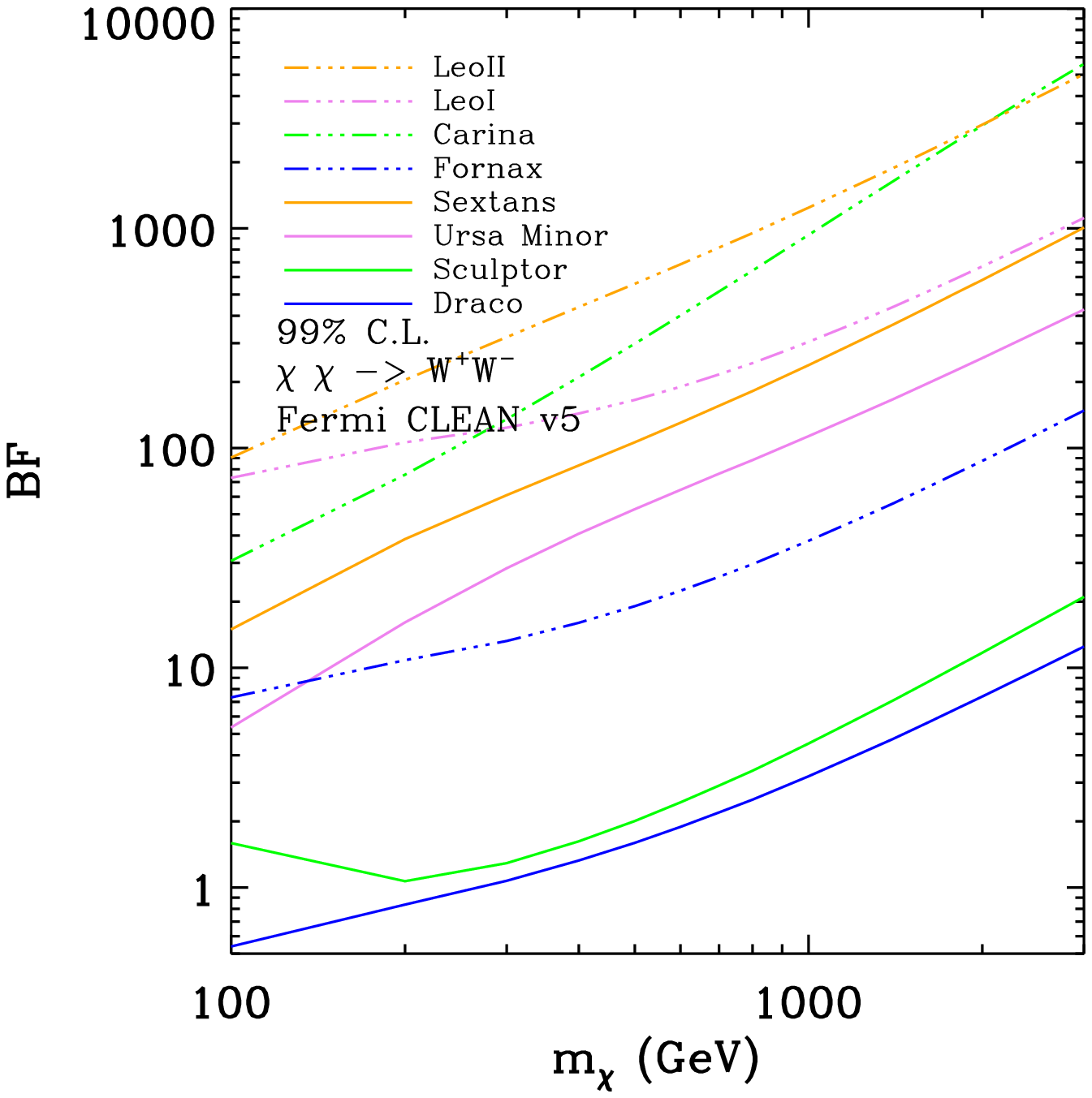}
\hspace{-0.35cm}
\includegraphics[width=2.35in,angle=0]{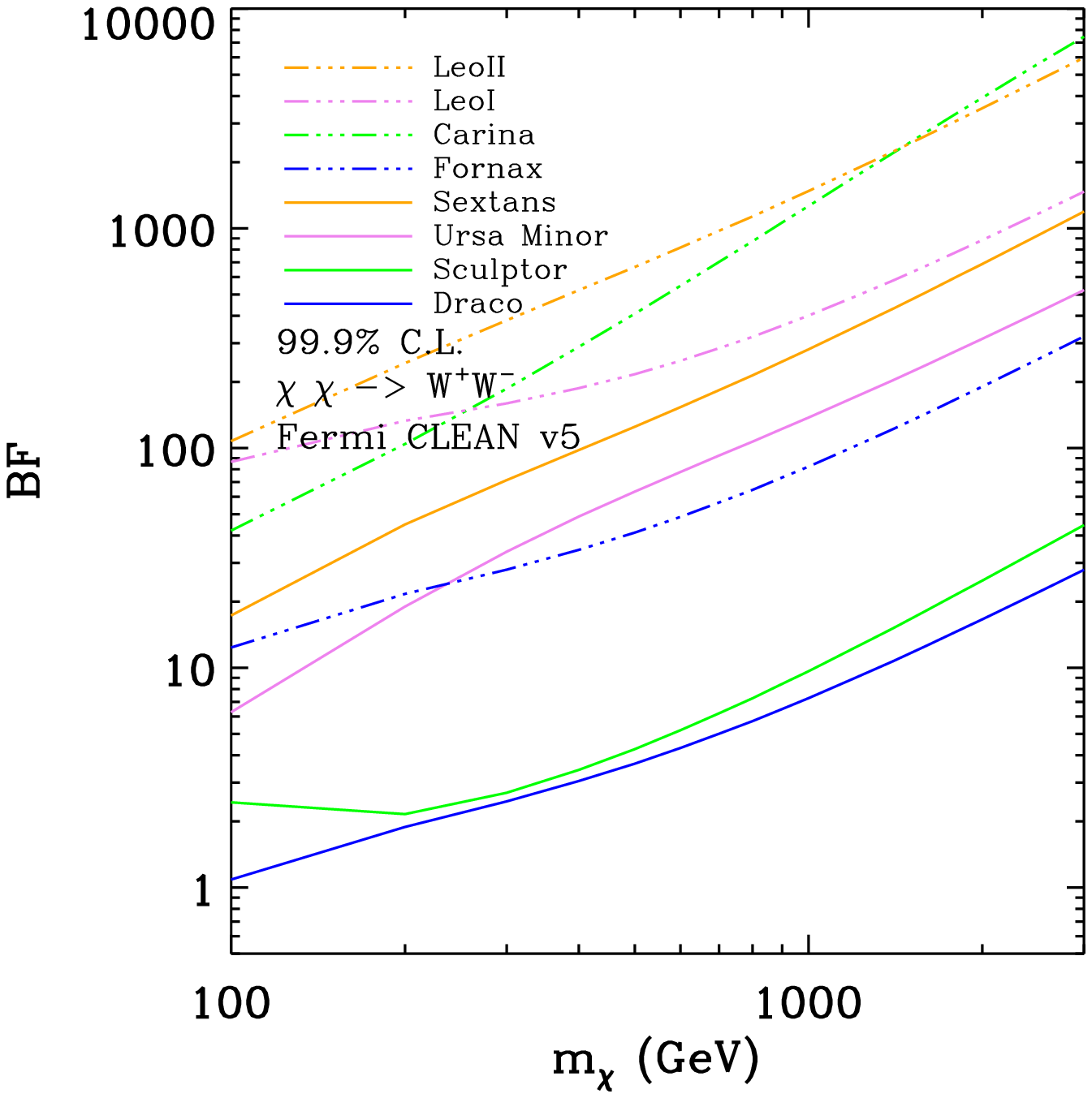}
\caption{Impact of different methods to estimate the background, on limits from dwarf spheroidals using \textit{Fermi} CLEAN class of data. \textit{Upper row}: using v4, \textit{lower row}: using v5. \textit{Left column:} 68$\%$ C.L., \textit{middle column:} 99$\%$ C.L., \textit{right column:} 99.9$\%$ C.L..}
\label{fig:targets}
\end{figure*}

As one can see from Fig.~\ref{fig:DMLimits_FermiColl}, the effect of different assumptions 
for the calculation of the residual flux can vary between targets and tends to give 
stronger constraints than our background models of version 4 and in some cases version 5 
(shown in Fig.~\ref{fig:targets}). Also for the case of $10^{\circ}$ ROI, a DM signal 
from Draco and Sculptor is excluded at 95$\%$ CL.

In Fig.~\ref{fig:targets}, we show the limits on DM annihilation rate,
 for our versions 4 and 5. We use as reference DM annihilating into $W^{+}W^{-}$
as in Fig.~\ref{fig:DMLimits_FermiColl}. We show 68$\%$, 99$\%$ and 99.9$\%$ CL 
limits from all eight dSphs at study. When lines are missing, that indicates that 
a DM annihilation signal for the given assumptions (DM mass, annihilation channel, 
and target assumptions) is excluded at the relevant CL. 
BF of less than 1 can be originating from an overestimation of the mean value of 
the $J$-factor, or an underestimation of the relevant error in calculating the 
$J$-factors (see discussion in section~\ref{sec:J-factors}). It can also be the 
case that DM annihilates to the relevant channel with a BR $< 1$, with the 
remaining channels of annihilation not giving a significant $\gamma$-ray signal 
at those energies.

In all cases the limits from the Draco dSph are the most stringent, with limits 
from Ursa Minor, Sextans, Sculptor and Fornax being the next most constraining 
set. Leo I, Leo II and Carina give limits systematically less stringent. That relevant 
power of dSph in setting limits on DM annihilation is also validated when using 
versions 1 to 3 (not shown here) and is also in rough agreement with results of 
other groups \cite{Abdo:2010ex, Ackermann:2011wa, Charbonnier:2011ft}. Yet the exact sequence of relative 
significance between dSphs changes between versions see for instance in 
Fig.~\ref{fig:targets} the relative change between Fornax and Sextans, and between 
Leo I and Carina.

We will consider that for our further discussion, it is enough to study the limits 
from four of those dSphs, namely Draco, Ursa Minor, Sextans and Sculptor, for which 
the residual spectra from our various versions have also been given in 
Figs.~\ref{fig:ExcessGammasVersions}-\ref{fig:ExcessGammas_v5v6}.
That choice is based both on the fact that these targets are more important in 
setting  DM constraints and also because they provide a good subset of dSphs to
show the issues that such an analysis on $\gamma$-ray data faces.

\begin{figure*}[t]
\centering\leavevmode
\includegraphics[width=3.40in,angle=0]{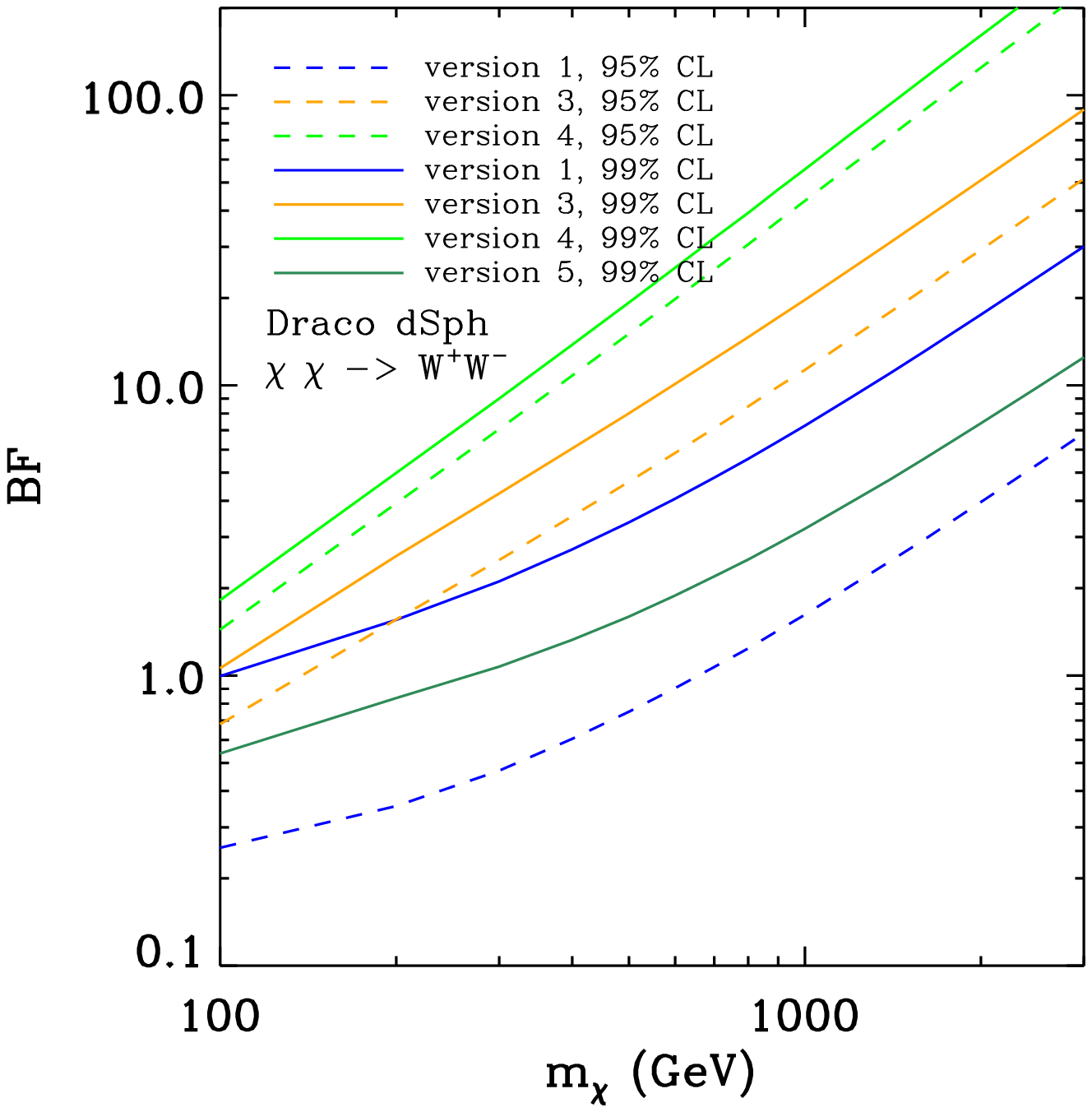}
\hspace{-0.1cm}
\includegraphics[width=3.40in,angle=0]{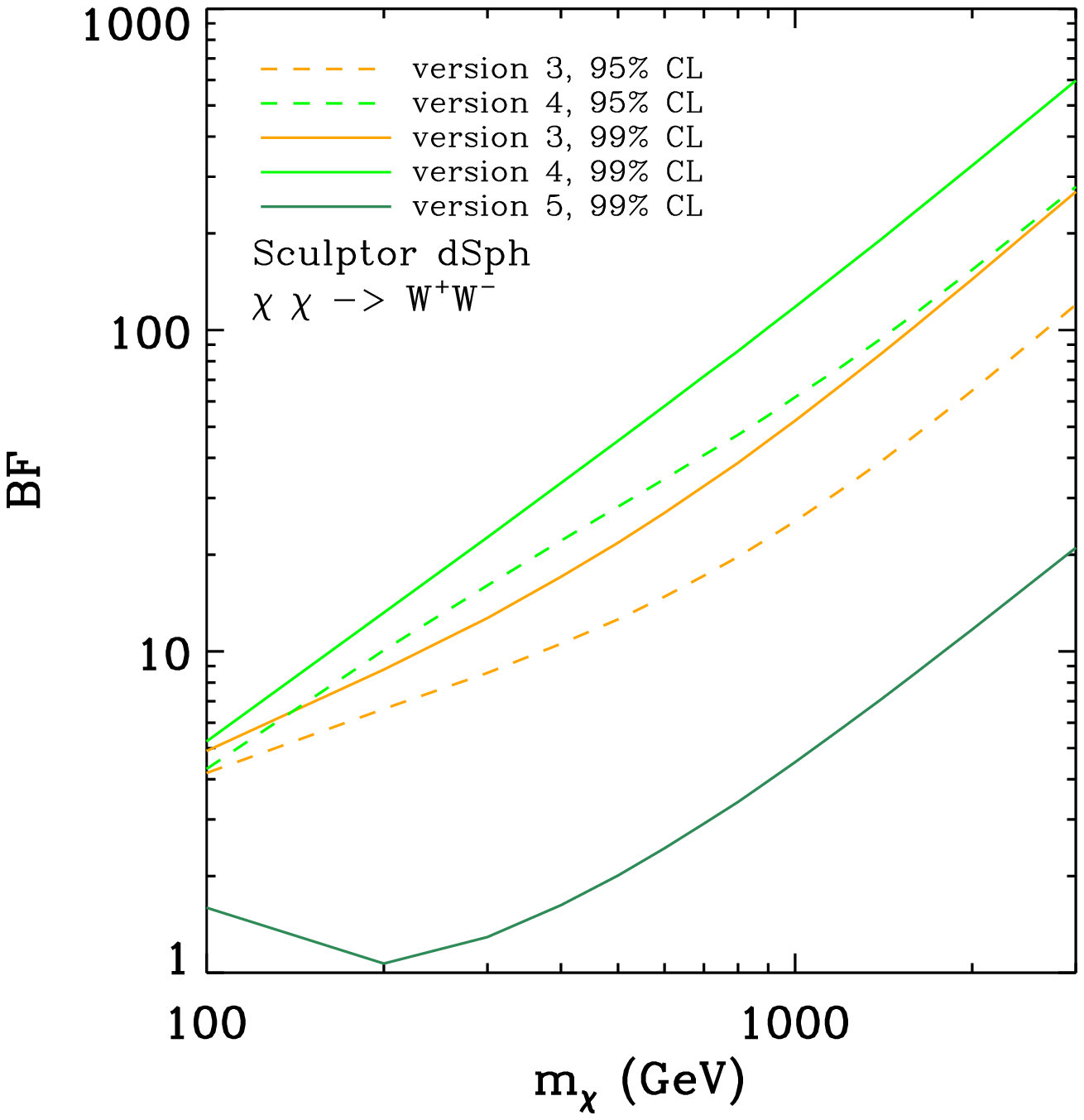}
\includegraphics[width=3.40in,angle=0]{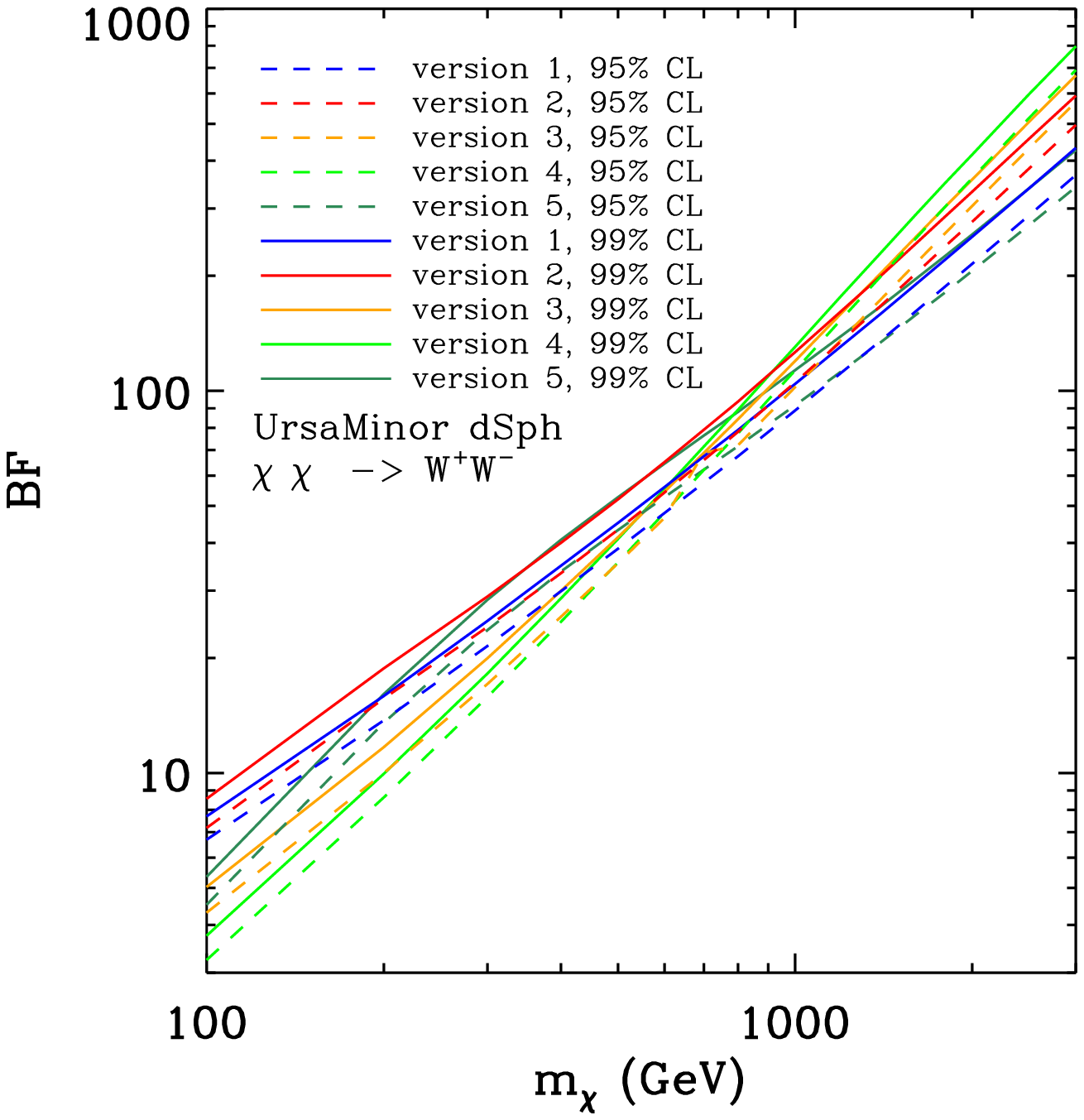}
\hspace{-0.1cm}
\includegraphics[width= 3.40in,angle=0]{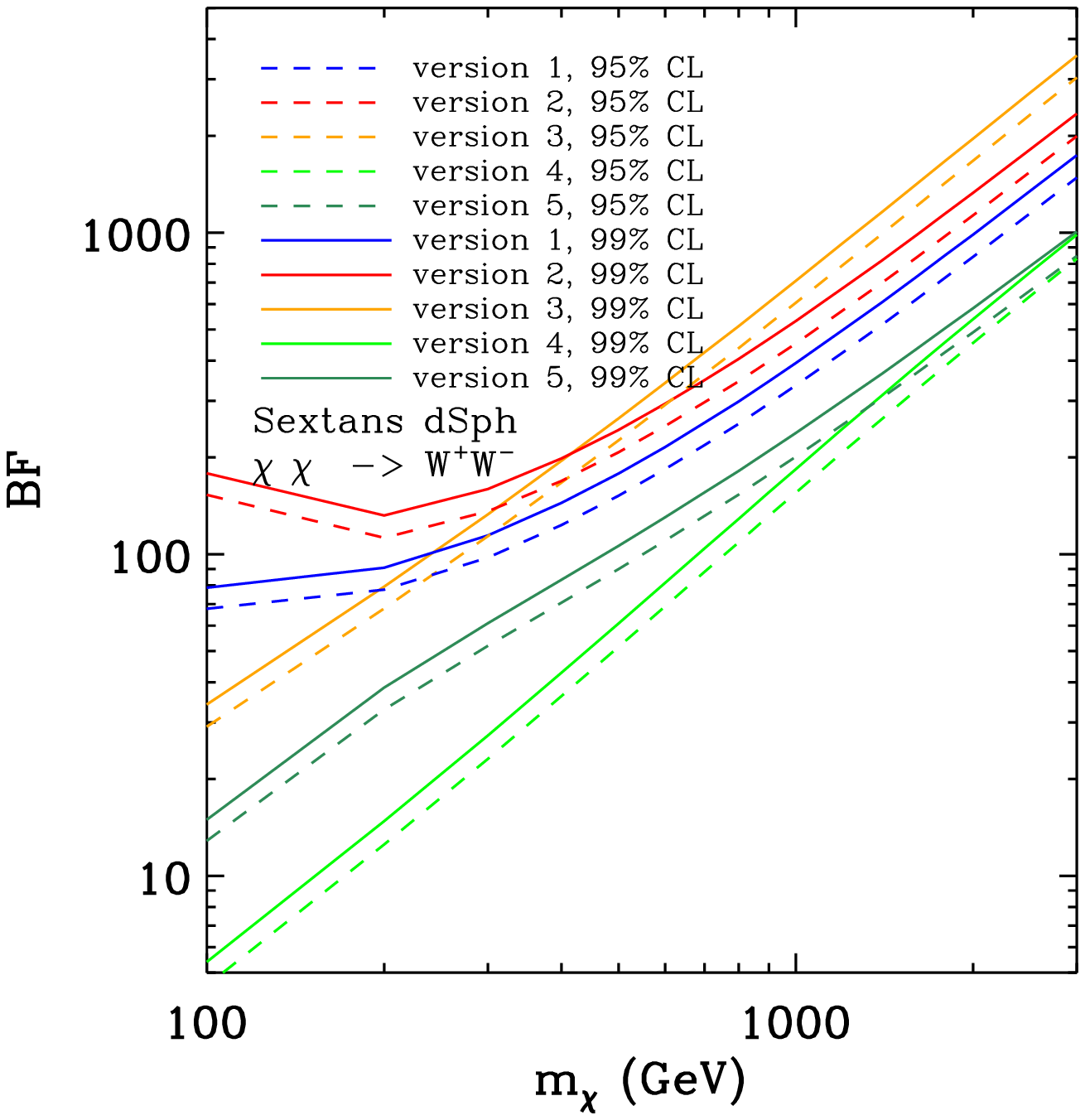}
\caption{95$\%$ C.L. and 99$\%$ C.L. limits from dwarf spheroidals for versions 1-5 using \textit{Fermi} CLEAN class data. \textit{Upper left:} Draco, \textit{upper right:} Sculptor, \textit{lower left:} Ursa Minor, \textit{lower right:} Sextans. For certain targets as Ursa Minor and Sextans, much more robust limits on DM annihilation can be achieved.}
\label{fig:5versions}
\end{figure*}

In Fig.~\ref{fig:5versions} we show the limits on DM annihilation from these 
dSph galaxies, using the five different versions of deriving residual fluxes.
While the limits from Draco dSph are more constraining as we show in 
Fig.~\ref{fig:targets}, they vary significantly between different methods. In 
fact, for Draco, using our version 2 we even get to exclude DM annihilating to 
$W^{+}W^{-}$ at $99\%$ CL. Of interest is to consider what is the variation 
in the DM limits between version 4 (light green) and version 5 (dark green) 
in Fig.~\ref{fig:5versions}. Both versions assume the same set of signal ROIs, 
and consider background ROIs that as we described in section~\ref{sec:FermiGamma} 
ensure proximity to the targets. Yet the change in the exact set of background 
ROIs is enough to influence the limits on Draco between v4 and v5 by a factor 
of 4 at $m_{\chi}=100 GeV$ to a factor of 20 at $m_{\chi}=3 TeV$. Similar is the 
case with limits from Sculptor. On the contrary, Ursa Minor gives much more 
consistent limits between different versions, all of them agreeing within a 
factor of 3. Specifically versions 4 and 5 are consistent with each other by
a factor of 2 or less. This fact suggests that while the limits from Ursa Minor 
are less constraining, they are much more robust. Finally Sextans, gives very 
robust limits at high masses (all versions agree within a factor of 3 for 
masses heavier than $m_{\chi} = 500$ GeV), 
but suggests greater relative differences in the DM limits at lower masses.   
Still even for Sextans our preferable versions 4 and 5, agree within a factor 
of 2 or less; proving an other target to derive robust limits.

There are various reasons why some targets provide more consistent limits 
than others when changing the background ROIs. One is that increasing the 
angular size of background ROIs point sources that may be far from the 
center of the dSph start having an impact. This is the case of Draco and
Sextans at low $E_{\gamma}$. An other issue can be the existence of some 
structure in the galactic diffuse background within the ROIs, usually 
due to the presence of an ISM gas structure. 

For the case of an ISM gas structure within the ROIs of interest, its 
contribution to the background $\gamma$-rays comes from bremsstrahlung 
off CR $e^{\pm}$ or pi0 (and other mesons) decay, produced from nucleon nucleon
 collisions (mainly pp) (see discussion in section~\ref{sec:FermiGamma}). 
To calculate that contribution, one needs to know the actual location of the 
gas in the Galaxy and to have a good understanding of CR density distribution
 in the Galaxy. The latter has been addressed in various works 
\cite{Strong:2004de, Strong:2007nh, 2011A&A...531A..37D, Cholis:2011un, 2012arXiv1202.4039T} where to ensure agreement with $\gamma$-ray 
data, large windows of the sky are used (for a recent analysis see 
\cite{2012arXiv1202.4039T}). 
The former, knowing the actual location of the gas structure is usually based 
on assumptions on rotation curves\cite{2012arXiv1202.4039T, Strong:2004de, Nakanishi:2006zf, Nakanishi:2003eb} 
(or some velocity fields from stars motions\cite{2008ApJ...677..283P}), which are generally 
correct for low latitudes since the gasses there are expected to follow well, the 
general motion. Yet at high distances from the galactic disk (and thus high 
latitudes), as are all the dSphs at study\footnote{dSphs at low latitudes have been 
generally avoided in studies in order not to have too much galactic $\gamma$-ray 
contribution into the ROI.}, the peculiar velocities of 
these ISM gasses can have a strong impact in setting their position in the Galaxy (see also discussion in \cite{Moskalenko:2006zy}).  
Placing the small structures ISM gasses in the wrong position in the Galaxy (without 
changing its longitude and latitude coordinates), would result in making wrong 
assumptions on the CR electron and proton density environment that these structures 
exist in. For instance the steady state CR protons density at $b = 30^{\circ}$,
may vary by a factor of $50 \%$ from moving a gas from 2 kpc to 4 kpc away from us
(see \cite{Cholis:2011un} for cases of CR protons distribution profiles)
\footnote{Only if all gas at high latitudes is local and taken as local
can we trust the galactic diffuse $\gamma$-ray component. This is not always
the case.}.

The bremsstrahlung and pi0 components depend proportionally to the CR $e^{\pm}$
and the CR $p$ densities (equivalently). Moreover the CR $e^{\pm}$ spectra
due to fast energy losses, can vary significantly between different locations
in the Galaxy.    
While these errors may seem insignificant we remind the reader that the residuals 
that we are in search of, are
usually of $O(0.1)$ of the expected galactic diffuse component.   
Models such as that of \cite{2012arXiv1202.4039T} or the "gal\_2yearp7v6\_v0" while 
they can be consistent with 
$\gamma$-rays at large windows, can not be considered correct at small
angular windows as those discussed.

The method of \cite{Ackermann:2011wa}, that models the background components 
within the ROI, as discussed in section~\ref{sec:FermiGamma}, assumes 
too many degrees of freedom, since the p.s. flux normalizations and 
spectral power laws, the isotropic component normalization and galactic 
diffuse component normalization, are let to vary freely. The isotropic 
flux normalization should not be let to vary, especially between ROIs 
for different dSphs (see discussion~\ref{sec:FermiGamma}).

For those reasons we consider preferable to use in deriving limits from
dSphs, only the targets that give consistent limits between versions 4 
and 5 and for which we consider that the uncertainties in the J-factors 
are properly taken into account.
From that class of targets the strongest limits come from Ursa Minor and
Sextans. 

An alternative method of analysis of the $\gamma$-ray data from these 
targets could be, masking out the point sources and the possible galactic
features. Given that there are typically $O(100)$ $\gamma$-rays above 
10 GeV within $5^{\circ}-10^{\circ}$ window around these targets and for some
targets like Draco $O(10)$ known point sources, such an analysis may 
not be optimal due to too few $\gamma$-rays remaining (especially for cases 
like Draco dSph). Yet that would be a question for a separate rigorous 
analysis.       
                                                                                           
\section{Constraints}
\label{sec:constraints}

\begin{figure*}[t]
\centering\leavevmode
\includegraphics[width=3.40in,angle=0]{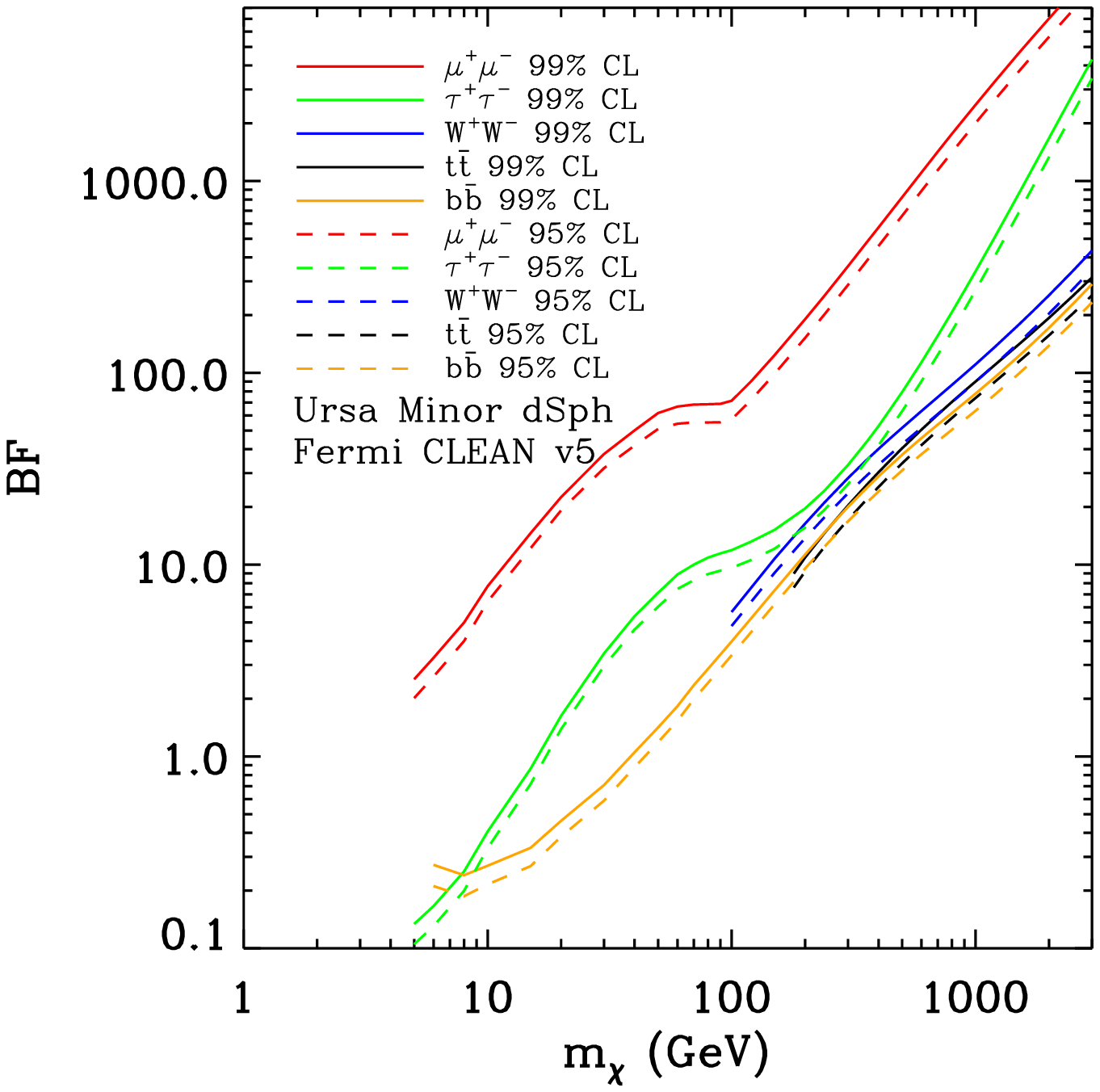}
\hspace{-0.1cm}
\includegraphics[width=3.40in,angle=0]{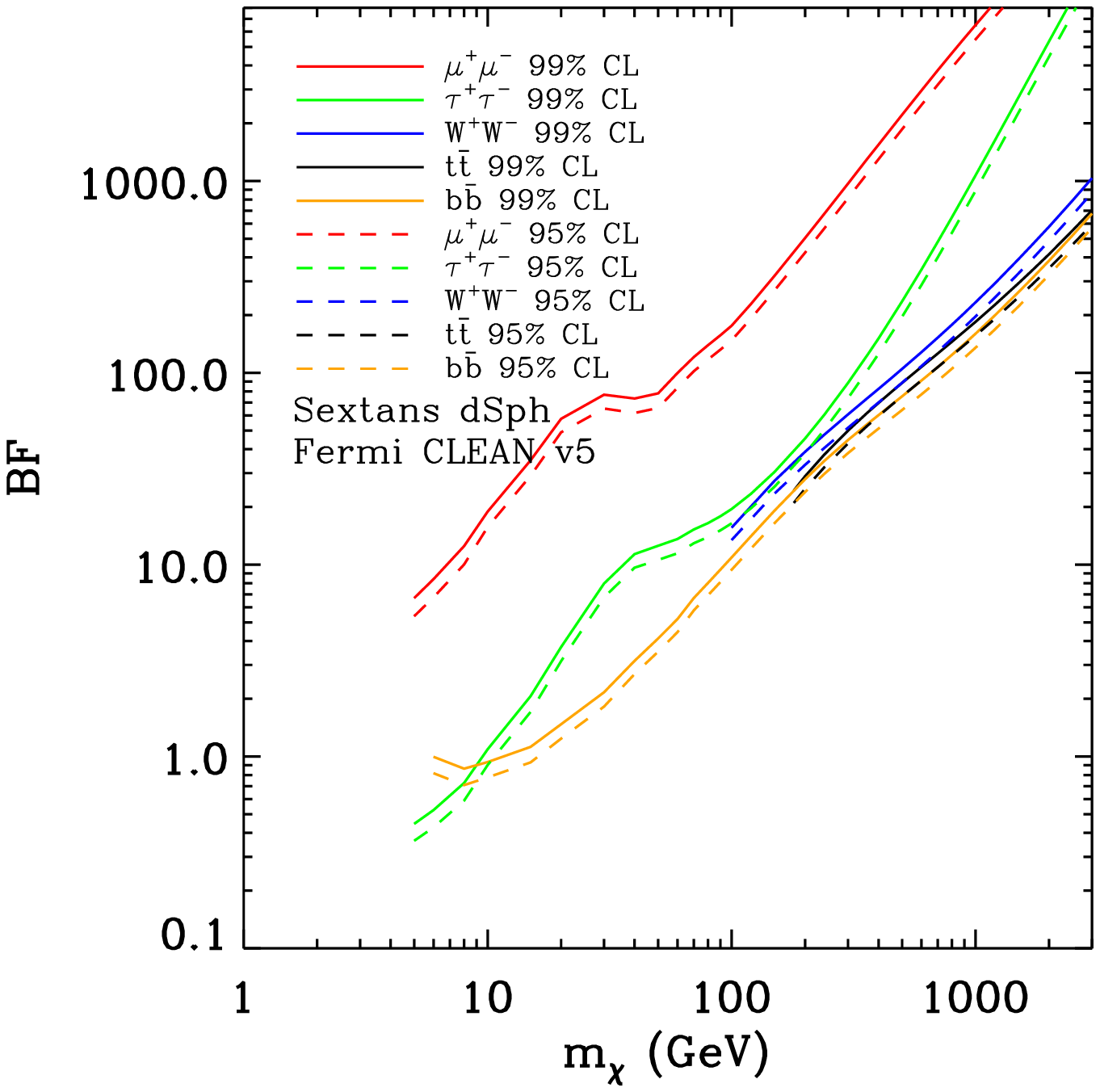}
\caption{99$\%$ and 95$\%$ C.L. limits from Ursa Minor (\textit{left}) and Sextans (\textit{right}) for different annihilation channels. Red: $\chi \chi \longrightarrow \mu^{+}\mu^{-}$, green: $\chi \chi \longrightarrow \tau^{+}\tau^{-}$, orange: $\chi \chi \longrightarrow b\bar{b}$, blue: $\chi \chi \longrightarrow W^{+}W^{-}$, black: $\chi \chi \longrightarrow t\bar{t}$.}
\label{fig:DMchannels}
\end{figure*}

Having shown in Fig.~\ref{fig:5versions}, versions 4 and 5 and using the 
$W^{+}W^{-}$ channel as a guide, we notice that version 5 gives slightly weaker
limits. In Fig.~\ref{fig:DMchannels} we show the limits on DM 
annihilation BF only from version 5, for Ursa Minor (left column), 
and Sextans (right column). These limits are true conservative limits 
since apart from being slightly weaker than those of version 4, they 
come from targets that can give robust limits. These limits also come 
from a version that for the reasons described in 
section~\ref{sec:FermiGamma} addresses all the issues that can arise
in calculating residual spectra and DM limits from dSphs.
Finally, the limits are conservative, since we used only the prompt 
part of $\gamma$-rays from DM annihilations, including electroweak corrections
\cite{Ciafaloni:2010ti, Cirelli:2010xx}, but ignoring ICS from and 
bremsstrahlung off the $e^{\pm}$ that are also final products of DM 
annihilations in all the channels shown.

Apart from the $\chi \chi \longrightarrow W^{+}W^{-}$, in
Fig~\ref{fig:DMchannels} we give the $95\%$ and $99\%$ CL limits for 
general phenomenological channels as
$\chi \chi \longrightarrow \tau^{+}\tau^{-}$, 
$\chi \chi \longrightarrow \mu^{+}\mu^{-}$, 
$\chi \chi \longrightarrow b\bar{b}$ and $\chi \chi \longrightarrow t\bar{t}$.

As is clear Ursa Minor gives stronger limits than Sextans by a 
factor of 2.5 at 1TeV to a factor of 4 at 10 GeV for all channels shown.
Comparing our limits to those of \cite{Ackermann:2011wa} we get for
Ursa Minor a factor of 5 stronger limits at 10 GeV and a factor of 2 at 
1TeV. Although given that there is an uncertainty by a factor of 3 in the
actual limits derived from Ursa Minor between 100 GeV and 3 TeV (shown 
in Fig~\ref{fig:5versions} bottom left), and that at the 10 GeV range we 
consider -using the $b\bar{b}$ channel- those limits to be consistent to that of \cite{Ackermann:2011wa}
for the same dSph. 
For Sextans our limits compared to those of \cite{Ackermann:2011wa}, 
are at the TeV range a factor of 2 stronger. At low masses, as is 
shown in Fig.~\ref{fig:5versions} (bottom right) it is unsafe to 
make a claim based on the robustness of these limits. 
Finally, when comparing our strongest limits from Ursa Minor to the limits 
from the joint likelihood of \cite{Ackermann:2011wa},  we get that 
our constraints between $m_{\chi}$ of 100 GeV and 1 TeV are a factor of 2 to 3 
weaker for all channels.
Between $m_{\chi}$ of 10-100 GeV the relative difference of our limits to those of 
\cite{Ackermann:2011wa} fluctuates for $b\bar{b}$ and $\tau^{+}\tau^{-}$
with our limits being slightly tighter at 10 GeV. For the $\mu^{+}\mu^{-}$ 
channel our limits remain always a factor of 2 more weak. 

Limits from VERITAS \cite{collaboration:2011sm, Aliu:2012ga} are most 
competitive at the TeV range of masses. Yet even at $m_{\chi} = 1$ TeV, 
our limits are tighter by a factor of 10 (for the $b \bar{b}$ channel). 
Limits  from MAGIC-I telescope \cite{Aleksic:2011jx, Paiano:2011uq} 
give slightly weaker limits to those of VERITAS but are expected to get 
more competitive with MAGIC-II up-grate 
\cite{Bringmann:2008kj, Tibolla:2012zn}. 

Given that in our analysis we study separately each dSph galaxy/target 
in order to understand the robustness of the DM annihilation limits derived
and then based \textit{only} on the targets that give the most robust limits, claim
constraints on DM annihilation rates, our results are safer, while providing 
also similarly tight limits. Since in our process it is of equal importance, 
to understand well the relevant uncertainties on the $J$-factors, we have left 
the study of dSphs such as the Ursa Magor II, Seque I and Coma Berenices for 
future work where the uncertainty in the relevant $J$-factors is well modeled.

\begin{figure*}[t]
\centering\leavevmode
\includegraphics[width=3.40in,angle=0]{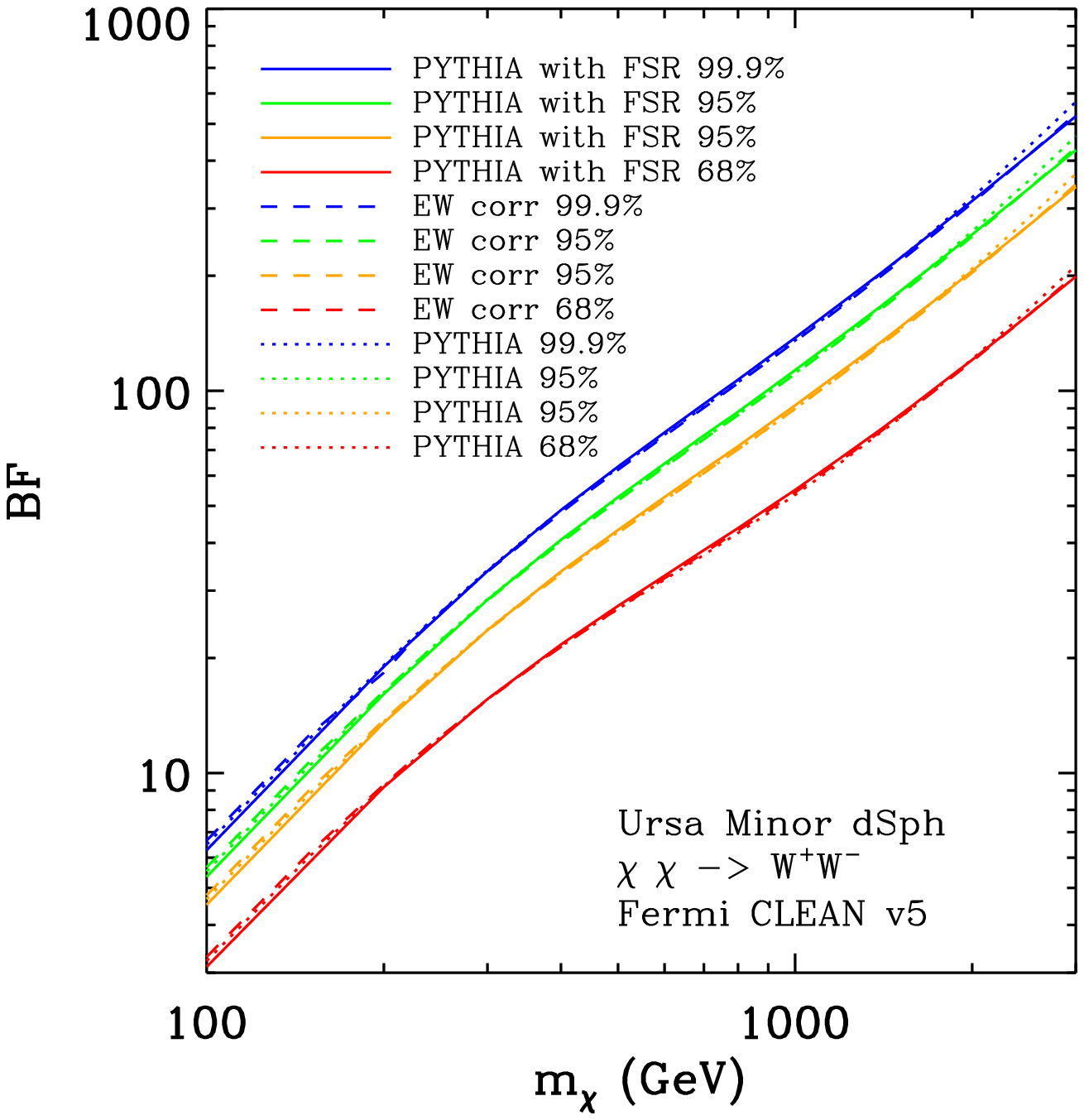}
\hspace{-0.1cm}
\includegraphics[width=3.40in,angle=0]{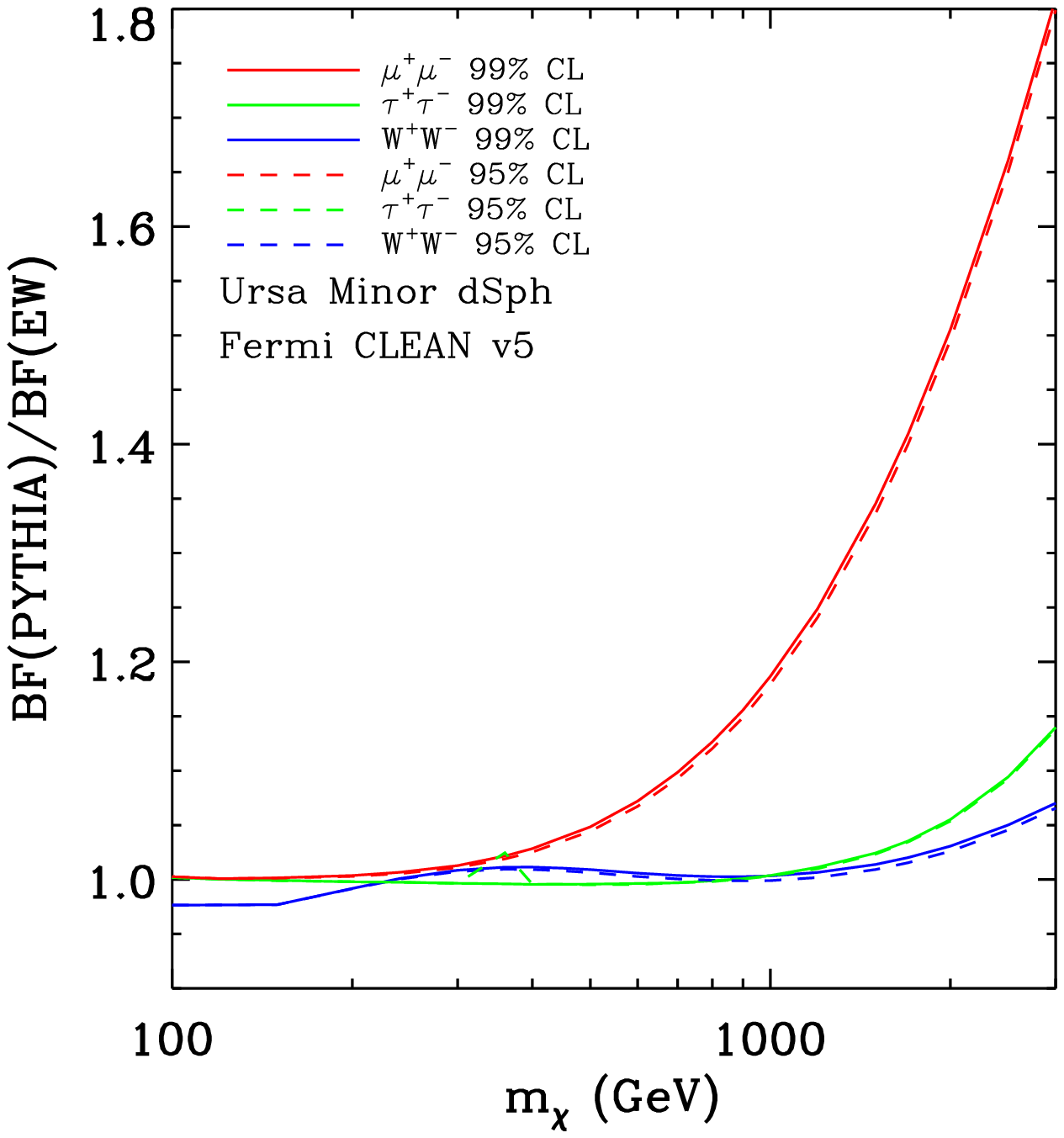}
\caption{\textit{Left}: limits from Ursa Minor, on DM annihilation to 
$W^{+}W^{-}$, \textit{blue} 99.9$\%$, \textit{green}: 99$\%$, 
\textit{orange}: 95$\%$ and \textit{red}: 68$\%$ C.L. using \textit{Fermi} 
CLEAN class data. \textit{Solid lines}: PYTHIA with FSR, \textit{dotted lines}:
 PYTHIA without FSR, \textit{dashed lines}: including EW corrections. 
\textit{Right}: for channels of DM annihilation to $\mu^{+}\mu^{-}$, 
$\tau^{+}\tau^{-}$ and $W^{+}W^{-}$ we show the ratio of limits using the 
PYTHIA spectra to the limits using EW corrected spectra.}
\label{fig:DMlimits}
\end{figure*}

In Fig.~\ref{fig:DMlimits} (left), we also give the $68\%$ to
$99.9\%$ CL limits for $\chi \chi \longrightarrow W^{+}W^{-}$.
We note that the $99.9\%$ CL are only by a factor of 2 more stringent
than the $68\%$ for both dSphs. We show limits using either only the PYTHIA
simulation without final state radiation (FSR), with FSR (used in 
Fig.~\ref{fig:DMLimits_FermiColl}-\ref{fig:5versions}) and including 
electroweak (EW) (Fig.~\ref{fig:DMchannels})corrections. 

For the $W^{+}W^{-}$ channel, the difference between the three 
cases of prompt $\gamma$-ray spectra calculation, is not significant.
Yet as we show in Fig.~\ref{fig:DMlimits} (right), EW corrections are 
more important for channels as the leptonic ones, as 
is the  $\mu^{+}\mu^{-}$ (see also discussion in
\cite{Ciafaloni:2010ti, Cirelli:2010xx, DeSimone:2012hj, Ciafaloni:2012gs}).

While our limits from Ursa Minor are slightly weaker than the joint 
likelihood of \cite{Ackermann:2011wa}, they are still stronger than 
limits using $\gamma$-rays at medium and high latitudes \cite{Cirelli:2009dv}
and can put constraints on DM at $\gamma$-rays from regions sarounding 
the GC where an annihilation signal may lay especially in the leptophilic DM 
case \cite{Cholis:2008wq, Cholis:2009gv, Cirelli:2009dv, Papucci:2009gd}.
 
Limits from the galactic center region 
\cite{Meade:2009rb, Cirelli:2009dv, Papucci:2009gd, Abramowski:2011hc, Abazajian:2011ak}
are more competitive above 100 GeV, but suffer strongly form uncertainties 
in the exact profile assumed and for the Sommerfeld enhancement cases also from 
the different velocity dispersion locally than in the GC 
\cite{Cholis:2009va} and the extend of DM substructures in the Galaxy 
\cite{Cholis:2010px, Slatyer:2011kg}.
A case of a DM particle with mass $\approx 10$ GeV annihilating mainly into leptons
\cite{Hooper:2010mq, Hooper:2011ti, Hooper:2012ft} to account for the suggested 
excess of $\gamma$-rays towards the GC \cite{Hooper:2010mq}, can not be ruled 
out since as we described in section~\ref{sec:FermiGamma}  
our limits bellow $m_{\chi}<10$ GeV depend only on the low energy $\gamma$-ray 
bins and are more sensitive to background assumptions. 

Observations of galaxy clusters at $\gamma$-rays have also given limits on 
annihilating DM \cite{Pinzke:2011ek, Ando:2012vu, Han:2012au} which though 
in some cases are stronger, depend greatly (up to 3 orders of magnitude in 
the strength of the limits \cite{Han:2012au}) on the significance of substructure 
in the outer parts of the clusters and also on taking into account the 
$\gamma$-ray production from SM particles, which is significant for those 
targets of DM annihilation signals.    

Finally limits on DM annihilation from antiprotons 
\cite{Cirelli:2008pk, Donato:2008jk, Cholis:2010xb, Evoli:2011id}, are 
sensitive on galactic propagation assumptions \cite{Evoli:2011id}. 
Yet when comparing to limits assuming conventional propagation assumptions 
our limits for the 
$\chi \chi \longrightarrow W^{+}W^{-}$, $b \bar{b}$ are weaker at $\approx$100 
GeV and as strong as at 1 TeV, as those from antiprotons. Also for the 
$\chi \chi \longrightarrow \mu^{+}\mu^{-}$ including EW corrections that are
responsible for the emission of $\bar{p}$\cite{Ciafaloni:2010ti, Cirelli:2010xx}
our limits are comparable to those from antiprotons \cite{Evoli:2011id} 
above a TeV.  

 
\section{Conclusions}
\label{sec:conclusions}

In this work we have revisited the constraints on DM annihilation 
from dSph galaxies which have been claimed to provide very tight limits 
on DM annihilation cross-sections \cite{Abdo:2010ex, Ackermann:2011wa}. 

In doing so, we have discussed why cored DM 
halo profiles are preferable to the NFW profile, or in general cuspy 
profiles for dSphs. The main argument in favor of the cored profiles 
comes from observations as we discussed in section~\ref{sec:J-factors}. 
There is also theoretical
motivation though: some early strong outflow event depleted the 
region of the dSphs from baryons suppressing also the baryonic infall 
at later stages, thus making the dSphs less cuspy than standard 
galaxies or galaxy clusters \cite{Governato:2012fa}, suggested also 
by observations.
Choosing the correct DM profile can have a result on the limits on 
DM annihilation cross-sections vs mass that are usually presented. 
What we actually observe and constrain is the total annihilation 
rate which is proportional to the product of the annihilation cross-section 
and the $J$-factor that accounts for the observed DM annihilation profile
integrated along the lines of sight within the observation angular window. 

Apart from the profile itself, in setting the correct limits on 
DM annihilation cross-section it is also important to have a 
good understanding of the relevant uncertainty in the $J$-factors, which
in many cases of dSphs is not insignificant. We choose to study eight
dSphs for which we have accurate measurements of both the $J$-factors 
and their errors \cite{2012MNRAS.tmp.2161S}; namely Carina, Draco, 
Fornax, Leo I, Leo II, Sculptor, Sextans and Ursa Minor. 

For every target we model the background using alternative methods,
in order  to discriminate targets for which we have good understanding 
of the background and can give robust residuals leading to robust limits 
on DM annihilation.

All alternative background estimation/residual estimation methods, avoid 
having many degrees of freedom that could result in hiding any small excess. 
The only degrees of freedom are actually related to the borders of 
the regions 
of interest. These in turn are related (as are the borders of the signal ROIs)
to the PSF at different energies. In fact we use we use the fact that all dSphs 
are practically 
point sources, thus their size (which defines the size of the signal ROIs) is 
\textit{not} constant but depends sensitively on the energy of the observed
$\gamma$-rays. By using $\gamma$-rays with energies between 1 and 100 GeV we 
can avoid having low energy $\gamma$-rays dominating our results. Above 100 
GeV very few photons are observed around each dSph. In extracting residuals 
our methods cancel out the CR contamination and isotropic gamma-rays components,
which at higher latitudes and energies can be significant leaving only their
Poisson noise to have an impact on our limits.

Having calculated our residual spectra from alternative methods, by comparing 
them, we can find for which dSphs close-by (in angle) point sources and 
galactic diffuse features can be significant in setting DM limits. The 
dSphs for which by changing among alternative methods (alternative choices for 
the ROIs), the limits on DM annihilating change dramatically 
(see Figs.~\ref{fig:targets} and~\ref{fig:5versions}) are excluded from further 
analysis. For such targets a physical model properly accounting for the 
$\gamma$-ray fluxes from point sources and galactic diffuse features is necessary. 
Such targets that are excluded are Draco and Sculptor dSphs even though 
they do provide tight limits in some cases.
 
Among the "clean" background dSph galaxies that we studied, Ursa Minor provides the 
best target to set robust and tight limits on DM annihilation. 
We give limits for $\chi \chi \longrightarrow t \bar{t}$, $W^{+}W^{-}$, $b \bar{b}$,
$\tau^{+}\tau^{-}$ and $\mu^{+}\mu^{-}$ annihilation channels, shown in 
Fig.~\ref{fig:DMchannels}.
Our limits form Ursa Minor are stronger than those of 
\cite{Ackermann:2011wa} for Ursa Minor by a factor of 5 (at DM mass 
$m_{\chi} \approx 10$ GeV) to 2 at $m_{\chi} \approx 1$ TeV for $\chi \chi \longrightarrow b \bar{b}$. 
When comparing our limits from Ursa Minor to those from the joint 
likelihood of \cite{Ackermann:2011wa} then we get comparable 
limits at 10 GeV for the  $b \bar{b}$, $\tau^{+}\tau^{-}$ channels,
a factor of 2 weaker limits for the $\mu^{+}\mu^{-}$ channel and a factor of
2-3 weaker limits for all channels above 100 GeV. 
Yet the joint 
likelihood method does not take into account our arguments about robustness
of DM limits between the individual targets.

Finally dSph galaxies such as Ursa Major II, Seque I and Coma Berenices
are possibly the most interesting targets among those not discussed in this 
paper, to look for future work. Such a work will have to include for them 
as well, accurate estimates on the $J$-factors best values and uncertainties. 

\vskip 0.2 in
\section*{Acknowledgments}

We warmly thank Sam Leach, Maryam Tavakoli, Piero Ullio, Neal Weiner and Christoph Weniger
for the valuable discussions that we have shared.    
\vskip 0.05in


\bibliography{DMdwarfSpheroidal}

\begin{thebibliography}{149}
\expandafter\ifx\csname natexlab\endcsname\relax\def\natexlab#1{#1}\fi
\expandafter\ifx\csname bibnamefont\endcsname\relax
  \def\bibnamefont#1{#1}\fi
\expandafter\ifx\csname bibfnamefont\endcsname\relax
  \def\bibfnamefont#1{#1}\fi
\expandafter\ifx\csname citenamefont\endcsname\relax
  \def\citenamefont#1{#1}\fi
\expandafter\ifx\csname url\endcsname\relax
  \def\url#1{\texttt{#1}}\fi
\expandafter\ifx\csname urlprefix\endcsname\relax\def\urlprefix{URL }\fi
\providecommand{\bibinfo}[2]{#2}
\providecommand{\eprint}[2][]{\url{#2}}

\bibitem[{\citenamefont{Boezio et~al.}(2008)}]{Boezio:2008mp}
\bibinfo{author}{\bibfnamefont{M.}~\bibnamefont{Boezio}} \bibnamefont{et~al.}
  (\bibinfo{year}{2008}), \eprint{0810.3508}.

\bibitem[{\citenamefont{Adriani et~al.}(2009{\natexlab{a}})}]{Adriani:2008zr}
\bibinfo{author}{\bibfnamefont{O.}~\bibnamefont{Adriani}} \bibnamefont{et~al.}
  (\bibinfo{collaboration}{PAMELA}), \bibinfo{journal}{Nature}
  \textbf{\bibinfo{volume}{458}}, \bibinfo{pages}{607}
  (\bibinfo{year}{2009}{\natexlab{a}}), \eprint{0810.4995}.

\bibitem[{\citenamefont{Adriani et~al.}(2009{\natexlab{b}})}]{Adriani:2008zq}
\bibinfo{author}{\bibfnamefont{O.}~\bibnamefont{Adriani}} \bibnamefont{et~al.},
  \bibinfo{journal}{Phys. Rev. Lett.} \textbf{\bibinfo{volume}{102}},
  \bibinfo{pages}{051101} (\bibinfo{year}{2009}{\natexlab{b}}),
  \eprint{0810.4994}.

\bibitem[{\citenamefont{Aharonian et~al.}(2008)}]{Collaboration:2008aaa}
\bibinfo{author}{\bibfnamefont{F.}~\bibnamefont{Aharonian}}
  \bibnamefont{et~al.} (\bibinfo{collaboration}{H.E.S.S.}),
  \bibinfo{journal}{Phys. Rev. Lett.} \textbf{\bibinfo{volume}{101}},
  \bibinfo{pages}{261104} (\bibinfo{year}{2008}), \eprint{0811.3894}.

\bibitem[{\citenamefont{Aharonian et~al.}(2009)}]{Aharonian:2009ah}
\bibinfo{author}{\bibfnamefont{F.}~\bibnamefont{Aharonian}}
  \bibnamefont{et~al.} (\bibinfo{collaboration}{H.E.S.S.}),
  \bibinfo{journal}{Astron. Astrophys.} \textbf{\bibinfo{volume}{508}},
  \bibinfo{pages}{561} (\bibinfo{year}{2009}), \eprint{0905.0105}.

\bibitem[{\citenamefont{Abdo et~al.}(2009)}]{Abdo:2009zk}
\bibinfo{author}{\bibfnamefont{A.~A.} \bibnamefont{Abdo}} \bibnamefont{et~al.}
  (\bibinfo{collaboration}{The Fermi LAT}), \bibinfo{journal}{Phys. Rev. Lett.}
  \textbf{\bibinfo{volume}{102}}, \bibinfo{pages}{181101}
  (\bibinfo{year}{2009}), \eprint{0905.0025}.

\bibitem[{\citenamefont{Chang et~al.}(2008)}]{ATIClatest}
\bibinfo{author}{\bibfnamefont{J.}~\bibnamefont{Chang}} \bibnamefont{et~al.},
  \bibinfo{journal}{Nature} \textbf{\bibinfo{volume}{456}},
  \bibinfo{pages}{362} (\bibinfo{year}{2008}).

\bibitem[{\citenamefont{Arkani-Hamed et~al.}(2009)\citenamefont{Arkani-Hamed,
  Finkbeiner, Slatyer, and Weiner}}]{ArkaniHamed:2008qn}
\bibinfo{author}{\bibfnamefont{N.}~\bibnamefont{Arkani-Hamed}},
  \bibinfo{author}{\bibfnamefont{D.~P.} \bibnamefont{Finkbeiner}},
  \bibinfo{author}{\bibfnamefont{T.~R.} \bibnamefont{Slatyer}},
  \bibnamefont{and} \bibinfo{author}{\bibfnamefont{N.}~\bibnamefont{Weiner}},
  \bibinfo{journal}{Phys. Rev.} \textbf{\bibinfo{volume}{D79}},
  \bibinfo{pages}{015014} (\bibinfo{year}{2009}), \eprint{0810.0713}.

\bibitem[{\citenamefont{Cholis et~al.}(2009{\natexlab{a}})\citenamefont{Cholis,
  Finkbeiner, Goodenough, and Weiner}}]{Cholis:2008qq}
\bibinfo{author}{\bibfnamefont{I.}~\bibnamefont{Cholis}},
  \bibinfo{author}{\bibfnamefont{D.~P.} \bibnamefont{Finkbeiner}},
  \bibinfo{author}{\bibfnamefont{L.}~\bibnamefont{Goodenough}},
  \bibnamefont{and} \bibinfo{author}{\bibfnamefont{N.}~\bibnamefont{Weiner}},
  \bibinfo{journal}{JCAP} \textbf{\bibinfo{volume}{0912}}, \bibinfo{pages}{007}
  (\bibinfo{year}{2009}{\natexlab{a}}), \eprint{0810.5344}.

\bibitem[{\citenamefont{Harnik and Kribs}(2009)}]{Harnik:2008uu}
\bibinfo{author}{\bibfnamefont{R.}~\bibnamefont{Harnik}} \bibnamefont{and}
  \bibinfo{author}{\bibfnamefont{G.~D.} \bibnamefont{Kribs}},
  \bibinfo{journal}{Phys.Rev.} \textbf{\bibinfo{volume}{D79}},
  \bibinfo{pages}{095007} (\bibinfo{year}{2009}), \eprint{0810.5557}.

\bibitem[{\citenamefont{Bai and Han}(2009)}]{Bai:2008jt}
\bibinfo{author}{\bibfnamefont{Y.}~\bibnamefont{Bai}} \bibnamefont{and}
  \bibinfo{author}{\bibfnamefont{Z.}~\bibnamefont{Han}},
  \bibinfo{journal}{Phys.Rev.} \textbf{\bibinfo{volume}{D79}},
  \bibinfo{pages}{095023} (\bibinfo{year}{2009}), \eprint{0811.0387}.

\bibitem[{\citenamefont{Fox and Poppitz}(2009)}]{Fox:2008kb}
\bibinfo{author}{\bibfnamefont{P.~J.} \bibnamefont{Fox}} \bibnamefont{and}
  \bibinfo{author}{\bibfnamefont{E.}~\bibnamefont{Poppitz}},
  \bibinfo{journal}{Phys.Rev.} \textbf{\bibinfo{volume}{D79}},
  \bibinfo{pages}{083528} (\bibinfo{year}{2009}), \eprint{0811.0399}.

\bibitem[{\citenamefont{Grajek et~al.}(2009)\citenamefont{Grajek, Kane, Phalen,
  Pierce, and Watson}}]{Grajek:2008pg}
\bibinfo{author}{\bibfnamefont{P.}~\bibnamefont{Grajek}},
  \bibinfo{author}{\bibfnamefont{G.}~\bibnamefont{Kane}},
  \bibinfo{author}{\bibfnamefont{D.}~\bibnamefont{Phalen}},
  \bibinfo{author}{\bibfnamefont{A.}~\bibnamefont{Pierce}}, \bibnamefont{and}
  \bibinfo{author}{\bibfnamefont{S.}~\bibnamefont{Watson}},
  \bibinfo{journal}{Phys.Rev.} \textbf{\bibinfo{volume}{D79}},
  \bibinfo{pages}{043506} (\bibinfo{year}{2009}), \eprint{0812.4555}.

\bibitem[{\citenamefont{Pospelov and Ritz}(2009)}]{Pospelov:2008jd}
\bibinfo{author}{\bibfnamefont{M.}~\bibnamefont{Pospelov}} \bibnamefont{and}
  \bibinfo{author}{\bibfnamefont{A.}~\bibnamefont{Ritz}},
  \bibinfo{journal}{Phys.Lett.} \textbf{\bibinfo{volume}{B671}},
  \bibinfo{pages}{391} (\bibinfo{year}{2009}), \eprint{0810.1502}.

\bibitem[{\citenamefont{March-Russell and West}(2009)}]{MarchRussell:2008tu}
\bibinfo{author}{\bibfnamefont{J.~D.} \bibnamefont{March-Russell}}
  \bibnamefont{and} \bibinfo{author}{\bibfnamefont{S.~M.} \bibnamefont{West}},
  \bibinfo{journal}{Phys.Lett.} \textbf{\bibinfo{volume}{B676}},
  \bibinfo{pages}{133} (\bibinfo{year}{2009}), \eprint{0812.0559}.

\bibitem[{\citenamefont{Cirelli and Strumia}(2009)}]{Cirelli:2009uv}
\bibinfo{author}{\bibfnamefont{M.}~\bibnamefont{Cirelli}} \bibnamefont{and}
  \bibinfo{author}{\bibfnamefont{A.}~\bibnamefont{Strumia}},
  \bibinfo{journal}{New J.Phys.} \textbf{\bibinfo{volume}{11}},
  \bibinfo{pages}{105005} (\bibinfo{year}{2009}), \eprint{0903.3381}.

\bibitem[{\citenamefont{Shepherd et~al.}(2009)\citenamefont{Shepherd, Tait, and
  Zaharijas}}]{Shepherd:2009sa}
\bibinfo{author}{\bibfnamefont{W.}~\bibnamefont{Shepherd}},
  \bibinfo{author}{\bibfnamefont{T.~M.} \bibnamefont{Tait}}, \bibnamefont{and}
  \bibinfo{author}{\bibfnamefont{G.}~\bibnamefont{Zaharijas}},
  \bibinfo{journal}{Phys.Rev.} \textbf{\bibinfo{volume}{D79}},
  \bibinfo{pages}{055022} (\bibinfo{year}{2009}), \eprint{0901.2125}.

\bibitem[{\citenamefont{Phalen et~al.}(2009)\citenamefont{Phalen, Pierce, and
  Weiner}}]{Phalen:2009xw}
\bibinfo{author}{\bibfnamefont{D.~J.} \bibnamefont{Phalen}},
  \bibinfo{author}{\bibfnamefont{A.}~\bibnamefont{Pierce}}, \bibnamefont{and}
  \bibinfo{author}{\bibfnamefont{N.}~\bibnamefont{Weiner}},
  \bibinfo{journal}{Phys.Rev.} \textbf{\bibinfo{volume}{D80}},
  \bibinfo{pages}{063513} (\bibinfo{year}{2009}), \eprint{0901.3165}.

\bibitem[{\citenamefont{Hooper and Zurek}(2009)}]{Hooper:2009fj}
\bibinfo{author}{\bibfnamefont{D.}~\bibnamefont{Hooper}} \bibnamefont{and}
  \bibinfo{author}{\bibfnamefont{K.~M.} \bibnamefont{Zurek}},
  \bibinfo{journal}{Phys.Rev.} \textbf{\bibinfo{volume}{D79}},
  \bibinfo{pages}{103529} (\bibinfo{year}{2009}), \eprint{0902.0593}.

\bibitem[{\citenamefont{Goh et~al.}(2009)\citenamefont{Goh, Hall, and
  Kumar}}]{Goh:2009wg}
\bibinfo{author}{\bibfnamefont{H.-S.} \bibnamefont{Goh}},
  \bibinfo{author}{\bibfnamefont{L.~J.} \bibnamefont{Hall}}, \bibnamefont{and}
  \bibinfo{author}{\bibfnamefont{P.}~\bibnamefont{Kumar}},
  \bibinfo{journal}{JHEP} \textbf{\bibinfo{volume}{0905}}, \bibinfo{pages}{097}
  (\bibinfo{year}{2009}), \eprint{0902.0814}.

\bibitem[{\citenamefont{Cholis and Weiner}(2009)}]{Cholis:2009va}
\bibinfo{author}{\bibfnamefont{I.}~\bibnamefont{Cholis}} \bibnamefont{and}
  \bibinfo{author}{\bibfnamefont{N.}~\bibnamefont{Weiner}}
  (\bibinfo{year}{2009}), \eprint{0911.4954}.

\bibitem[{\citenamefont{Hooper and Tait}(2009)}]{Hooper:2009gm}
\bibinfo{author}{\bibfnamefont{D.}~\bibnamefont{Hooper}} \bibnamefont{and}
  \bibinfo{author}{\bibfnamefont{T.~M.} \bibnamefont{Tait}},
  \bibinfo{journal}{Phys.Rev.} \textbf{\bibinfo{volume}{D80}},
  \bibinfo{pages}{055028} (\bibinfo{year}{2009}), \eprint{0906.0362}.

\bibitem[{\citenamefont{Cirelli et~al.}(2009)\citenamefont{Cirelli, Kadastik,
  Raidal, and Strumia}}]{Cirelli:2008pk}
\bibinfo{author}{\bibfnamefont{M.}~\bibnamefont{Cirelli}},
  \bibinfo{author}{\bibfnamefont{M.}~\bibnamefont{Kadastik}},
  \bibinfo{author}{\bibfnamefont{M.}~\bibnamefont{Raidal}}, \bibnamefont{and}
  \bibinfo{author}{\bibfnamefont{A.}~\bibnamefont{Strumia}},
  \bibinfo{journal}{Nucl.Phys.} \textbf{\bibinfo{volume}{B813}},
  \bibinfo{pages}{1} (\bibinfo{year}{2009}), \eprint{0809.2409}.

\bibitem[{\citenamefont{Donato et~al.}(2009{\natexlab{a}})\citenamefont{Donato,
  Maurin, Brun, Delahaye, and Salati}}]{Donato:2008jk}
\bibinfo{author}{\bibfnamefont{F.}~\bibnamefont{Donato}},
  \bibinfo{author}{\bibfnamefont{D.}~\bibnamefont{Maurin}},
  \bibinfo{author}{\bibfnamefont{P.}~\bibnamefont{Brun}},
  \bibinfo{author}{\bibfnamefont{T.}~\bibnamefont{Delahaye}}, \bibnamefont{and}
  \bibinfo{author}{\bibfnamefont{P.}~\bibnamefont{Salati}},
  \bibinfo{journal}{Phys. Rev. Lett.} \textbf{\bibinfo{volume}{102}},
  \bibinfo{pages}{071301} (\bibinfo{year}{2009}{\natexlab{a}}),
  \eprint{0810.5292}.

\bibitem[{\citenamefont{Cholis}(2011)}]{Cholis:2010xb}
\bibinfo{author}{\bibfnamefont{I.}~\bibnamefont{Cholis}},
  \bibinfo{journal}{JCAP} \textbf{\bibinfo{volume}{1109}}, \bibinfo{pages}{007}
  (\bibinfo{year}{2011}), \eprint{1007.1160}.

\bibitem[{\citenamefont{Evoli et~al.}(2012)\citenamefont{Evoli, Cholis, Grasso,
  Maccione, and Ullio}}]{Evoli:2011id}
\bibinfo{author}{\bibfnamefont{C.}~\bibnamefont{Evoli}},
  \bibinfo{author}{\bibfnamefont{I.}~\bibnamefont{Cholis}},
  \bibinfo{author}{\bibfnamefont{D.}~\bibnamefont{Grasso}},
  \bibinfo{author}{\bibfnamefont{L.}~\bibnamefont{Maccione}}, \bibnamefont{and}
  \bibinfo{author}{\bibfnamefont{P.}~\bibnamefont{Ullio}},
  \bibinfo{journal}{Phys.Rev.} \textbf{\bibinfo{volume}{D85}},
  \bibinfo{pages}{123511} (\bibinfo{year}{2012}), \eprint{1108.0664}.

\bibitem[{\citenamefont{Bergstrom et~al.}(2009)\citenamefont{Bergstrom, Edsjo,
  and Zaharijas}}]{Bergstrom:2009fa}
\bibinfo{author}{\bibfnamefont{L.}~\bibnamefont{Bergstrom}},
  \bibinfo{author}{\bibfnamefont{J.}~\bibnamefont{Edsjo}}, \bibnamefont{and}
  \bibinfo{author}{\bibfnamefont{G.}~\bibnamefont{Zaharijas}},
  \bibinfo{journal}{Phys.Rev.Lett.} \textbf{\bibinfo{volume}{103}},
  \bibinfo{pages}{031103} (\bibinfo{year}{2009}), \eprint{0905.0333}.

\bibitem[{\citenamefont{Baudis}(2007)}]{Baudis:2007ew}
\bibinfo{author}{\bibfnamefont{L.}~\bibnamefont{Baudis}}, \bibinfo{journal}{J.
  Phys. Conf. Ser.} \textbf{\bibinfo{volume}{65}}, \bibinfo{pages}{012015}
  (\bibinfo{year}{2007}), \eprint{astro-ph/0703183}.

\bibitem[{\citenamefont{Barbeau et~al.}(2007)\citenamefont{Barbeau, Collar, and
  Tench}}]{Barbeau:2007qi}
\bibinfo{author}{\bibfnamefont{P.}~\bibnamefont{Barbeau}},
  \bibinfo{author}{\bibfnamefont{J.}~\bibnamefont{Collar}}, \bibnamefont{and}
  \bibinfo{author}{\bibfnamefont{O.}~\bibnamefont{Tench}},
  \bibinfo{journal}{JCAP} \textbf{\bibinfo{volume}{0709}}, \bibinfo{pages}{009}
  (\bibinfo{year}{2007}), \bibinfo{note}{submitted to Phys. Rev. C},
  \eprint{nucl-ex/0701012}.

\bibitem[{\citenamefont{Aprile et~al.}(2008)\citenamefont{Aprile, Baudis, and
  Collaboration}}]{Aprile:2009yh}
\bibinfo{author}{\bibfnamefont{E.}~\bibnamefont{Aprile}},
  \bibinfo{author}{\bibfnamefont{L.}~\bibnamefont{Baudis}}, \bibnamefont{and}
  \bibinfo{author}{\bibfnamefont{f.~t.~X.} \bibnamefont{Collaboration}},
  \bibinfo{journal}{PoS} \textbf{\bibinfo{volume}{IDM2008}},
  \bibinfo{pages}{018} (\bibinfo{year}{2008}), \eprint{0902.4253}.

\bibitem[{\citenamefont{Ahmed et~al.}(2010)}]{Ahmed:2009zw}
\bibinfo{author}{\bibfnamefont{Z.}~\bibnamefont{Ahmed}} \bibnamefont{et~al.}
  (\bibinfo{collaboration}{The CDMS-II Collaboration}),
  \bibinfo{journal}{Science} \textbf{\bibinfo{volume}{327}},
  \bibinfo{pages}{1619} (\bibinfo{year}{2010}), \eprint{0912.3592}.

\bibitem[{\citenamefont{Bernabei et~al.}(2008)}]{Bernabei:2008yh}
\bibinfo{author}{\bibfnamefont{R.}~\bibnamefont{Bernabei}} \bibnamefont{et~al.}
  (\bibinfo{collaboration}{DAMA}), \bibinfo{journal}{Nucl. Instrum. Meth.}
  \textbf{\bibinfo{volume}{A592}}, \bibinfo{pages}{297} (\bibinfo{year}{2008}),
  \eprint{0804.2738}.

\bibitem[{\citenamefont{Angloher et~al.}(2008)\citenamefont{Angloher, Bauer,
  Bavykina, Bento, Brown et~al.}}]{Angloher:2008jj}
\bibinfo{author}{\bibfnamefont{G.}~\bibnamefont{Angloher}},
  \bibinfo{author}{\bibfnamefont{M.}~\bibnamefont{Bauer}},
  \bibinfo{author}{\bibfnamefont{I.}~\bibnamefont{Bavykina}},
  \bibinfo{author}{\bibfnamefont{A.}~\bibnamefont{Bento}},
  \bibinfo{author}{\bibfnamefont{A.}~\bibnamefont{Brown}}, \bibnamefont{et~al.}
  (\bibinfo{year}{2008}), \eprint{0809.1829}.

\bibitem[{\citenamefont{Angle et~al.}(2008)}]{Angle:2008we}
\bibinfo{author}{\bibfnamefont{J.}~\bibnamefont{Angle}} \bibnamefont{et~al.},
  \bibinfo{journal}{Phys. Rev. Lett.} \textbf{\bibinfo{volume}{101}},
  \bibinfo{pages}{091301} (\bibinfo{year}{2008}), \eprint{0805.2939}.

\bibitem[{\citenamefont{Kopp et~al.}(2010)\citenamefont{Kopp, Schwetz, and
  Zupan}}]{Kopp:2009qt}
\bibinfo{author}{\bibfnamefont{J.}~\bibnamefont{Kopp}},
  \bibinfo{author}{\bibfnamefont{T.}~\bibnamefont{Schwetz}}, \bibnamefont{and}
  \bibinfo{author}{\bibfnamefont{J.}~\bibnamefont{Zupan}},
  \bibinfo{journal}{JCAP} \textbf{\bibinfo{volume}{1002}}, \bibinfo{pages}{014}
  (\bibinfo{year}{2010}), \eprint{0912.4264}.

\bibitem[{\citenamefont{Bernabei et~al.}(2010)}]{Bernabei:2010mq}
\bibinfo{author}{\bibfnamefont{R.}~\bibnamefont{Bernabei}}
  \bibnamefont{et~al.}, \bibinfo{journal}{Eur. Phys. J.}
  \textbf{\bibinfo{volume}{C67}}, \bibinfo{pages}{39} (\bibinfo{year}{2010}),
  \eprint{1002.1028}.

\bibitem[{\citenamefont{Aprile et~al.}(2010)}]{Aprile:2010um}
\bibinfo{author}{\bibfnamefont{E.}~\bibnamefont{Aprile}} \bibnamefont{et~al.}
  (\bibinfo{collaboration}{XENON100 Collaboration}),
  \bibinfo{journal}{Phys.Rev.Lett.} \textbf{\bibinfo{volume}{105}},
  \bibinfo{pages}{131302} (\bibinfo{year}{2010}), \eprint{1005.0380}.

\bibitem[{\citenamefont{Aalseth et~al.}(2011{\natexlab{a}})}]{Aalseth:2010vx}
\bibinfo{author}{\bibfnamefont{C.}~\bibnamefont{Aalseth}} \bibnamefont{et~al.}
  (\bibinfo{collaboration}{CoGeNT collaboration}),
  \bibinfo{journal}{Phys.Rev.Lett.} \textbf{\bibinfo{volume}{106}},
  \bibinfo{pages}{131301} (\bibinfo{year}{2011}{\natexlab{a}}),
  \eprint{1002.4703}.

\bibitem[{\citenamefont{Aalseth
  et~al.}(2011{\natexlab{b}})\citenamefont{Aalseth, Barbeau, Colaresi, Collar,
  Diaz~Leon et~al.}}]{Aalseth:2011wp}
\bibinfo{author}{\bibfnamefont{C.}~\bibnamefont{Aalseth}},
  \bibinfo{author}{\bibfnamefont{P.}~\bibnamefont{Barbeau}},
  \bibinfo{author}{\bibfnamefont{J.}~\bibnamefont{Colaresi}},
  \bibinfo{author}{\bibfnamefont{J.}~\bibnamefont{Collar}},
  \bibinfo{author}{\bibfnamefont{J.}~\bibnamefont{Diaz~Leon}},
  \bibnamefont{et~al.}, \bibinfo{journal}{Phys.Rev.Lett.}
  \textbf{\bibinfo{volume}{107}}, \bibinfo{pages}{141301}
  (\bibinfo{year}{2011}{\natexlab{b}}), \bibinfo{note}{published version,
  slightly expanded discussion of ROI uncertainties, one reference added},
  \eprint{1106.0650}.

\bibitem[{\citenamefont{Angloher et~al.}(2011)\citenamefont{Angloher, Bauer,
  Bavykina, Bento, Bucci et~al.}}]{Angloher:2011uu}
\bibinfo{author}{\bibfnamefont{G.}~\bibnamefont{Angloher}},
  \bibinfo{author}{\bibfnamefont{M.}~\bibnamefont{Bauer}},
  \bibinfo{author}{\bibfnamefont{I.}~\bibnamefont{Bavykina}},
  \bibinfo{author}{\bibfnamefont{A.}~\bibnamefont{Bento}},
  \bibinfo{author}{\bibfnamefont{C.}~\bibnamefont{Bucci}}, \bibnamefont{et~al.}
  (\bibinfo{year}{2011}), \eprint{1109.0702}.

\bibitem[{\citenamefont{Collar and McKinsey}(2010)}]{Collar:2010gg}
\bibinfo{author}{\bibfnamefont{J.~I.} \bibnamefont{Collar}} \bibnamefont{and}
  \bibinfo{author}{\bibfnamefont{D.~N.} \bibnamefont{McKinsey}}
  (\bibinfo{year}{2010}), \eprint{1005.0838}.

\bibitem[{\citenamefont{Fox et~al.}(2012)\citenamefont{Fox, Kopp, Lisanti, and
  Weiner}}]{Fox:2011px}
\bibinfo{author}{\bibfnamefont{P.~J.} \bibnamefont{Fox}},
  \bibinfo{author}{\bibfnamefont{J.}~\bibnamefont{Kopp}},
  \bibinfo{author}{\bibfnamefont{M.}~\bibnamefont{Lisanti}}, \bibnamefont{and}
  \bibinfo{author}{\bibfnamefont{N.}~\bibnamefont{Weiner}},
  \bibinfo{journal}{Phys.Rev.} \textbf{\bibinfo{volume}{D85}},
  \bibinfo{pages}{036008} (\bibinfo{year}{2012}), \bibinfo{note}{28 pages, 14
  figures, 3 tables/ version 2 has minor clarifications in the text},
  \eprint{1107.0717}.

\bibitem[{\citenamefont{Hooper and Kelso}(2011)}]{Hooper:2011hd}
\bibinfo{author}{\bibfnamefont{D.}~\bibnamefont{Hooper}} \bibnamefont{and}
  \bibinfo{author}{\bibfnamefont{C.}~\bibnamefont{Kelso}},
  \bibinfo{journal}{Phys.Rev.} \textbf{\bibinfo{volume}{D84}},
  \bibinfo{pages}{083001} (\bibinfo{year}{2011}), \eprint{1106.1066}.

\bibitem[{\citenamefont{Fitzpatrick et~al.}(2010)\citenamefont{Fitzpatrick,
  Hooper, and Zurek}}]{Fitzpatrick:2010em}
\bibinfo{author}{\bibfnamefont{A.}~\bibnamefont{Fitzpatrick}},
  \bibinfo{author}{\bibfnamefont{D.}~\bibnamefont{Hooper}}, \bibnamefont{and}
  \bibinfo{author}{\bibfnamefont{K.~M.} \bibnamefont{Zurek}},
  \bibinfo{journal}{Phys.Rev.} \textbf{\bibinfo{volume}{D81}},
  \bibinfo{pages}{115005} (\bibinfo{year}{2010}), \bibinfo{note}{16 pages, 14
  figures. v2: references added, fig 4 and surrounding discussion modified.},
  \eprint{1003.0014}.

\bibitem[{\citenamefont{Chang et~al.}(2010{\natexlab{a}})\citenamefont{Chang,
  Liu, Pierce, Weiner, and Yavin}}]{Chang:2010yk}
\bibinfo{author}{\bibfnamefont{S.}~\bibnamefont{Chang}},
  \bibinfo{author}{\bibfnamefont{J.}~\bibnamefont{Liu}},
  \bibinfo{author}{\bibfnamefont{A.}~\bibnamefont{Pierce}},
  \bibinfo{author}{\bibfnamefont{N.}~\bibnamefont{Weiner}}, \bibnamefont{and}
  \bibinfo{author}{\bibfnamefont{I.}~\bibnamefont{Yavin}},
  \bibinfo{journal}{JCAP} \textbf{\bibinfo{volume}{1008}}, \bibinfo{pages}{018}
  (\bibinfo{year}{2010}{\natexlab{a}}), \eprint{1004.0697}.

\bibitem[{\citenamefont{Chang et~al.}(2010{\natexlab{b}})\citenamefont{Chang,
  Weiner, and Yavin}}]{Chang:2010en}
\bibinfo{author}{\bibfnamefont{S.}~\bibnamefont{Chang}},
  \bibinfo{author}{\bibfnamefont{N.}~\bibnamefont{Weiner}}, \bibnamefont{and}
  \bibinfo{author}{\bibfnamefont{I.}~\bibnamefont{Yavin}},
  \bibinfo{journal}{Phys.Rev.} \textbf{\bibinfo{volume}{D82}},
  \bibinfo{pages}{125011} (\bibinfo{year}{2010}{\natexlab{b}}),
  \eprint{1007.4200}.

\bibitem[{\citenamefont{Buckley et~al.}(2011)\citenamefont{Buckley, Hooper, and
  Tait}}]{Buckley:2010ve}
\bibinfo{author}{\bibfnamefont{M.~R.} \bibnamefont{Buckley}},
  \bibinfo{author}{\bibfnamefont{D.}~\bibnamefont{Hooper}}, \bibnamefont{and}
  \bibinfo{author}{\bibfnamefont{T.~M.} \bibnamefont{Tait}},
  \bibinfo{journal}{Phys.Lett.} \textbf{\bibinfo{volume}{B702}},
  \bibinfo{pages}{216} (\bibinfo{year}{2011}), \eprint{1011.1499}.

\bibitem[{\citenamefont{Belikov et~al.}(2011)\citenamefont{Belikov, Gunion,
  Hooper, and Tait}}]{Belikov:2010yi}
\bibinfo{author}{\bibfnamefont{A.~V.} \bibnamefont{Belikov}},
  \bibinfo{author}{\bibfnamefont{J.~F.} \bibnamefont{Gunion}},
  \bibinfo{author}{\bibfnamefont{D.}~\bibnamefont{Hooper}}, \bibnamefont{and}
  \bibinfo{author}{\bibfnamefont{T.~M.} \bibnamefont{Tait}},
  \bibinfo{journal}{Phys.Lett.} \textbf{\bibinfo{volume}{B705}},
  \bibinfo{pages}{82} (\bibinfo{year}{2011}), \bibinfo{note}{6 pages, published
  version}, \eprint{1009.0549}.

\bibitem[{\citenamefont{Del~Nobile et~al.}(2011)\citenamefont{Del~Nobile,
  Kouvaris, and Sannino}}]{DelNobile:2011je}
\bibinfo{author}{\bibfnamefont{E.}~\bibnamefont{Del~Nobile}},
  \bibinfo{author}{\bibfnamefont{C.}~\bibnamefont{Kouvaris}}, \bibnamefont{and}
  \bibinfo{author}{\bibfnamefont{F.}~\bibnamefont{Sannino}},
  \bibinfo{journal}{Phys.Rev.} \textbf{\bibinfo{volume}{D84}},
  \bibinfo{pages}{027301} (\bibinfo{year}{2011}), \eprint{1105.5431}.

\bibitem[{\citenamefont{Abdo et~al.}(2010{\natexlab{a}})\citenamefont{Abdo,
  Ackermann, Ajello, Atwood, Baldini et~al.}}]{Abdo:2010ex}
\bibinfo{author}{\bibfnamefont{A.}~\bibnamefont{Abdo}},
  \bibinfo{author}{\bibfnamefont{M.}~\bibnamefont{Ackermann}},
  \bibinfo{author}{\bibfnamefont{M.}~\bibnamefont{Ajello}},
  \bibinfo{author}{\bibfnamefont{W.}~\bibnamefont{Atwood}},
  \bibinfo{author}{\bibfnamefont{L.}~\bibnamefont{Baldini}},
  \bibnamefont{et~al.}, \bibinfo{journal}{Astrophys.J.}
  \textbf{\bibinfo{volume}{712}}, \bibinfo{pages}{147}
  (\bibinfo{year}{2010}{\natexlab{a}}), \eprint{1001.4531}.

\bibitem[{\citenamefont{Ackermann et~al.}(2011)}]{Ackermann:2011wa}
\bibinfo{author}{\bibfnamefont{M.}~\bibnamefont{Ackermann}}
  \bibnamefont{et~al.} (\bibinfo{collaboration}{Fermi-LAT collaboration}),
  \bibinfo{journal}{Phys.Rev.Lett.} \textbf{\bibinfo{volume}{107}},
  \bibinfo{pages}{241302} (\bibinfo{year}{2011}), \eprint{1108.3546}.

\bibitem[{\citenamefont{{Evans} et~al.}(2004)\citenamefont{{Evans}, {Ferrer},
  and {Sarkar}}}]{2004PhRvD..69l3501E}
\bibinfo{author}{\bibfnamefont{N.~W.} \bibnamefont{{Evans}}},
  \bibinfo{author}{\bibfnamefont{F.}~\bibnamefont{{Ferrer}}}, \bibnamefont{and}
  \bibinfo{author}{\bibfnamefont{S.}~\bibnamefont{{Sarkar}}},
  \bibinfo{journal}{\prd} \textbf{\bibinfo{volume}{69}}, \bibinfo{eid}{123501}
  (\bibinfo{year}{2004}), \eprint{arXiv:astro-ph/0311145}.

\bibitem[{\citenamefont{Colafrancesco et~al.}(2007)\citenamefont{Colafrancesco,
  Profumo, and Ullio}}]{Colafrancesco:2006he}
\bibinfo{author}{\bibfnamefont{S.}~\bibnamefont{Colafrancesco}},
  \bibinfo{author}{\bibfnamefont{S.}~\bibnamefont{Profumo}}, \bibnamefont{and}
  \bibinfo{author}{\bibfnamefont{P.}~\bibnamefont{Ullio}},
  \bibinfo{journal}{Phys.Rev.} \textbf{\bibinfo{volume}{D75}},
  \bibinfo{pages}{023513} (\bibinfo{year}{2007}), \eprint{astro-ph/0607073}.

\bibitem[{\citenamefont{Strigari et~al.}(2007)\citenamefont{Strigari,
  Koushiappas, Bullock, and Kaplinghat}}]{Strigari:2006rd}
\bibinfo{author}{\bibfnamefont{L.~E.} \bibnamefont{Strigari}},
  \bibinfo{author}{\bibfnamefont{S.~M.} \bibnamefont{Koushiappas}},
  \bibinfo{author}{\bibfnamefont{J.~S.} \bibnamefont{Bullock}},
  \bibnamefont{and}
  \bibinfo{author}{\bibfnamefont{M.}~\bibnamefont{Kaplinghat}},
  \bibinfo{journal}{Phys.Rev.} \textbf{\bibinfo{volume}{D75}},
  \bibinfo{pages}{083526} (\bibinfo{year}{2007}), \eprint{astro-ph/0611925}.

\bibitem[{\citenamefont{Bovy}(2009)}]{Bovy:2009zs}
\bibinfo{author}{\bibfnamefont{J.}~\bibnamefont{Bovy}},
  \bibinfo{journal}{Phys.Rev.} \textbf{\bibinfo{volume}{D79}},
  \bibinfo{pages}{083539} (\bibinfo{year}{2009}), \eprint{0903.0413}.

\bibitem[{\citenamefont{Scott et~al.}(2010)\citenamefont{Scott, Conrad, Edsjo,
  Bergstrom, Farnier et~al.}}]{Scott:2009jn}
\bibinfo{author}{\bibfnamefont{P.}~\bibnamefont{Scott}},
  \bibinfo{author}{\bibfnamefont{J.}~\bibnamefont{Conrad}},
  \bibinfo{author}{\bibfnamefont{J.}~\bibnamefont{Edsjo}},
  \bibinfo{author}{\bibfnamefont{L.}~\bibnamefont{Bergstrom}},
  \bibinfo{author}{\bibfnamefont{C.}~\bibnamefont{Farnier}},
  \bibnamefont{et~al.}, \bibinfo{journal}{JCAP}
  \textbf{\bibinfo{volume}{1001}}, \bibinfo{pages}{031} (\bibinfo{year}{2010}),
  \eprint{0909.3300}.

\bibitem[{\citenamefont{Perelstein and Shakya}(2010)}]{Perelstein:2010at}
\bibinfo{author}{\bibfnamefont{M.}~\bibnamefont{Perelstein}} \bibnamefont{and}
  \bibinfo{author}{\bibfnamefont{B.}~\bibnamefont{Shakya}},
  \bibinfo{journal}{JCAP} \textbf{\bibinfo{volume}{1010}}, \bibinfo{pages}{016}
  (\bibinfo{year}{2010}), \eprint{1007.0018}.

\bibitem[{\citenamefont{{Persic} et~al.}(1996)\citenamefont{{Persic},
  {Salucci}, and {Stel}}}]{PSS96}
\bibinfo{author}{\bibfnamefont{M.}~\bibnamefont{{Persic}}},
  \bibinfo{author}{\bibfnamefont{P.}~\bibnamefont{{Salucci}}},
  \bibnamefont{and} \bibinfo{author}{\bibfnamefont{F.}~\bibnamefont{{Stel}}},
  \bibinfo{journal}{\mnras} \textbf{\bibinfo{volume}{281}}, \bibinfo{pages}{27}
  (\bibinfo{year}{1996}), \eprint{arXiv:astro-ph/9506004}.

\bibitem[{\citenamefont{{Salucci} et~al.}(2007)\citenamefont{{Salucci}, {Lapi},
  {Tonini}, {Gentile}, {Yegorova}, and {Klein}}}]{Salucci07}
\bibinfo{author}{\bibfnamefont{P.}~\bibnamefont{{Salucci}}},
  \bibinfo{author}{\bibfnamefont{A.}~\bibnamefont{{Lapi}}},
  \bibinfo{author}{\bibfnamefont{C.}~\bibnamefont{{Tonini}}},
  \bibinfo{author}{\bibfnamefont{G.}~\bibnamefont{{Gentile}}},
  \bibinfo{author}{\bibfnamefont{I.}~\bibnamefont{{Yegorova}}},
  \bibnamefont{and} \bibinfo{author}{\bibfnamefont{U.}~\bibnamefont{{Klein}}},
  \bibinfo{journal}{\mnras} \textbf{\bibinfo{volume}{378}}, \bibinfo{pages}{41}
  (\bibinfo{year}{2007}), \eprint{arXiv:astro-ph/0703115}.

\bibitem[{\citenamefont{{de Blok} et~al.}(2008)\citenamefont{{de Blok},
  {Walter}, {Brinks}, {Trachternach}, {Oh}, and {Kennicutt}}}]{deblok08}
\bibinfo{author}{\bibfnamefont{W.~J.~G.} \bibnamefont{{de Blok}}},
  \bibinfo{author}{\bibfnamefont{F.}~\bibnamefont{{Walter}}},
  \bibinfo{author}{\bibfnamefont{E.}~\bibnamefont{{Brinks}}},
  \bibinfo{author}{\bibfnamefont{C.}~\bibnamefont{{Trachternach}}},
  \bibinfo{author}{\bibfnamefont{S.-H.} \bibnamefont{{Oh}}}, \bibnamefont{and}
  \bibinfo{author}{\bibfnamefont{R.~C.} \bibnamefont{{Kennicutt}},
  \bibfnamefont{Jr.}}, \bibinfo{journal}{\aj} \textbf{\bibinfo{volume}{136}},
  \bibinfo{eid}{2648} (\bibinfo{year}{2008}), \eprint{0810.2100}.

\bibitem[{\citenamefont{{Chemin} et~al.}(2011)\citenamefont{{Chemin}, {de
  Blok}, and {Mamon}}}]{chemin11}
\bibinfo{author}{\bibfnamefont{L.}~\bibnamefont{{Chemin}}},
  \bibinfo{author}{\bibfnamefont{W.~J.~G.} \bibnamefont{{de Blok}}},
  \bibnamefont{and} \bibinfo{author}{\bibfnamefont{G.~A.}
  \bibnamefont{{Mamon}}}, \bibinfo{journal}{\aj}
  \textbf{\bibinfo{volume}{142}}, \bibinfo{eid}{109} (\bibinfo{year}{2011}),
  \eprint{1109.4247}.

\bibitem[{\citenamefont{{Gentile} et~al.}(2004)\citenamefont{{Gentile},
  {Salucci}, {Klein}, {Vergani}, and {Kalberla}}}]{gentile04}
\bibinfo{author}{\bibfnamefont{G.}~\bibnamefont{{Gentile}}},
  \bibinfo{author}{\bibfnamefont{P.}~\bibnamefont{{Salucci}}},
  \bibinfo{author}{\bibfnamefont{U.}~\bibnamefont{{Klein}}},
  \bibinfo{author}{\bibfnamefont{D.}~\bibnamefont{{Vergani}}},
  \bibnamefont{and}
  \bibinfo{author}{\bibfnamefont{P.}~\bibnamefont{{Kalberla}}},
  \bibinfo{journal}{\mnras} \textbf{\bibinfo{volume}{351}},
  \bibinfo{pages}{903} (\bibinfo{year}{2004}), \eprint{arXiv:astro-ph/0403154}.

\bibitem[{\citenamefont{{Gentile} et~al.}(2005)\citenamefont{{Gentile},
  {Burkert}, {Salucci}, {Klein}, and {Walter}}}]{g05}
\bibinfo{author}{\bibfnamefont{G.}~\bibnamefont{{Gentile}}},
  \bibinfo{author}{\bibfnamefont{A.}~\bibnamefont{{Burkert}}},
  \bibinfo{author}{\bibfnamefont{P.}~\bibnamefont{{Salucci}}},
  \bibinfo{author}{\bibfnamefont{U.}~\bibnamefont{{Klein}}}, \bibnamefont{and}
  \bibinfo{author}{\bibfnamefont{F.}~\bibnamefont{{Walter}}},
  \bibinfo{journal}{\apjl} \textbf{\bibinfo{volume}{634}},
  \bibinfo{pages}{L145} (\bibinfo{year}{2005}),
  \eprint{arXiv:astro-ph/0506538}.

\bibitem[{\citenamefont{{Gentile} et~al.}(2007)\citenamefont{{Gentile},
  {Salucci}, {Klein}, and {Granato}}}]{g07}
\bibinfo{author}{\bibfnamefont{G.}~\bibnamefont{{Gentile}}},
  \bibinfo{author}{\bibfnamefont{P.}~\bibnamefont{{Salucci}}},
  \bibinfo{author}{\bibfnamefont{U.}~\bibnamefont{{Klein}}}, \bibnamefont{and}
  \bibinfo{author}{\bibfnamefont{G.~L.} \bibnamefont{{Granato}}},
  \bibinfo{journal}{\mnras} \textbf{\bibinfo{volume}{375}},
  \bibinfo{pages}{199} (\bibinfo{year}{2007}), \eprint{arXiv:astro-ph/0611355}.

\bibitem[{\citenamefont{{Kormendy}}(1985)}]{Kormendy85}
\bibinfo{author}{\bibfnamefont{J.}~\bibnamefont{{Kormendy}}},
  \bibinfo{journal}{\apj} \textbf{\bibinfo{volume}{295}}, \bibinfo{pages}{73}
  (\bibinfo{year}{1985}).

\bibitem[{\citenamefont{{Burkert}}(1995)}]{Burkert95}
\bibinfo{author}{\bibfnamefont{A.}~\bibnamefont{{Burkert}}},
  \bibinfo{journal}{\apjl} \textbf{\bibinfo{volume}{447}}, \bibinfo{pages}{L25}
  (\bibinfo{year}{1995}), \eprint{arXiv:astro-ph/9504041}.

\bibitem[{\citenamefont{{Kormendy} and {Freeman}}(2004)}]{KF04}
\bibinfo{author}{\bibfnamefont{J.}~\bibnamefont{{Kormendy}}} \bibnamefont{and}
  \bibinfo{author}{\bibfnamefont{K.~C.} \bibnamefont{{Freeman}}}, in
  \emph{\bibinfo{booktitle}{Dark Matter in Galaxies}}, edited by
  \bibinfo{editor}{\bibnamefont{{S.~Ryder, D.~Pisano, M.~Walker, \&
  K.~Freeman}}} (\bibinfo{year}{2004}), vol. \bibinfo{volume}{220} of
  \emph{\bibinfo{series}{IAU Symposium}}, p. \bibinfo{pages}{377},
  \eprint{arXiv:astro-ph/0407321}.

\bibitem[{\citenamefont{Salucci and Burkert}(2000)}]{Salucci:2000ps}
\bibinfo{author}{\bibfnamefont{P.}~\bibnamefont{Salucci}} \bibnamefont{and}
  \bibinfo{author}{\bibfnamefont{A.}~\bibnamefont{Burkert}},
  \bibinfo{journal}{Astrophys.J.} \textbf{\bibinfo{volume}{537}},
  \bibinfo{pages}{L9} (\bibinfo{year}{2000}), \eprint{astro-ph/0004397}.

\bibitem[{\citenamefont{{Pe{\~n}arrubia}
  et~al.}(2008)\citenamefont{{Pe{\~n}arrubia}, {McConnachie}, and
  {Navarro}}}]{pena08}
\bibinfo{author}{\bibfnamefont{J.}~\bibnamefont{{Pe{\~n}arrubia}}},
  \bibinfo{author}{\bibfnamefont{A.~W.} \bibnamefont{{McConnachie}}},
  \bibnamefont{and} \bibinfo{author}{\bibfnamefont{J.~F.}
  \bibnamefont{{Navarro}}}, \bibinfo{journal}{\apj}
  \textbf{\bibinfo{volume}{672}}, \bibinfo{pages}{904} (\bibinfo{year}{2008}),
  \eprint{arXiv:astro-ph/0701780}.

\bibitem[{\citenamefont{{Walker}
  et~al.}(2009{\natexlab{a}})\citenamefont{{Walker}, {Mateo}, and
  {Olszewski}}}]{walker09a}
\bibinfo{author}{\bibfnamefont{M.~G.} \bibnamefont{{Walker}}},
  \bibinfo{author}{\bibfnamefont{M.}~\bibnamefont{{Mateo}}}, \bibnamefont{and}
  \bibinfo{author}{\bibfnamefont{E.~W.} \bibnamefont{{Olszewski}}},
  \bibinfo{journal}{\aj} \textbf{\bibinfo{volume}{137}}, \bibinfo{pages}{3100}
  (\bibinfo{year}{2009}{\natexlab{a}}), \eprint{0811.0118}.

\bibitem[{\citenamefont{{Kleyna} et~al.}(2002)\citenamefont{{Kleyna},
  {Wilkinson}, {Evans}, {Gilmore}, and {Frayn}}}]{kleyna02}
\bibinfo{author}{\bibfnamefont{J.}~\bibnamefont{{Kleyna}}},
  \bibinfo{author}{\bibfnamefont{M.~I.} \bibnamefont{{Wilkinson}}},
  \bibinfo{author}{\bibfnamefont{N.~W.} \bibnamefont{{Evans}}},
  \bibinfo{author}{\bibfnamefont{G.}~\bibnamefont{{Gilmore}}},
  \bibnamefont{and} \bibinfo{author}{\bibfnamefont{C.}~\bibnamefont{{Frayn}}},
  \bibinfo{journal}{\mnras} \textbf{\bibinfo{volume}{330}},
  \bibinfo{pages}{792} (\bibinfo{year}{2002}), \eprint{arXiv:astro-ph/0109450}.

\bibitem[{\citenamefont{{Mateo}}(1998)}]{Mateo98}
\bibinfo{author}{\bibfnamefont{M.~L.} \bibnamefont{{Mateo}}},
  \bibinfo{journal}{\araa} \textbf{\bibinfo{volume}{36}}, \bibinfo{pages}{435}
  (\bibinfo{year}{1998}), \eprint{arXiv:astro-ph/9810070}.

\bibitem[{\citenamefont{{Gilmore} et~al.}(2007)\citenamefont{{Gilmore},
  {Wilkinson}, {Wyse}, {Kleyna}, {Koch}, {Evans}, and {Grebel}}}]{Gilmore07}
\bibinfo{author}{\bibfnamefont{G.}~\bibnamefont{{Gilmore}}},
  \bibinfo{author}{\bibfnamefont{M.~I.} \bibnamefont{{Wilkinson}}},
  \bibinfo{author}{\bibfnamefont{R.~F.~G.} \bibnamefont{{Wyse}}},
  \bibinfo{author}{\bibfnamefont{J.~T.} \bibnamefont{{Kleyna}}},
  \bibinfo{author}{\bibfnamefont{A.}~\bibnamefont{{Koch}}},
  \bibinfo{author}{\bibfnamefont{N.~W.} \bibnamefont{{Evans}}},
  \bibnamefont{and} \bibinfo{author}{\bibfnamefont{E.~K.}
  \bibnamefont{{Grebel}}}, \bibinfo{journal}{\apj}
  \textbf{\bibinfo{volume}{663}}, \bibinfo{pages}{948} (\bibinfo{year}{2007}),
  \eprint{arXiv:astro-ph/0703308}.

\bibitem[{\citenamefont{{Koch} et~al.}(2007{\natexlab{a}})\citenamefont{{Koch},
  {Wilkinson}, {Kleyna}, {Gilmore}, {Grebel}, {Mackey}, {Evans}, and
  {Wyse}}}]{Koch07a}
\bibinfo{author}{\bibfnamefont{A.}~\bibnamefont{{Koch}}},
  \bibinfo{author}{\bibfnamefont{M.~I.} \bibnamefont{{Wilkinson}}},
  \bibinfo{author}{\bibfnamefont{J.~T.} \bibnamefont{{Kleyna}}},
  \bibinfo{author}{\bibfnamefont{G.~F.} \bibnamefont{{Gilmore}}},
  \bibinfo{author}{\bibfnamefont{E.~K.} \bibnamefont{{Grebel}}},
  \bibinfo{author}{\bibfnamefont{A.~D.} \bibnamefont{{Mackey}}},
  \bibinfo{author}{\bibfnamefont{N.~W.} \bibnamefont{{Evans}}},
  \bibnamefont{and} \bibinfo{author}{\bibfnamefont{R.~F.~G.}
  \bibnamefont{{Wyse}}}, \bibinfo{journal}{\apj}
  \textbf{\bibinfo{volume}{657}}, \bibinfo{pages}{241}
  (\bibinfo{year}{2007}{\natexlab{a}}), \eprint{arXiv:astro-ph/0611372}.

\bibitem[{\citenamefont{{Strigari} et~al.}(2008)\citenamefont{{Strigari},
  {Bullock}, {Kaplinghat}, {Simon}, {Geha}, {Willman}, and
  {Walker}}}]{strigari08}
\bibinfo{author}{\bibfnamefont{L.~E.} \bibnamefont{{Strigari}}},
  \bibinfo{author}{\bibfnamefont{J.~S.} \bibnamefont{{Bullock}}},
  \bibinfo{author}{\bibfnamefont{M.}~\bibnamefont{{Kaplinghat}}},
  \bibinfo{author}{\bibfnamefont{J.~D.} \bibnamefont{{Simon}}},
  \bibinfo{author}{\bibfnamefont{M.}~\bibnamefont{{Geha}}},
  \bibinfo{author}{\bibfnamefont{B.}~\bibnamefont{{Willman}}},
  \bibnamefont{and} \bibinfo{author}{\bibfnamefont{M.~G.}
  \bibnamefont{{Walker}}}, \bibinfo{journal}{\nat}
  \textbf{\bibinfo{volume}{454}}, \bibinfo{pages}{1096} (\bibinfo{year}{2008}),
  \eprint{0808.3772}.

\bibitem[{\citenamefont{{Walker}
  et~al.}(2009{\natexlab{b}})\citenamefont{{Walker}, {Mateo}, {Olszewski},
  {Pe{\~n}arrubia}, {Wyn Evans}, and {Gilmore}}}]{walker09b}
\bibinfo{author}{\bibfnamefont{M.~G.} \bibnamefont{{Walker}}},
  \bibinfo{author}{\bibfnamefont{M.}~\bibnamefont{{Mateo}}},
  \bibinfo{author}{\bibfnamefont{E.~W.} \bibnamefont{{Olszewski}}},
  \bibinfo{author}{\bibfnamefont{J.}~\bibnamefont{{Pe{\~n}arrubia}}},
  \bibinfo{author}{\bibfnamefont{N.}~\bibnamefont{{Wyn Evans}}},
  \bibnamefont{and}
  \bibinfo{author}{\bibfnamefont{G.}~\bibnamefont{{Gilmore}}},
  \bibinfo{journal}{\apj} \textbf{\bibinfo{volume}{704}}, \bibinfo{pages}{1274}
  (\bibinfo{year}{2009}{\natexlab{b}}), \eprint{0906.0341}.

\bibitem[{\citenamefont{{Mu{\~n}oz} et~al.}(2006)\citenamefont{{Mu{\~n}oz},
  {Majewski}, {Zaggia}, {Kunkel}, {Frinchaboy}, {Nidever}, {Crnojevic},
  {Patterson}, {Crane}, {Johnston} et~al.}}]{Munoz06}
\bibinfo{author}{\bibfnamefont{R.~R.} \bibnamefont{{Mu{\~n}oz}}},
  \bibinfo{author}{\bibfnamefont{S.~R.} \bibnamefont{{Majewski}}},
  \bibinfo{author}{\bibfnamefont{S.}~\bibnamefont{{Zaggia}}},
  \bibinfo{author}{\bibfnamefont{W.~E.} \bibnamefont{{Kunkel}}},
  \bibinfo{author}{\bibfnamefont{P.~M.} \bibnamefont{{Frinchaboy}}},
  \bibinfo{author}{\bibfnamefont{D.~L.} \bibnamefont{{Nidever}}},
  \bibinfo{author}{\bibfnamefont{D.}~\bibnamefont{{Crnojevic}}},
  \bibinfo{author}{\bibfnamefont{R.~J.} \bibnamefont{{Patterson}}},
  \bibinfo{author}{\bibfnamefont{J.~D.} \bibnamefont{{Crane}}},
  \bibinfo{author}{\bibfnamefont{K.~V.} \bibnamefont{{Johnston}}},
  \bibnamefont{et~al.}, \bibinfo{journal}{\apj} \textbf{\bibinfo{volume}{649}},
  \bibinfo{pages}{201} (\bibinfo{year}{2006}), \eprint{arXiv:astro-ph/0605098}.

\bibitem[{\citenamefont{{Koch} et~al.}(2007{\natexlab{b}})\citenamefont{{Koch},
  {Kleyna}, {Wilkinson}, {Grebel}, {Gilmore}, {Evans}, {Wyse}, and
  {Harbeck}}}]{Koch07b}
\bibinfo{author}{\bibfnamefont{A.}~\bibnamefont{{Koch}}},
  \bibinfo{author}{\bibfnamefont{J.~T.} \bibnamefont{{Kleyna}}},
  \bibinfo{author}{\bibfnamefont{M.~I.} \bibnamefont{{Wilkinson}}},
  \bibinfo{author}{\bibfnamefont{E.~K.} \bibnamefont{{Grebel}}},
  \bibinfo{author}{\bibfnamefont{G.~F.} \bibnamefont{{Gilmore}}},
  \bibinfo{author}{\bibfnamefont{N.~W.} \bibnamefont{{Evans}}},
  \bibinfo{author}{\bibfnamefont{R.~F.~G.} \bibnamefont{{Wyse}}},
  \bibnamefont{and} \bibinfo{author}{\bibfnamefont{D.~R.}
  \bibnamefont{{Harbeck}}}, \bibinfo{journal}{\aj}
  \textbf{\bibinfo{volume}{134}}, \bibinfo{pages}{566}
  (\bibinfo{year}{2007}{\natexlab{b}}), \eprint{0704.3437}.

\bibitem[{\citenamefont{{Battaglia} et~al.}(2008)\citenamefont{{Battaglia},
  {Helmi}, {Tolstoy}, {Irwin}, {Hill}, and {Jablonka}}}]{Battaglia08}
\bibinfo{author}{\bibfnamefont{G.}~\bibnamefont{{Battaglia}}},
  \bibinfo{author}{\bibfnamefont{A.}~\bibnamefont{{Helmi}}},
  \bibinfo{author}{\bibfnamefont{E.}~\bibnamefont{{Tolstoy}}},
  \bibinfo{author}{\bibfnamefont{M.}~\bibnamefont{{Irwin}}},
  \bibinfo{author}{\bibfnamefont{V.}~\bibnamefont{{Hill}}}, \bibnamefont{and}
  \bibinfo{author}{\bibfnamefont{P.}~\bibnamefont{{Jablonka}}},
  \bibinfo{journal}{\apjl} \textbf{\bibinfo{volume}{681}}, \bibinfo{pages}{L13}
  (\bibinfo{year}{2008}), \eprint{0802.4220}.

\bibitem[{\citenamefont{{Walker} et~al.}(2007)\citenamefont{{Walker}, {Mateo},
  {Olszewski}, {Gnedin}, {Wang}, {Sen}, and {Woodroofe}}}]{Walker07}
\bibinfo{author}{\bibfnamefont{M.~G.} \bibnamefont{{Walker}}},
  \bibinfo{author}{\bibfnamefont{M.}~\bibnamefont{{Mateo}}},
  \bibinfo{author}{\bibfnamefont{E.~W.} \bibnamefont{{Olszewski}}},
  \bibinfo{author}{\bibfnamefont{O.~Y.} \bibnamefont{{Gnedin}}},
  \bibinfo{author}{\bibfnamefont{X.}~\bibnamefont{{Wang}}},
  \bibinfo{author}{\bibfnamefont{B.}~\bibnamefont{{Sen}}}, \bibnamefont{and}
  \bibinfo{author}{\bibfnamefont{M.}~\bibnamefont{{Woodroofe}}},
  \bibinfo{journal}{\apjl} \textbf{\bibinfo{volume}{667}}, \bibinfo{pages}{L53}
  (\bibinfo{year}{2007}), \eprint{0708.0010}.

\bibitem[{\citenamefont{{Navarro} et~al.}(1997)\citenamefont{{Navarro},
  {Frenk}, and {White}}}]{navarro97}
\bibinfo{author}{\bibfnamefont{J.~F.} \bibnamefont{{Navarro}}},
  \bibinfo{author}{\bibfnamefont{C.~S.} \bibnamefont{{Frenk}}},
  \bibnamefont{and} \bibinfo{author}{\bibfnamefont{S.~D.~M.}
  \bibnamefont{{White}}}, \bibinfo{journal}{\apj}
  \textbf{\bibinfo{volume}{490}}, \bibinfo{pages}{493} (\bibinfo{year}{1997}),
  \eprint{arXiv:astro-ph/9611107}.

\bibitem[{\citenamefont{{Evans} et~al.}(2009)\citenamefont{{Evans}, {An}, and
  {Walker}}}]{Evans09}
\bibinfo{author}{\bibfnamefont{N.~W.} \bibnamefont{{Evans}}},
  \bibinfo{author}{\bibfnamefont{J.}~\bibnamefont{{An}}}, \bibnamefont{and}
  \bibinfo{author}{\bibfnamefont{M.~G.} \bibnamefont{{Walker}}},
  \bibinfo{journal}{\mnras} \textbf{\bibinfo{volume}{393}},
  \bibinfo{pages}{L50} (\bibinfo{year}{2009}), \eprint{0811.1488}.

\bibitem[{\citenamefont{{Goerdt} et~al.}(2006)\citenamefont{{Goerdt}, {Moore},
  {Read}, {Stadel}, and {Zemp}}}]{Goerdt06}
\bibinfo{author}{\bibfnamefont{T.}~\bibnamefont{{Goerdt}}},
  \bibinfo{author}{\bibfnamefont{B.}~\bibnamefont{{Moore}}},
  \bibinfo{author}{\bibfnamefont{J.~I.} \bibnamefont{{Read}}},
  \bibinfo{author}{\bibfnamefont{J.}~\bibnamefont{{Stadel}}}, \bibnamefont{and}
  \bibinfo{author}{\bibfnamefont{M.}~\bibnamefont{{Zemp}}},
  \bibinfo{journal}{\mnras} \textbf{\bibinfo{volume}{368}},
  \bibinfo{pages}{1073} (\bibinfo{year}{2006}),
  \eprint{arXiv:astro-ph/0601404}.

\bibitem[{\citenamefont{Amorisco and Evans}(2011)}]{Amorisco11}
\bibinfo{author}{\bibfnamefont{N.}~\bibnamefont{Amorisco}} \bibnamefont{and}
  \bibinfo{author}{\bibfnamefont{N.}~\bibnamefont{Evans}},
  \bibinfo{journal}{Mon.Not.Roy.Astron.Soc.} \textbf{\bibinfo{volume}{411}},
  \bibinfo{pages}{2118} (\bibinfo{year}{2011}), \eprint{1009.1813}.

\bibitem[{\citenamefont{Walker and Penarrubia}(2011)}]{Walker11}
\bibinfo{author}{\bibfnamefont{M.~G.} \bibnamefont{Walker}} \bibnamefont{and}
  \bibinfo{author}{\bibfnamefont{J.}~\bibnamefont{Penarrubia}},
  \bibinfo{journal}{Astrophys.J.} \textbf{\bibinfo{volume}{742}},
  \bibinfo{pages}{20} (\bibinfo{year}{2011}), \eprint{1108.2404}.

\bibitem[{\citenamefont{Gentile et~al.}(2005)\citenamefont{Gentile, Burkert,
  Salucci, Klein, and Walter}}]{Gentile:2005de}
\bibinfo{author}{\bibfnamefont{G.}~\bibnamefont{Gentile}},
  \bibinfo{author}{\bibfnamefont{A.}~\bibnamefont{Burkert}},
  \bibinfo{author}{\bibfnamefont{P.}~\bibnamefont{Salucci}},
  \bibinfo{author}{\bibfnamefont{U.}~\bibnamefont{Klein}}, \bibnamefont{and}
  \bibinfo{author}{\bibfnamefont{F.}~\bibnamefont{Walter}},
  \bibinfo{journal}{Astrophys.J.Lett.}  (\bibinfo{year}{2005}),
  \eprint{astro-ph/0506538}.

\bibitem[{\citenamefont{Maccio' et~al.}(2011)\citenamefont{Maccio', Stinson,
  Brook, Wadsley, Couchman et~al.}}]{Maccio':2011eh}
\bibinfo{author}{\bibfnamefont{A.~V.} \bibnamefont{Maccio'}},
  \bibinfo{author}{\bibfnamefont{G.}~\bibnamefont{Stinson}},
  \bibinfo{author}{\bibfnamefont{C.~B.} \bibnamefont{Brook}},
  \bibinfo{author}{\bibfnamefont{J.}~\bibnamefont{Wadsley}},
  \bibinfo{author}{\bibfnamefont{H.}~\bibnamefont{Couchman}},
  \bibnamefont{et~al.} (\bibinfo{year}{2011}), \eprint{1111.5620}.

\bibitem[{\citenamefont{Ragone-Figueroa
  et~al.}(2012)\citenamefont{Ragone-Figueroa, Granato, and
  Abadi}}]{RagoneFigueroa:2012vc}
\bibinfo{author}{\bibfnamefont{C.}~\bibnamefont{Ragone-Figueroa}},
  \bibinfo{author}{\bibfnamefont{G.~L.} \bibnamefont{Granato}},
  \bibnamefont{and} \bibinfo{author}{\bibfnamefont{M.~G.} \bibnamefont{Abadi}}
  (\bibinfo{year}{2012}), \eprint{1202.1527}.

\bibitem[{\citenamefont{{Salucci} et~al.}(2012)\citenamefont{{Salucci},
  {Wilkinson}, {Walker}, {Gilmore}, {Grebel}, {Koch}, {Frigerio Martins}, and
  {Wyse}}}]{2012MNRAS.tmp.2161S}
\bibinfo{author}{\bibfnamefont{P.}~\bibnamefont{{Salucci}}},
  \bibinfo{author}{\bibfnamefont{M.~I.} \bibnamefont{{Wilkinson}}},
  \bibinfo{author}{\bibfnamefont{M.~G.} \bibnamefont{{Walker}}},
  \bibinfo{author}{\bibfnamefont{G.~F.} \bibnamefont{{Gilmore}}},
  \bibinfo{author}{\bibfnamefont{E.~K.} \bibnamefont{{Grebel}}},
  \bibinfo{author}{\bibfnamefont{A.}~\bibnamefont{{Koch}}},
  \bibinfo{author}{\bibfnamefont{C.}~\bibnamefont{{Frigerio Martins}}},
  \bibnamefont{and} \bibinfo{author}{\bibfnamefont{R.~F.~G.}
  \bibnamefont{{Wyse}}}, \bibinfo{journal}{\mnras} p. \bibinfo{pages}{2161}
  (\bibinfo{year}{2012}), \eprint{1111.1165}.

\bibitem[{\citenamefont{{Binney} and {Tremaine}}(2008)}]{bt08}
\bibinfo{author}{\bibfnamefont{J.}~\bibnamefont{{Binney}}} \bibnamefont{and}
  \bibinfo{author}{\bibfnamefont{S.}~\bibnamefont{{Tremaine}}},
  \emph{\bibinfo{title}{{Galactic Dynamics: Second Edition}}}
  (\bibinfo{publisher}{Princeton University Press}, \bibinfo{year}{2008}).

\bibitem[{\citenamefont{{Mamon} and {{\L}okas}}(2005)}]{mamon05}
\bibinfo{author}{\bibfnamefont{G.~A.} \bibnamefont{{Mamon}}} \bibnamefont{and}
  \bibinfo{author}{\bibfnamefont{E.~L.} \bibnamefont{{{\L}okas}}},
  \bibinfo{journal}{\mnras} \textbf{\bibinfo{volume}{363}},
  \bibinfo{pages}{705} (\bibinfo{year}{2005}), \eprint{arXiv:astro-ph/0405491}.

\bibitem[{\citenamefont{Donato et~al.}(2009{\natexlab{b}})\citenamefont{Donato,
  Gentile, Salucci, Martins, Wilkinson et~al.}}]{Donato:2009ab}
\bibinfo{author}{\bibfnamefont{F.}~\bibnamefont{Donato}},
  \bibinfo{author}{\bibfnamefont{G.}~\bibnamefont{Gentile}},
  \bibinfo{author}{\bibfnamefont{P.}~\bibnamefont{Salucci}},
  \bibinfo{author}{\bibfnamefont{C.}~\bibnamefont{Martins}},
  \bibinfo{author}{\bibfnamefont{M.}~\bibnamefont{Wilkinson}},
  \bibnamefont{et~al.} (\bibinfo{year}{2009}{\natexlab{b}}),
  \eprint{0904.4054}.

\bibitem[{\citenamefont{Charbonnier et~al.}(2011)\citenamefont{Charbonnier,
  Combet, Daniel, Funk, Hinton et~al.}}]{Charbonnier:2011ft}
\bibinfo{author}{\bibfnamefont{A.}~\bibnamefont{Charbonnier}},
  \bibinfo{author}{\bibfnamefont{C.}~\bibnamefont{Combet}},
  \bibinfo{author}{\bibfnamefont{M.}~\bibnamefont{Daniel}},
  \bibinfo{author}{\bibfnamefont{S.}~\bibnamefont{Funk}},
  \bibinfo{author}{\bibfnamefont{J.}~\bibnamefont{Hinton}},
  \bibnamefont{et~al.}, \bibinfo{journal}{Mon.Not.Roy.Astron.Soc.}
  \textbf{\bibinfo{volume}{418}}, \bibinfo{pages}{1526} (\bibinfo{year}{2011}),
  \eprint{1104.0412}.

\bibitem[{\citenamefont{Abramowski et~al.}(2011{\natexlab{a}})}]{:2010zzt}
\bibinfo{author}{\bibfnamefont{A.}~\bibnamefont{Abramowski}}
  \bibnamefont{et~al.} (\bibinfo{collaboration}{HESS Collaboration}),
  \bibinfo{journal}{Astropart.Phys.} \textbf{\bibinfo{volume}{34}},
  \bibinfo{pages}{608} (\bibinfo{year}{2011}{\natexlab{a}}),
  \eprint{1012.5602}.

\bibitem[{\citenamefont{Abdo et~al.}(2010{\natexlab{b}})}]{Abdo:2010nz}
\bibinfo{author}{\bibfnamefont{A.}~\bibnamefont{Abdo}} \bibnamefont{et~al.}
  (\bibinfo{collaboration}{The Fermi-LAT collaboration}),
  \bibinfo{journal}{Phys.Rev.Lett.} \textbf{\bibinfo{volume}{104}},
  \bibinfo{pages}{101101} (\bibinfo{year}{2010}{\natexlab{b}}),
  \eprint{1002.3603}.

\bibitem[{\citenamefont{Strong et~al.}(2004)\citenamefont{Strong, Moskalenko,
  and Reimer}}]{Strong:2004de}
\bibinfo{author}{\bibfnamefont{A.~W.} \bibnamefont{Strong}},
  \bibinfo{author}{\bibfnamefont{I.~V.} \bibnamefont{Moskalenko}},
  \bibnamefont{and} \bibinfo{author}{\bibfnamefont{O.}~\bibnamefont{Reimer}},
  \bibinfo{journal}{Astrophys. J.} \textbf{\bibinfo{volume}{613}},
  \bibinfo{pages}{962} (\bibinfo{year}{2004}), \eprint{astro-ph/0406254}.

\bibitem[{\citenamefont{Cholis et~al.}(2012)\citenamefont{Cholis, Tavakoli,
  Evoli, Maccione, and Ullio}}]{Cholis:2011un}
\bibinfo{author}{\bibfnamefont{I.}~\bibnamefont{Cholis}},
  \bibinfo{author}{\bibfnamefont{M.}~\bibnamefont{Tavakoli}},
  \bibinfo{author}{\bibfnamefont{C.}~\bibnamefont{Evoli}},
  \bibinfo{author}{\bibfnamefont{L.}~\bibnamefont{Maccione}}, \bibnamefont{and}
  \bibinfo{author}{\bibfnamefont{P.}~\bibnamefont{Ullio}},
  \bibinfo{journal}{JCAP} \textbf{\bibinfo{volume}{1205}}, \bibinfo{pages}{004}
  (\bibinfo{year}{2012}), \eprint{1106.5073}.

\bibitem[{\citenamefont{{Abdo} et~al.}(2009)\citenamefont{{Abdo}, {Ackermann},
  {Ajello}, {Atwood}, {Axelsson}, {Baldini}, {Ballet}, {Barbiellini}, {Baring},
  {Bastieri} et~al.}}]{2009Sci...325..848A}
\bibinfo{author}{\bibfnamefont{A.~A.} \bibnamefont{{Abdo}}},
  \bibinfo{author}{\bibfnamefont{M.}~\bibnamefont{{Ackermann}}},
  \bibinfo{author}{\bibfnamefont{M.}~\bibnamefont{{Ajello}}},
  \bibinfo{author}{\bibfnamefont{W.~B.} \bibnamefont{{Atwood}}},
  \bibinfo{author}{\bibfnamefont{M.}~\bibnamefont{{Axelsson}}},
  \bibinfo{author}{\bibfnamefont{L.}~\bibnamefont{{Baldini}}},
  \bibinfo{author}{\bibfnamefont{J.}~\bibnamefont{{Ballet}}},
  \bibinfo{author}{\bibfnamefont{G.}~\bibnamefont{{Barbiellini}}},
  \bibinfo{author}{\bibfnamefont{M.~G.} \bibnamefont{{Baring}}},
  \bibinfo{author}{\bibfnamefont{D.}~\bibnamefont{{Bastieri}}},
  \bibnamefont{et~al.}, \bibinfo{journal}{Science}
  \textbf{\bibinfo{volume}{325}}, \bibinfo{pages}{848} (\bibinfo{year}{2009}).

\bibitem[{\citenamefont{Siegal-Gaskins
  et~al.}(2010)\citenamefont{Siegal-Gaskins, Reesman, Pavlidou, Profumo, and
  Walker}}]{SiegalGaskins:2010mp}
\bibinfo{author}{\bibfnamefont{J.~M.} \bibnamefont{Siegal-Gaskins}},
  \bibinfo{author}{\bibfnamefont{R.}~\bibnamefont{Reesman}},
  \bibinfo{author}{\bibfnamefont{V.}~\bibnamefont{Pavlidou}},
  \bibinfo{author}{\bibfnamefont{S.}~\bibnamefont{Profumo}}, \bibnamefont{and}
  \bibinfo{author}{\bibfnamefont{T.~P.} \bibnamefont{Walker}}
  (\bibinfo{year}{2010}), \eprint{1011.5501}.

\bibitem[{\citenamefont{Calore et~al.}(2012)\citenamefont{Calore, De~Romeri,
  and Donato}}]{Calore:2011bt}
\bibinfo{author}{\bibfnamefont{F.}~\bibnamefont{Calore}},
  \bibinfo{author}{\bibfnamefont{V.}~\bibnamefont{De~Romeri}},
  \bibnamefont{and} \bibinfo{author}{\bibfnamefont{F.}~\bibnamefont{Donato}},
  \bibinfo{journal}{Phys.Rev.} \textbf{\bibinfo{volume}{D85}},
  \bibinfo{pages}{023004} (\bibinfo{year}{2012}), \bibinfo{note}{10 pages, 6
  figures Version updated, as sent to PRD}, \eprint{1105.4230}.

\bibitem[{\citenamefont{Malyshev et~al.}(2010)\citenamefont{Malyshev, Cholis,
  and Gelfand}}]{Malyshev:2010xc}
\bibinfo{author}{\bibfnamefont{D.}~\bibnamefont{Malyshev}},
  \bibinfo{author}{\bibfnamefont{I.}~\bibnamefont{Cholis}}, \bibnamefont{and}
  \bibinfo{author}{\bibfnamefont{J.~D.} \bibnamefont{Gelfand}},
  \bibinfo{journal}{Astrophys.J.} \textbf{\bibinfo{volume}{722}},
  \bibinfo{pages}{1939} (\bibinfo{year}{2010}), \eprint{1002.0587}.

\bibitem[{\citenamefont{{The~Fermi-LAT~Collaboration}}(2010)}]{Collaboration:2010ru}
\bibinfo{author}{\bibnamefont{{The~Fermi-LAT~Collaboration}}},
  \bibinfo{journal}{Astrophys.J.Suppl.} \textbf{\bibinfo{volume}{188}},
  \bibinfo{pages}{405} (\bibinfo{year}{2010}), \eprint{1002.2280}.

\bibitem[{\citenamefont{{The~Fermi-LAT~Collaboration}}(2012)}]{Collaboration:2011bm}
\bibinfo{author}{\bibnamefont{{The~Fermi-LAT~Collaboration}}}
  (\bibinfo{collaboration}{Fermi-LAT Collaboration}),
  \bibinfo{journal}{Astrophys.J.Suppl.} \textbf{\bibinfo{volume}{199}},
  \bibinfo{pages}{31} (\bibinfo{year}{2012}), \eprint{1108.1435}.

\bibitem[{\citenamefont{Abdo et~al.}(2009)}]{Abdo:2009mg}
\bibinfo{author}{\bibfnamefont{A.~A.} \bibnamefont{Abdo}} \bibnamefont{et~al.}
  (\bibinfo{collaboration}{Fermi LAT Collaboration}),
  \bibinfo{journal}{Astrophys.J.Suppl.} \textbf{\bibinfo{volume}{183}},
  \bibinfo{pages}{46} (\bibinfo{year}{2009}), \eprint{0902.1340}.

\bibitem[{\citenamefont{Abramowski
  et~al.}(2011{\natexlab{b}})}]{Abramowski:2011hc}
\bibinfo{author}{\bibfnamefont{A.}~\bibnamefont{Abramowski}}
  \bibnamefont{et~al.} (\bibinfo{collaboration}{H.E.S.S.Collaboration}),
  \bibinfo{journal}{Phys.Rev.Lett.} \textbf{\bibinfo{volume}{106}},
  \bibinfo{pages}{161301} (\bibinfo{year}{2011}{\natexlab{b}}),
  \eprint{1103.3266}.

\bibitem[{\citenamefont{Aharonian
  et~al.}(2006{\natexlab{a}})}]{Aharonian:2006au}
\bibinfo{author}{\bibfnamefont{F.}~\bibnamefont{Aharonian}}
  \bibnamefont{et~al.} (\bibinfo{collaboration}{H.E.S.S. Collaboration}),
  \bibinfo{journal}{Nature} \textbf{\bibinfo{volume}{439}},
  \bibinfo{pages}{695} (\bibinfo{year}{2006}{\natexlab{a}}),
  \eprint{astro-ph/0603021}.

\bibitem[{\citenamefont{Aharonian
  et~al.}(2006{\natexlab{b}})}]{Aharonian:2006wh}
\bibinfo{author}{\bibfnamefont{F.}~\bibnamefont{Aharonian}}
  \bibnamefont{et~al.} (\bibinfo{collaboration}{H.E.S.S. Collaboration}),
  \bibinfo{journal}{Phys.Rev.Lett.} \textbf{\bibinfo{volume}{97}},
  \bibinfo{pages}{221102} (\bibinfo{year}{2006}{\natexlab{b}}),
  \eprint{astro-ph/0610509}.

\bibitem[{Pla(2011)}]{Planck}
\bibinfo{journal}{http://www.sciops.esa.int/SA/PLANCK/docs/ERCSC\_explanatory\_supplement.pdf}
   (\bibinfo{year}{2011}).

\bibitem[{\citenamefont{{Planck Collaboration}
  et~al.}(2011)\citenamefont{{Planck Collaboration}, {Ade}, {Aghanim},
  {Arnaud}, {Ashdown}, {Aumont}, {Baccigalupi}, {Balbi}, {Banday}, {Barreiro}
  et~al.}}]{2011A&A...536A...7P}
\bibinfo{author}{\bibnamefont{{Planck Collaboration}}},
  \bibinfo{author}{\bibfnamefont{P.~A.~R.} \bibnamefont{{Ade}}},
  \bibinfo{author}{\bibfnamefont{N.}~\bibnamefont{{Aghanim}}},
  \bibinfo{author}{\bibfnamefont{M.}~\bibnamefont{{Arnaud}}},
  \bibinfo{author}{\bibfnamefont{M.}~\bibnamefont{{Ashdown}}},
  \bibinfo{author}{\bibfnamefont{J.}~\bibnamefont{{Aumont}}},
  \bibinfo{author}{\bibfnamefont{C.}~\bibnamefont{{Baccigalupi}}},
  \bibinfo{author}{\bibfnamefont{A.}~\bibnamefont{{Balbi}}},
  \bibinfo{author}{\bibfnamefont{A.~J.} \bibnamefont{{Banday}}},
  \bibinfo{author}{\bibfnamefont{R.~B.} \bibnamefont{{Barreiro}}},
  \bibnamefont{et~al.}, \bibinfo{journal}{\aap} \textbf{\bibinfo{volume}{536}},
  \bibinfo{eid}{A7} (\bibinfo{year}{2011}), \eprint{1101.2041}.

\bibitem[{\citenamefont{Haakonsen and Rutledge}(2009)}]{Haakonsen:2009mg}
\bibinfo{author}{\bibfnamefont{C.~B.} \bibnamefont{Haakonsen}}
  \bibnamefont{and} \bibinfo{author}{\bibfnamefont{R.~E.}
  \bibnamefont{Rutledge}}, \bibinfo{journal}{Astrophys.J.Suppl.}
  \textbf{\bibinfo{volume}{184}}, \bibinfo{pages}{138} (\bibinfo{year}{2009}),
  \eprint{0910.3229}.

\bibitem[{\citenamefont{Shirahata et~al.}(2009)\citenamefont{Shirahata,
  Matsuura, Hasegawa, Ootsubo, Makiuti et~al.}}]{Shirahata:2009ui}
\bibinfo{author}{\bibfnamefont{M.}~\bibnamefont{Shirahata}},
  \bibinfo{author}{\bibfnamefont{S.}~\bibnamefont{Matsuura}},
  \bibinfo{author}{\bibfnamefont{S.}~\bibnamefont{Hasegawa}},
  \bibinfo{author}{\bibfnamefont{T.}~\bibnamefont{Ootsubo}},
  \bibinfo{author}{\bibfnamefont{S.}~\bibnamefont{Makiuti}},
  \bibnamefont{et~al.} (\bibinfo{year}{2009}), \eprint{0904.3788}.

\bibitem[{\citenamefont{Sjostrand et~al.}(2006)\citenamefont{Sjostrand, Mrenna,
  and Skands}}]{Sjostrand:2006za}
\bibinfo{author}{\bibfnamefont{T.}~\bibnamefont{Sjostrand}},
  \bibinfo{author}{\bibfnamefont{S.}~\bibnamefont{Mrenna}}, \bibnamefont{and}
  \bibinfo{author}{\bibfnamefont{P.~Z.} \bibnamefont{Skands}},
  \bibinfo{journal}{JHEP} \textbf{\bibinfo{volume}{0605}}, \bibinfo{pages}{026}
  (\bibinfo{year}{2006}), \eprint{hep-ph/0603175}.

\bibitem[{\citenamefont{et~al.}(2010)}]{PDG2010}
\bibinfo{author}{\bibfnamefont{K.~N.} \bibnamefont{et~al.}},
  \bibinfo{journal}{J. Phys. G 37, 075021 (2010) and 2011 partial update for
  the 2012 edition}  (\bibinfo{year}{2010}).

\bibitem[{\citenamefont{Ciafaloni et~al.}(2011)\citenamefont{Ciafaloni,
  Comelli, Riotto, Sala, Strumia et~al.}}]{Ciafaloni:2010ti}
\bibinfo{author}{\bibfnamefont{P.}~\bibnamefont{Ciafaloni}},
  \bibinfo{author}{\bibfnamefont{D.}~\bibnamefont{Comelli}},
  \bibinfo{author}{\bibfnamefont{A.}~\bibnamefont{Riotto}},
  \bibinfo{author}{\bibfnamefont{F.}~\bibnamefont{Sala}},
  \bibinfo{author}{\bibfnamefont{A.}~\bibnamefont{Strumia}},
  \bibnamefont{et~al.}, \bibinfo{journal}{JCAP}
  \textbf{\bibinfo{volume}{1103}}, \bibinfo{pages}{019} (\bibinfo{year}{2011}),
  \eprint{1009.0224}.

\bibitem[{\citenamefont{Hryczuk and Iengo}(2012)}]{Hryczuk:2011vi}
\bibinfo{author}{\bibfnamefont{A.}~\bibnamefont{Hryczuk}} \bibnamefont{and}
  \bibinfo{author}{\bibfnamefont{R.}~\bibnamefont{Iengo}},
  \bibinfo{journal}{JHEP} \textbf{\bibinfo{volume}{1201}}, \bibinfo{pages}{163}
  (\bibinfo{year}{2012}), \eprint{1111.2916}.

\bibitem[{\citenamefont{Ciafaloni et~al.}(2012)\citenamefont{Ciafaloni,
  Comelli, De~Simone, Riotto, and Urbano}}]{Ciafaloni:2012gs}
\bibinfo{author}{\bibfnamefont{P.}~\bibnamefont{Ciafaloni}},
  \bibinfo{author}{\bibfnamefont{D.}~\bibnamefont{Comelli}},
  \bibinfo{author}{\bibfnamefont{A.}~\bibnamefont{De~Simone}},
  \bibinfo{author}{\bibfnamefont{A.}~\bibnamefont{Riotto}}, \bibnamefont{and}
  \bibinfo{author}{\bibfnamefont{A.}~\bibnamefont{Urbano}},
  \bibinfo{journal}{JCAP} \textbf{\bibinfo{volume}{1206}}, \bibinfo{pages}{016}
  (\bibinfo{year}{2012}), \eprint{1202.0692}.

\bibitem[{\citenamefont{Strong et~al.}(2007)\citenamefont{Strong, Moskalenko,
  and Ptuskin}}]{Strong:2007nh}
\bibinfo{author}{\bibfnamefont{A.~W.} \bibnamefont{Strong}},
  \bibinfo{author}{\bibfnamefont{I.~V.} \bibnamefont{Moskalenko}},
  \bibnamefont{and} \bibinfo{author}{\bibfnamefont{V.~S.}
  \bibnamefont{Ptuskin}}, \bibinfo{journal}{Ann. Rev. Nucl. Part. Sci.}
  \textbf{\bibinfo{volume}{57}}, \bibinfo{pages}{285} (\bibinfo{year}{2007}),
  \eprint{astro-ph/0701517}.

\bibitem[{\citenamefont{{Delahaye} et~al.}(2011)\citenamefont{{Delahaye},
  {Fiasson}, {Pohl}, and {Salati}}}]{2011A&A...531A..37D}
\bibinfo{author}{\bibfnamefont{T.}~\bibnamefont{{Delahaye}}},
  \bibinfo{author}{\bibfnamefont{A.}~\bibnamefont{{Fiasson}}},
  \bibinfo{author}{\bibfnamefont{M.}~\bibnamefont{{Pohl}}}, \bibnamefont{and}
  \bibinfo{author}{\bibfnamefont{P.}~\bibnamefont{{Salati}}},
  \bibinfo{journal}{\aap} \textbf{\bibinfo{volume}{531}}, \bibinfo{eid}{A37}
  (\bibinfo{year}{2011}), \eprint{1102.0744}.

\bibitem[{\citenamefont{{The Fermi-LAT
  Collaboration}}(2012)}]{2012arXiv1202.4039T}
\bibinfo{author}{\bibnamefont{{The Fermi-LAT Collaboration}}}
  (\bibinfo{collaboration}{Fermi-LAT Collaboration}),
  \bibinfo{journal}{Astrophys.J.} \textbf{\bibinfo{volume}{750}},
  \bibinfo{pages}{3} (\bibinfo{year}{2012}), \eprint{1202.4039}.

\bibitem[{\citenamefont{Nakanishi and Sofue}(2006)}]{Nakanishi:2006zf}
\bibinfo{author}{\bibfnamefont{H.}~\bibnamefont{Nakanishi}} \bibnamefont{and}
  \bibinfo{author}{\bibfnamefont{Y.}~\bibnamefont{Sofue}},
  \bibinfo{journal}{Publ.Astron.Soc.Jap.}  (\bibinfo{year}{2006}),
  \eprint{astro-ph/0610769}.

\bibitem[{\citenamefont{Nakanishi and Sofue}(2003)}]{Nakanishi:2003eb}
\bibinfo{author}{\bibfnamefont{H.}~\bibnamefont{Nakanishi}} \bibnamefont{and}
  \bibinfo{author}{\bibfnamefont{Y.}~\bibnamefont{Sofue}},
  \bibinfo{journal}{Publ.Astron.Soc.Jap.} \textbf{\bibinfo{volume}{55}},
  \bibinfo{pages}{191} (\bibinfo{year}{2003}), \eprint{astro-ph/0304338}.

\bibitem[{\citenamefont{{Pohl} et~al.}(2008)\citenamefont{{Pohl}, {Englmaier},
  and {Bissantz}}}]{2008ApJ...677..283P}
\bibinfo{author}{\bibfnamefont{M.}~\bibnamefont{{Pohl}}},
  \bibinfo{author}{\bibfnamefont{P.}~\bibnamefont{{Englmaier}}},
  \bibnamefont{and}
  \bibinfo{author}{\bibfnamefont{N.}~\bibnamefont{{Bissantz}}},
  \bibinfo{journal}{\apj} \textbf{\bibinfo{volume}{677}}, \bibinfo{pages}{283}
  (\bibinfo{year}{2008}), \eprint{0712.4264}.

\bibitem[{\citenamefont{Moskalenko et~al.}(2007)\citenamefont{Moskalenko,
  Digel, Porter, Reimer, and Strong}}]{Moskalenko:2006zy}
\bibinfo{author}{\bibfnamefont{I.~V.} \bibnamefont{Moskalenko}},
  \bibinfo{author}{\bibfnamefont{S.}~\bibnamefont{Digel}},
  \bibinfo{author}{\bibfnamefont{T.}~\bibnamefont{Porter}},
  \bibinfo{author}{\bibfnamefont{O.}~\bibnamefont{Reimer}}, \bibnamefont{and}
  \bibinfo{author}{\bibfnamefont{A.}~\bibnamefont{Strong}},
  \bibinfo{journal}{Nucl.Phys.Proc.Suppl.} \textbf{\bibinfo{volume}{173}},
  \bibinfo{pages}{44} (\bibinfo{year}{2007}), \eprint{astro-ph/0609768}.

\bibitem[{\citenamefont{Cirelli et~al.}(2011)\citenamefont{Cirelli, Corcella,
  Hektor, Hutsi, Kadastik et~al.}}]{Cirelli:2010xx}
\bibinfo{author}{\bibfnamefont{M.}~\bibnamefont{Cirelli}},
  \bibinfo{author}{\bibfnamefont{G.}~\bibnamefont{Corcella}},
  \bibinfo{author}{\bibfnamefont{A.}~\bibnamefont{Hektor}},
  \bibinfo{author}{\bibfnamefont{G.}~\bibnamefont{Hutsi}},
  \bibinfo{author}{\bibfnamefont{M.}~\bibnamefont{Kadastik}},
  \bibnamefont{et~al.}, \bibinfo{journal}{JCAP}
  \textbf{\bibinfo{volume}{1103}}, \bibinfo{pages}{051} (\bibinfo{year}{2011}),
  \bibinfo{note}{57 pages with many figures and tables. v2: several discussions
  and references added, some figures improved, matches version published on
  JCAP. v3: a few typos corrected and some references updated. All results are
  available at http://www.marcocirelli.net/PPPC4DMID.html}, \eprint{1012.4515}.

\bibitem[{\citenamefont{{{Vivier},
  M.~for~the~VERITAS~Collaboration}}(2011)}]{collaboration:2011sm}
\bibinfo{author}{\bibnamefont{{{Vivier}, M.~for~the~VERITAS~Collaboration}}}
  (\bibinfo{year}{2011}), \eprint{1110.6615}.

\bibitem[{\citenamefont{Aliu et~al.}(2012)}]{Aliu:2012ga}
\bibinfo{author}{\bibfnamefont{E.}~\bibnamefont{Aliu}} \bibnamefont{et~al.}
  (\bibinfo{collaboration}{VERITAS Collaboration}),
  \bibinfo{journal}{Phys.Rev.} \textbf{\bibinfo{volume}{D85}},
  \bibinfo{pages}{062001} (\bibinfo{year}{2012}), \eprint{1202.2144}.

\bibitem[{\citenamefont{Aleksic et~al.}(2011)}]{Aleksic:2011jx}
\bibinfo{author}{\bibfnamefont{J.}~\bibnamefont{Aleksic}} \bibnamefont{et~al.}
  (\bibinfo{collaboration}{The MAGIC Collaboration}), \bibinfo{journal}{JCAP}
  \textbf{\bibinfo{volume}{1106}}, \bibinfo{pages}{035} (\bibinfo{year}{2011}),
  \eprint{1103.0477}.

\bibitem[{\citenamefont{Paiano et~al.}(2011)\citenamefont{Paiano, Lombardi,
  Doro, Nieto, Collaboration et~al.}}]{Paiano:2011uq}
\bibinfo{author}{\bibfnamefont{S.}~\bibnamefont{Paiano}},
  \bibinfo{author}{\bibfnamefont{S.}~\bibnamefont{Lombardi}},
  \bibinfo{author}{\bibfnamefont{M.}~\bibnamefont{Doro}},
  \bibinfo{author}{\bibfnamefont{D.}~\bibnamefont{Nieto}},
  \bibinfo{author}{\bibfnamefont{f.~t.~M.} \bibnamefont{Collaboration}},
  \bibnamefont{et~al.} (\bibinfo{year}{2011}), \eprint{1110.6775}.

\bibitem[{\citenamefont{Bringmann et~al.}(2009)\citenamefont{Bringmann, Doro,
  and Fornasa}}]{Bringmann:2008kj}
\bibinfo{author}{\bibfnamefont{T.}~\bibnamefont{Bringmann}},
  \bibinfo{author}{\bibfnamefont{M.}~\bibnamefont{Doro}}, \bibnamefont{and}
  \bibinfo{author}{\bibfnamefont{M.}~\bibnamefont{Fornasa}},
  \bibinfo{journal}{JCAP} \textbf{\bibinfo{volume}{0901}}, \bibinfo{pages}{016}
  (\bibinfo{year}{2009}), \eprint{0809.2269}.

\bibitem[{\citenamefont{Tibolla}(2012)}]{Tibolla:2012zn}
\bibinfo{author}{\bibfnamefont{O.}~\bibnamefont{Tibolla}}
  (\bibinfo{collaboration}{for the MAGIC collaboration})
  (\bibinfo{year}{2012}), \eprint{1201.2295}.

\bibitem[{\citenamefont{De~Simone}(2012)}]{DeSimone:2012hj}
\bibinfo{author}{\bibfnamefont{A.}~\bibnamefont{De~Simone}}
  (\bibinfo{year}{2012}), \eprint{1201.1443}.

\bibitem[{\citenamefont{Cirelli et~al.}(2010)\citenamefont{Cirelli, Panci, and
  Serpico}}]{Cirelli:2009dv}
\bibinfo{author}{\bibfnamefont{M.}~\bibnamefont{Cirelli}},
  \bibinfo{author}{\bibfnamefont{P.}~\bibnamefont{Panci}}, \bibnamefont{and}
  \bibinfo{author}{\bibfnamefont{P.~D.} \bibnamefont{Serpico}},
  \bibinfo{journal}{Nucl.Phys.} \textbf{\bibinfo{volume}{B840}},
  \bibinfo{pages}{284} (\bibinfo{year}{2010}), \eprint{0912.0663}.

\bibitem[{\citenamefont{Cholis et~al.}(2009{\natexlab{b}})\citenamefont{Cholis,
  Dobler, Finkbeiner, Goodenough, and Weiner}}]{Cholis:2008wq}
\bibinfo{author}{\bibfnamefont{I.}~\bibnamefont{Cholis}},
  \bibinfo{author}{\bibfnamefont{G.}~\bibnamefont{Dobler}},
  \bibinfo{author}{\bibfnamefont{D.~P.} \bibnamefont{Finkbeiner}},
  \bibinfo{author}{\bibfnamefont{L.}~\bibnamefont{Goodenough}},
  \bibnamefont{and} \bibinfo{author}{\bibfnamefont{N.}~\bibnamefont{Weiner}},
  \bibinfo{journal}{Phys. Rev.} \textbf{\bibinfo{volume}{D80}},
  \bibinfo{pages}{123518} (\bibinfo{year}{2009}{\natexlab{b}}),
  \eprint{0811.3641}.

\bibitem[{\citenamefont{Cholis et~al.}(2009{\natexlab{c}})\citenamefont{Cholis,
  Dobler, Finkbeiner, Goodenough, Slatyer et~al.}}]{Cholis:2009gv}
\bibinfo{author}{\bibfnamefont{I.}~\bibnamefont{Cholis}},
  \bibinfo{author}{\bibfnamefont{G.}~\bibnamefont{Dobler}},
  \bibinfo{author}{\bibfnamefont{D.~P.} \bibnamefont{Finkbeiner}},
  \bibinfo{author}{\bibfnamefont{L.}~\bibnamefont{Goodenough}},
  \bibinfo{author}{\bibfnamefont{T.~R.} \bibnamefont{Slatyer}},
  \bibnamefont{et~al.} (\bibinfo{year}{2009}{\natexlab{c}}),
  \eprint{0907.3953}.

\bibitem[{\citenamefont{Papucci and Strumia}(2010)}]{Papucci:2009gd}
\bibinfo{author}{\bibfnamefont{M.}~\bibnamefont{Papucci}} \bibnamefont{and}
  \bibinfo{author}{\bibfnamefont{A.}~\bibnamefont{Strumia}},
  \bibinfo{journal}{JCAP} \textbf{\bibinfo{volume}{1003}}, \bibinfo{pages}{014}
  (\bibinfo{year}{2010}), \eprint{0912.0742}.

\bibitem[{\citenamefont{Meade et~al.}(2009)\citenamefont{Meade, Papucci, and
  Volansky}}]{Meade:2009rb}
\bibinfo{author}{\bibfnamefont{P.}~\bibnamefont{Meade}},
  \bibinfo{author}{\bibfnamefont{M.}~\bibnamefont{Papucci}}, \bibnamefont{and}
  \bibinfo{author}{\bibfnamefont{T.}~\bibnamefont{Volansky}},
  \bibinfo{journal}{JHEP} \textbf{\bibinfo{volume}{0912}}, \bibinfo{pages}{052}
  (\bibinfo{year}{2009}), \eprint{0901.2925}.

\bibitem[{\citenamefont{Abazajian and Harding}(2012)}]{Abazajian:2011ak}
\bibinfo{author}{\bibfnamefont{K.~N.} \bibnamefont{Abazajian}}
  \bibnamefont{and} \bibinfo{author}{\bibfnamefont{J.}~\bibnamefont{Harding}},
  \bibinfo{journal}{JCAP} \textbf{\bibinfo{volume}{1201}}, \bibinfo{pages}{041}
  (\bibinfo{year}{2012}), \bibinfo{note}{19 pages, 5 figures/ v3: Matches JCAP
  version/ includes discussion of numerical studies of the density profile of
  MW-type halos, updated references and comparisons}, \eprint{1110.6151}.

\bibitem[{\citenamefont{Cholis and Goodenough}(2010)}]{Cholis:2010px}
\bibinfo{author}{\bibfnamefont{I.}~\bibnamefont{Cholis}} \bibnamefont{and}
  \bibinfo{author}{\bibfnamefont{L.}~\bibnamefont{Goodenough}},
  \bibinfo{journal}{JCAP} \textbf{\bibinfo{volume}{1009}}, \bibinfo{pages}{010}
  (\bibinfo{year}{2010}), \eprint{1006.2089}.

\bibitem[{\citenamefont{Slatyer et~al.}(2011)\citenamefont{Slatyer, Toro, and
  Weiner}}]{Slatyer:2011kg}
\bibinfo{author}{\bibfnamefont{T.~R.} \bibnamefont{Slatyer}},
  \bibinfo{author}{\bibfnamefont{N.}~\bibnamefont{Toro}}, \bibnamefont{and}
  \bibinfo{author}{\bibfnamefont{N.}~\bibnamefont{Weiner}}
  (\bibinfo{year}{2011}), \bibinfo{note}{17 pages, 7 figures, submitted to PRD.
  v2 adds the KITP preprint number and a comment on accelerator searches for
  light gauge bosons}, \eprint{1107.3546}.

\bibitem[{\citenamefont{Hooper and Goodenough}(2011)}]{Hooper:2010mq}
\bibinfo{author}{\bibfnamefont{D.}~\bibnamefont{Hooper}} \bibnamefont{and}
  \bibinfo{author}{\bibfnamefont{L.}~\bibnamefont{Goodenough}},
  \bibinfo{journal}{Phys.Lett.} \textbf{\bibinfo{volume}{B697}},
  \bibinfo{pages}{412} (\bibinfo{year}{2011}), \eprint{1010.2752}.

\bibitem[{\citenamefont{Hooper and Linden}(2011)}]{Hooper:2011ti}
\bibinfo{author}{\bibfnamefont{D.}~\bibnamefont{Hooper}} \bibnamefont{and}
  \bibinfo{author}{\bibfnamefont{T.}~\bibnamefont{Linden}},
  \bibinfo{journal}{Phys.Rev.} \textbf{\bibinfo{volume}{D84}},
  \bibinfo{pages}{123005} (\bibinfo{year}{2011}), \bibinfo{note}{13 pages, 11
  figures}, \eprint{1110.0006}.

\bibitem[{\citenamefont{Hooper}(2012)}]{Hooper:2012ft}
\bibinfo{author}{\bibfnamefont{D.}~\bibnamefont{Hooper}}
  (\bibinfo{year}{2012}), \eprint{1201.1303}.

\bibitem[{\citenamefont{Pinzke et~al.}(2011)\citenamefont{Pinzke, Pfrommer, and
  Bergstrom}}]{Pinzke:2011ek}
\bibinfo{author}{\bibfnamefont{A.}~\bibnamefont{Pinzke}},
  \bibinfo{author}{\bibfnamefont{C.}~\bibnamefont{Pfrommer}}, \bibnamefont{and}
  \bibinfo{author}{\bibfnamefont{L.}~\bibnamefont{Bergstrom}},
  \bibinfo{journal}{Phys.Rev.} \textbf{\bibinfo{volume}{D84}},
  \bibinfo{pages}{123509} (\bibinfo{year}{2011}), \bibinfo{note}{43 pages, 23
  figures, 10 tables. Accepted for publication in Phys. Rev. D: streamlined
  paper, added a paragraph about detectability to introduction, few references
  added, and few typos corrected}, \eprint{1105.3240}.

\bibitem[{\citenamefont{Ando and Nagai}(2012)}]{Ando:2012vu}
\bibinfo{author}{\bibfnamefont{S.}~\bibnamefont{Ando}} \bibnamefont{and}
  \bibinfo{author}{\bibfnamefont{D.}~\bibnamefont{Nagai}},
  \bibinfo{journal}{JCAP} \textbf{\bibinfo{volume}{1207}}, \bibinfo{pages}{017}
  (\bibinfo{year}{2012}), \eprint{1201.0753}.

\bibitem[{\citenamefont{Han et~al.}(2012)\citenamefont{Han, Frenk, Eke, Gao,
  and White}}]{Han:2012au}
\bibinfo{author}{\bibfnamefont{J.}~\bibnamefont{Han}},
  \bibinfo{author}{\bibfnamefont{C.~S.} \bibnamefont{Frenk}},
  \bibinfo{author}{\bibfnamefont{V.~R.} \bibnamefont{Eke}},
  \bibinfo{author}{\bibfnamefont{L.}~\bibnamefont{Gao}}, \bibnamefont{and}
  \bibinfo{author}{\bibfnamefont{S.~D.} \bibnamefont{White}}
  (\bibinfo{year}{2012}), \eprint{1201.1003}.

\bibitem[{\citenamefont{Governato et~al.}(2012)\citenamefont{Governato,
  Zolotov, Pontzen, Christensen, Oh et~al.}}]{Governato:2012fa}
\bibinfo{author}{\bibfnamefont{F.}~\bibnamefont{Governato}},
  \bibinfo{author}{\bibfnamefont{A.}~\bibnamefont{Zolotov}},
  \bibinfo{author}{\bibfnamefont{A.}~\bibnamefont{Pontzen}},
  \bibinfo{author}{\bibfnamefont{C.}~\bibnamefont{Christensen}},
  \bibinfo{author}{\bibfnamefont{S.}~\bibnamefont{Oh}}, \bibnamefont{et~al.},
  \bibinfo{journal}{Mon.Not.Roy.Astron.Soc.} \textbf{\bibinfo{volume}{422}},
  \bibinfo{pages}{1231} (\bibinfo{year}{2012}), \eprint{1202.0554}.

\bibitem[{\citenamefont{Hisano et~al.}(2007)\citenamefont{Hisano, Matsumoto,
  Nagai, Saito, and Senami}}]{Hisano:2006nn}
\bibinfo{author}{\bibfnamefont{J.}~\bibnamefont{Hisano}},
  \bibinfo{author}{\bibfnamefont{S.}~\bibnamefont{Matsumoto}},
  \bibinfo{author}{\bibfnamefont{M.}~\bibnamefont{Nagai}},
  \bibinfo{author}{\bibfnamefont{O.}~\bibnamefont{Saito}}, \bibnamefont{and}
  \bibinfo{author}{\bibfnamefont{M.}~\bibnamefont{Senami}},
  \bibinfo{journal}{Phys.Lett.} \textbf{\bibinfo{volume}{B646}},
  \bibinfo{pages}{34} (\bibinfo{year}{2007}), \eprint{hep-ph/0610249}.

\bibitem[{\citenamefont{Lattanzi and Silk}(2009)}]{Lattanzi:2008qa}
\bibinfo{author}{\bibfnamefont{M.}~\bibnamefont{Lattanzi}} \bibnamefont{and}
  \bibinfo{author}{\bibfnamefont{J.~I.} \bibnamefont{Silk}},
  \bibinfo{journal}{Phys. Rev.} \textbf{\bibinfo{volume}{D79}},
  \bibinfo{pages}{083523} (\bibinfo{year}{2009}), \eprint{0812.0360}.

\bibitem[{\citenamefont{{Sommerfeld}}(1931)}]{SommerfeldRef}
\bibinfo{author}{\bibfnamefont{A.}~\bibnamefont{{Sommerfeld}}},
  \bibinfo{journal}{Annalen der Physik} \textbf{\bibinfo{volume}{403}},
  \bibinfo{pages}{257} (\bibinfo{year}{1931}).

\end{thebibliography}
\bibliographystyle{apsrev}

\end{document}